\documentclass[useAMS,usenatbib,usegraphicx]{emulateapj}
\usepackage[usenames]{color}

\newcommand{\IFCA}{IFCA, Instituto de F\'isica de Cantabria (UC-CSIC), Av. de Los Castros s/n, 39005 Santander, Spain}
\newcommand{\HAWAII}{Institute for Astronomy, University of Hawaii, 2680 Woodlawn Drive, Honolulu, HI 96822-1839, USA}
\newcommand{\BILBAO}{Fisika Teorikoa, Zientzia eta Teknologia Fakultatea, Euskal Herriko Unibertsitatea UPV/EHU, E-48080 Bilbao, Spain}
\newcommand{\IKERBASQUE}{IKERBASQUE, Basque Foundation for Science, Alameda Urquijo, 36-5 48008 Bilbao, Spain}
\newcommand{\BERKELEY}{Department of Astronomy, University of California, Berkeley, CA 94720-3411, USA}
\newcommand{\MINNESOTA}{School of Physics and Astronomy, University of Minnesota, 116 Church Street SE, Minneapolis, MN 55455, USA}
\newcommand{\SCAROLINA}{Department of Physics and Astronomy, University of South Carolina, 712 Main St., Columbia, SC 29208, USA}
\newcommand{\UCLA}{Department of Physics and Astronomy, University of California, Los Angeles, CA 90095-1547, USA}
\newcommand{\TOKYOA}{Research Center for the Early Universe, University of Tokyo, 7-3-1 Hongo, Bunkyo-ku, Tokyo 113-0033, Japan}
\newcommand{\TOKYOB}{Department of Physics, University of Tokyo, 7-3-1 Hongo, Bunkyo-ku, Tokyo 113-0033, Japan}
\newcommand{\KAVLITOKYO}{Kavli Institute for the Physics and Mathematics of the Universe (Kavli IPMU, WPI), University of Tokyo, Kashiwa, Chiba 277-8583, Japan}
\newcommand{\NEGEV}{Physics Department, Ben-Gurion University of the Negev, P.O. Box 653, Be’er-Sheva 8410501, Israel}
\newcommand{\DURHAMA}{Centre for Extragalactic Astronomy, Department of Physics, Durham University, Durham DH1 3LE, U.K}
\newcommand{\DURHAMB}{Institute for Computational Cosmology, Durham University, South Road, Durham DH1 3LE, U.K}
\newcommand{\DURBAN}{Astrophysics and Cosmology Research Unit, School of Mathematical Sciences, University of KwaZulu-Natal, Durban 4041, South Africa}
\newcommand{\LYON}{Centre de Recherche Astrophysique de Lyon, Université Lyon 1, 9 avenue Charles André, 69561 Saint-Genis Laval Cedex, France}
\newcommand{\UPENN}{Department of Physics and Astronomy, University of Pennsylvania, 209 S. 33rd St, Philadelphia, PA 19104, USA}
\newcommand{\STEWARD}{Department of Astronomy/Steward Observatory, University of Arizona, 933 North Cherry Avenue, Tucson, AZ 85721, USA}
\newcommand{\MILLER}{Miller Senior Fellow, Miller Institute for Basic Research in Science, University of California, Berkeley, CA 94720, USA}

\newcounter{affilct}
\makeatletter
\newcommand{\affilref}[1]{%
  \@ifundefined{c@#1}%
    {\newcounter{#1}%
     \setcounter{#1}{\theaffilct}%
     \refstepcounter{affilct}%
     \label{#1}%
     }{}%
  \ref{#1}%
 }
\makeatother

\makeatletter
\newcommand*\affilreftxt[2]{%
  \@ifundefined{c@#1txt}
    {\newcounter{#1txt}%
     \setcounter{#1txt}{1}
     \altaffiltext{\ref{#1}}{#2}
     }{
     }
  }
\makeatother

\shorttitle{Microlenses near caustics}
\shortauthors{Diego et al.}

\begin{document}

\title{Dark matter under the microscope: Constraining compact dark matter with caustic crossing events}

\author{Jose~M.~Diego\altaffilmark{\affilref{IFCA}}}
\affilreftxt{IFCA}{\IFCA}   
\email{jdiego@ifca.unican.es}

\author{Nick~Kaiser\altaffilmark{\affilref{HAWAII}}}
\affilreftxt{HAWAII}{\HAWAII}   

\author{Tom~Broadhurst\altaffilmark{\affilref{BILBAO},\affilref{IKERBASQUE}}}
\affilreftxt{BILBAO}{\BILBAO}   
\affilreftxt{IKERBASQUE}{\IKERBASQUE}   

\author{Patrick~L.~Kelly\altaffilmark{\affilref{BERKELEY},\affilref{\MINNESOTA}}}
\affilreftxt{BERKELEY}{\BERKELEY}   
\affilreftxt{MINNESOTA}{\MINNESOTA}   

\author{Steve~Rodney\altaffilmark{\affilref{SCAROLINA}}}
\affilreftxt{SCAROLINA}{\SCAROLINA}   

\author{Takahiro~Morishita\altaffilmark{\affilref{UCLA}}}
\affilreftxt{UCLA}{\UCLA}   

\author{Masamune~Oguri\altaffilmark{\affilref{TOKYOA},\affilref{TOKYOB},\affilref{KAVLITOKYO}}}
\affilreftxt{TOKYOA}{\TOKYOA}   
\affilreftxt{TOKYOB}{\TOKYOB}   
\affilreftxt{KAVLITOKYO}{\KAVLITOKYO}   

\author{Timothy~W.~Ross\altaffilmark{\affilref{BERKELEY}}}
\affilreftxt{BERKELEY}{\BERKELEY}   

\author{Adi~Zitrin\altaffilmark{\affilref{NEGEV}}}
\affilreftxt{NEGEV}{\NEGEV}

\author{Mathilde~Jauzac\altaffilmark{\affilref{DURHAMA},\affilref{DURHAMB},\affilref{DURBAN}}}
\affilreftxt{DURHAMA}{\DURHAMA}   
\affilreftxt{DURHAMB}{\DURHAMB}   
\affilreftxt{DURBAN}{\DURBAN}   

\author{Johan~Richard\altaffilmark{\affilref{LYON}}}
\affilreftxt{LYON}{\LYON}   

\author{Liliya~Williams\altaffilmark{\affilref{MINNESOTA}}}
\affilreftxt{MINNESOTA}{\MINNESOTA}   

\author{Jesus~Vega-Ferrero\altaffilmark{\affilref{IFCA},\affilref{UPENN}}}
\affilreftxt{IFCA}{\IFCA}   
\affilreftxt{UPENN}{\UPENN}   

\author{Brenda~Frye\altaffilmark{\affilref{STEWARD}}}
\affilreftxt{STEWARD}{\STEWARD}

\author{Alexei~V.~Filippenko\altaffilmark{\affilref{BERKELEY},\affilref{MILLER}}}
\affilreftxt{BERKELEY}{\BERKELEY}
\affilreftxt{MILLER}{\MILLER}   



\begin{abstract}
A galaxy cluster acts as a cosmic telescope over background galaxies but also as a cosmic microscope magnifying the imperfections of the lens.
The diverging magnification of lensing caustics enhances the microlensing effect of substructure present within the lensing mass. 
Fine-scale structure can be accessed as a moving background source brightens and disappears when crossing these caustics.  The recent discovery of a distant lensed star  near the Einstein radius of the galaxy cluster MACSJ1149.5+2223 
allows the rare opportunity to reach subsolar-mass microlensing through a supercritical column of cluster matter. Here we compare these observations with high-resolution ray-tracing simulations that include stellar microlensing set by the observed intracluster starlight and also primordial black holes that may be responsible for the recently observed LIGO events. We explore different scenarios with microlenses from the intracluster medium and black holes, including primordial ones, and examine strategies to exploit these unique alignments. We find that the best constraints on the fraction of compact dark matter in the small-mass regime can be obtained in regions of the cluster where the intracluster medium plays a negligible role. This new lensing phenomenon should be widespread and can be detected within modest-redshift lensed galaxies so that the luminosity distance is not prohibitive for detecting individual magnified stars. High-cadence {\it Hubble Space Telescope} monitoring of several such optimal arcs will be rewarded by an unprecedented mass spectrum of compact objects that can contribute to uncovering the nature of dark matter.
\end{abstract}  

\section{Introduction}\label{sect_intro}  
\citet[][hereafter K17]{Kelly2017} present the first observations of a single high-redshift star in a background, lensed spiral galaxy at redshift $z = 1.49$ \citep{Smith2009,Zitrin2009a} being magnified by a factor of several thousand by a galaxy cluster MACSJ1149.5+2223 [hereafter MACS1149] at $z = 0.544$ \citep{Ebeling2007}. 

This event was discovered serendipitously while monitoring a lensed supernova (SN) behind the cluster \citep[SN Refsdal;][]{Kelly2015,Kelly2016,Rodney2016}. 
The light curve of the star shows at least one prominent peak in the spring of 2016 that lasted $\sim 2$ months. A first event, named \emph{Icarus} or LS1 / Lev~2016A by K17, produced a peak in the light curve that lasted several weeks; after the peak, the flux returned to its original value. 
This event is interpreted as a crossing of a bright background star through a microcaustic produced by one of the stars (or star remnant) in the intracluster medium.  
At a position separated by $0.26''$ from this initial peak, a second peak (named \emph{Iapyx}, or LS1 / Lev~2016B by K17) appeared between 1 and 2 months after the first event faded, and lasted less than 3 months. 
No object was observed at this second position in the previous 10~yr or in the months after it vanished. This second event is also interpreted as a microlensing event of the same background star 
(and a  different microlens in the intracluster medium). However, in this case one possible interpretation is that the low-magnification region around a microlens was hiding the background star for $>10$~yr with occasional brief periods of high magnification (see K17 for other possible interpretations, including binary stars). 
{A potential third event discussed by K17, {\it Perdix} or Ls1/ Lev~2017A, is found $0.1''$ away from the second event. If confirmed, this third event could be produced by the same network of microcaustics, but possibly involving a different background star.

As discussed by K17, the picture described above is consistent with the expected behavior of a background star traveling at typical relative 
velocities of $\sim 1000$\,km\,s$^{-1}$ and a lens plane populated with a density of stars that is compatible with the observed intracluster light (ICL) at the position of the two events. Microlensing events are expected to be produced by the stars responsible for the ICL (and their remnants). As shown in earlier work, the light curve of an object being lensed by a field of 
microlenses may contain high- and low-magnification periods. This behavior has been predicted in previous papers \citep[see, for instance,][]{Chang1984,Kayser1986,Paczynski1986}. In particular, \cite{Chang1979,Chang1984} were the first (to our knowledge) to recognize that counterimages 
may have very low fluxes (disappearing below the detection limit of a given instrument) for some periods of time, in agreement with the observed behavior of the Iapyx event (counterimage of the Icarus event of K17). 

A massive galaxy cluster acts as a cosmic telescope that enlarges the images of background galaxies. However, near a critical curve (CC), small changes in the deflection field  result in large changes in the magnification. These small changes in the deflection field can be produced by small masses in the range of a stellar mass or below. In this situation, as we will show, the galaxy cluster may act also as a cosmic microscope since it effectively enlarges any imperfection in the deflection field near the CC caused by microlenses.  Microlensing near a cluster CC has the interesting feature that the individual micro-CCs around the microlens (and corresponding microcaustics) get enlarged by a factor that is larger the closer they are to the main CC (see discussion of this effect in Section 2). This allows, in principle, probing small-mass microlenses as we approach the cluster CC. 

Earlier work has explored the behavior of counterimages during caustic-crossing events in smooth potentials (from galaxies to clusters). 
\citet{Jordi1991} considers, as in this work, the case of a single star crossing a caustic from a smooth lens model. 
He estimates the maximum magnitude of a lensed background star at the time of caustic crossing, as well as a rate of events based on the surface brightness of a background galaxy \citep[this case is also discussed by][]{Chang1979,Chang1984,Schneider1986}.
The combined effect of overlapping caustics from an ensemble of microlenses embedded in a stronger gravitational field has been also studied in detail 
\citep{Young1981,Gott1981,Chang1984,Kayser1986,Paczynski1986}, in particular in the context of quasar (hereafter QSO) microlensing \citep{Chang1979,Irwin1989,Witt1995,Metcalf2001}. 
\cite{Kayser1986} and \cite{Paczynski1986} show how a large number of microlenses embedded in a deep potential can redistribute the magnification, producing complex light curves 
of a background source. For certain configurations \citep[see, e.g., Figs. 9 and 10 of][]{Kayser1986}, the magnification splits into compact regions of large and low magnification. As shown in these papers, a source traveling across this field may {\it disappear} suddenly when entering one of the low-magnification regions, only to reappear at some time later as a bright source. 

This type of behavior resembles the observed flux in the Icarus and Iapyx events. However, when the microlenses are very close to the CC (a fraction of an arcsecond), the magnification pattern exhibits features that have not yet been studied in detail. \cite{Paczynski1986} investigated the general case of high optical depth of microlenses embedded in a galaxy or cluster potential, but he ignores the effect of shear and focuses on areas in the lens plane that are not close to the main CC.  
\cite{Kayser1986} include the shear term from the large deflector (cluster or galaxy) in their calculations, but again do not study the particular case of short distances to the 
main CC. As noted by \cite{Paczynski1986}, this regime is computationally very expensive (owing to the very large magnifications involved that require the mapping of a small field in the source plane into a very large field in the image plane), and could not be studied in detail in those early papers. 

Some authors have focused their attention on the {\it high magnification} regime \citep[see][and references therein]{Wambsganss1990,Schechter2002} in the context of QSO microlensing, but these high magnifications are still modest (a few tens at most) compared with the more extreme values (several hundred to several thousand) considered in this paper and do not reveal some of the properties of the lensed images that are accentuated with extreme magnification (see Section~\ref{SectSourceplane}). The smaller magnifications found in QSO microlensing are partially due to the larger intrinsic size of the background source. As we will show later, the maximum magnification attained by a background source scales as the inverse of the square root of its radius. For QSOs, the radius is related to the half-light radius of the accretion disk. These disks are typically of the order of ten light days, when observed in the optical, and about an order of magnitude smaller when observed in X-rays \citep{Chartas2009,Dai2010,Jimenez2012} for typical supermassive black holes. This radius is known to scale with the mass of the black hole \citep{Morgan2010,Jimenez2015}. When compared with the radius of a giant luminous star, the accretion disks around QSOs are approximately a factor $10^3$--$10^4$ times larger. Consequently, the maximum magnification attained by a lensed giant star can be up to two orders of magnitude larger than the corresponding one for QSOs. This is an important advantage that comes with the added bonus that the smaller stellar radii translate into shorter-lived events which are easier to monitor (days as opposed to years).  
Although QSO microlensing is not directly comparable to the work presented in this paper, there are also many similarities. 
Earlier papers focusing on the interpretation of QSO microlensing contain useful insights that are applicable to this work when the magnifications are significantly higher. \cite{Schechter2002} present interesting similarities with some of the results given here, in particular when discussing the statistics of the magnification around microminima and microsaddle points.

In this paper, we explore for the first time the regime of very short distances to the main CC (or, equivalently, very high magnification), motivated by the observation of the two (or possibly three) intriguing events discussed by K17\footnote{The reader will find also very interesting two recent publications that appeared after this manuscript was originally submitted and that are very closely related to this work \citep{Oguri2017,Venumadhav2017}.}.

In the case of a galaxy cluster, its larger size translates into a greater magnification of a background object. Also, if a significant fraction of dark matter (DM, hereafter) is made of compact objects like primordial black holes (PBHs),  galaxy clusters are ideal to study microlensing by these objects since it is possible to find CCs (with high optical depth for microlensing) relatively far away from member galaxies and reduce the impact of stars (or remnants) in these galaxies that could produce similar microlensing events. 
The case of PBHs is interesting since they are (still) a valid candidate for DM
(or at least a fraction of it) in some mass regimes 
\citep[see, for instance,][]{Carr2010,Clesse2015,Carr2016}. 
The fraction of DM that can be in the form of PBHs has been constrained for different PBH masses. The possibility that PBHs constitute a sizable fraction of the DM is interesting and has been studied extensively, although PBHs are excluded as the primary component of DM in virtually all mass ranges. 
\cite{Bird2016} proposed that at around $30\,{\rm M}_{\odot}$ there is still a range of masses that have not been convincingly ruled out \citep[see also][for a related result]{Sasaki2016,Clesse2017}. 
Interestingly, if a significant fraction of DM is in the form of PBHs with $M \approx 30 \, {\rm M}_{\odot}$, events like the collision of two black holes with these masses would be more common, facilitating the interpretation of the first LIGO detection \citep{LIGO2016}. This interpretation, however, is not supported by the second LIGO event with significantly smaller masses. On the other hand, more recently a new LIGO event as well as a LIGO/Virgo event imply detections of massive pairs of BHs ($M_{\rm BH}\approx 20$--30 M$_{\odot}$), implying a higher than expected abundance of BHs with $M_{\rm BH}\approx 30$ M$_{\odot})$ \citep{LIGO2017a,LIGO2017b}.

Analyses of multiply imaged QSOs have found that the observed microlensing signal is incompatible with the hypothesis that $\sim30\,{\rm M}_{\odot}$ PBHs make up most of the DM \citep[see][and references therein]{Mediavilla2017}. The same work concludes that the fraction of mass in the form of microlenses can still be as high as 20\% of the total mass, but with the most likely mass of microlenses being below 1 M$_\odot$. If confirmed, this key work leaves little room for the hypothesis that PBH with $\sim30\,{\rm M}_{\odot}$ can make a significant fraction of the DM ($\sim 10\%$) unless extended mass functions \citep[instead of the monochromatic or bimodal models considered by][]{Mediavilla2017} can have a significant impact on the results, or the size of accretion disks around QSO are an order of magnitude larger than what has been considered so far (the latter point being an important source of uncertainty in this and other work). Moreover, we should note that in \cite{Mediavilla2017}, the limit of high optical depth (for microlensing) does not seem to be explored, and as we will show later, in this regime the fluctuations in flux are smaller owing to the constant presence of multiple overlapping microcaustics that tend to average out the observed integrated flux. It would not be surprising to have constraints from QSO microlensing that differ from (or even contradict) those derived from microlensing of background stars (this work). If one finds that tensions between these regimes exist, some of the assumptions made in each regime will have to be reviewed.
Constraints from microlensing in our local environment (the Magellanic Clouds) are weaker, and recent work has shown that uncertainties in these constraints can be as high as one order of magnitude \citep{Green2017}. 

\cite{Carr2017} review the constraints for PBHs using more realistic extended mass functions and conclude that one could allow for as much as 10\% of the DM in the form of PBHs in the mass range  $M_{\rm BH}\approx 25$--100 M$_{\odot}$ \citep[although this work does not include the results of][]{Mediavilla2017}. This limit of 10\% will be adopted in this paper as an upper limit for the fraction of PBHs in this mass range.  
At smaller masses, constraints on the fraction of PBHs allow for a modest fraction of DM below  $M \approx 1\, {\rm M}_{\odot}$ \citep[see, however,][for the mass range $M_{\rm BH}\approx 10^{-10}$--$10^{-8}$ M$_{\odot}$]{Kuhnel2017,Inomata2017}. These constraints tighten at very low masses. A lower limit for the PBHs of  $M \approx 10^{11}$\,kg can already be established from theoretical grounds 
and  observations of $\gamma$-rays \citep[see, e.g.,][]{Kim1999}. Below this mass, no PBHs are expected to exist as they should have evaporated by now (down to the Planck mass). 
This limit can be increased a little from detailed observations of the $\gamma$-ray background, since sufficient PBHs with masses near the above limit would 
be a strong source of $\gamma$-rays in our vicinity, which is not observed.  
Continuous monitoring of background galaxies intersecting a cluster CC provides an excellent dataset for constraining the abundances of PBHs based on their lensing signature. Combining these data with models of the full stellar population in the lensing plane can address many of the systematic biases inherent in past measurements.

In this paper, we explore a different technique to constrain the fraction of compact DM, paying particular attention to the mass range relevant for the three most significant LIGO events. We show how microlensing events by relatively small masses can take place thousands of years before (or thousands of years after) a bright star in the background galaxy crosses the position of a cluster's main CC. Hence, the probability of observing a microcaustic crossing event is considerably increased when compared with earlier work that only considered the crossing of the main cluster caustic. As mentioned earlier, as the background star approaches the main cluster CC, the sensitivity to detect progressively smaller microlenses grows, offering a unique opportunity to probe masses that could not be tested otherwise. This provides an exciting opportunity to set limits on the fraction of DM in the form of compact objects in low-mass regimes that are difficult to study otherwise. 

This paper is organized as follows. In Section~\ref{sect_S2} we describe the basic properties of the magnification near a CC. 
Section~\ref{sect_S4} presents results based on numerical simulations with a focus on the structure of caustics in the source plane. 
In Section~\ref{sect_S5} we explore in detail the disruption of the CC in the image plane when microlenses populate the lens plane. We predict in Section~\ref{sect_S6} 
the behavior of the observed flux (light curve) of a background star traveling through a field of microcaustics. 
In Section~\ref{sect_S7}, we predict how events like Icarus will disappear (or first appear) once the last (or first) microcaustic is crossed.   
Section~\ref{sect_S8} considers the prospects for constraining compact DM with this type of observation. 
Some of our results are discussed in Section~\ref{sect_S9}, and we conclude in Section ~\ref{sect_S10}.  

This paper is very much related to K17. While this paper presents the theoretical (lensing) and numerical (simulations) background of K17, the reader is pointed to K17 for a detailed discussion of the particular Icarus and Iapyx events, including their interpretation. In this paper we refer to the Icarus and Iapyx events when appropriate or relevant for the discussion. 
Throughout the paper we assume a cosmological model with $\Omega_M=0.3$,
$\Omega_\Lambda=0.7$, and ${\rm H}_0=70$\,km\,s$^{-1}$\,Mpc$^{-1}$. For this model, 
$1\arcsec = 6.45$\,kpc at the distance of the cluster MACS1149 ($z=0.544$) and $1\arcsec = 8.4$\,kpc at the distance of the background source ($z=1.49$). 

Besides CC, several terms will be used often in this paper. We refer to the critical curves and caustics around microlenses as {\it micro-CCs} and {\it microcaustics}, respectively. 
The cluster CC and caustic that would form if there were no microlenses are the {\it main CC} and {\it main caustic}, respectively. {\it Macro-images} are 
the counterimages that would have formed if there were no microlenses and the lensing potential were sourced only by the cluster.   
A macro-image in a region filled with microlenses usually breaks up into smaller portions that we refer to as {\it micro-images} and also as {\it bits}. 
Then, around a CC we expect to find two macro-images, each composed of several smaller micro-images or bits. 
When the optical depth of microlenses is relatively small ($\Sigma_{\rm microlens}/\Sigma_{\rm crit} < 0.01$) and the macro-images form very close to the main CC, the resulting group of micro-images is usually stretched along a straight line, following the direction of the cluster deflection field. Because of this geometry, we refer to this group of micro-images as a {\it train of micro-images}, or simply as a {\it train}. At low optical depth of microlenses, a background source will form typically two trains (or macro-images), one on each side of the CC. At higher optical depth, a single background source can form more than two trains, and each train can contain multiple smaller micro-images. 
In this sense, we can think of Icarus and Iapyx as unresolved macro-images which consist of even smaller bits or micro-images. 
The surface mass density of microlenses, $\Sigma$, is used in two contexts. In its broader sense we simply use $\Sigma$. When $\Sigma$ takes the value of $7\, {\rm M}_{\odot}$\,pc$^{-2}$ (the one found by K17 at the position of Icarus/Iapyx\footnote{Although this value was recently updated by K17 to $\sim 12--19\, {\rm M}_{\odot}$\,pc$^{-2}$ which we also use in parts of this work}.), we refer to it as $\Sigma_o$. Sometimes we express $\Sigma$ in units of $\Sigma_o$ and use $f=\Sigma/\Sigma_o$. Toward the end of this work we use another variable, $F$, that should not be confused with $f$. We use $F$ to refer to the fraction of the total mass that is in the form of compact objects (whether this is made of stars from the ICL, PBHs, or both). By construction, $F$ is always smaller than 1 while $f$ can be larger than 1.

\section{Lensing properties near a critical region}\label{sect_S2}

\begin{figure}  
 \centerline{ \includegraphics[width=9cm]{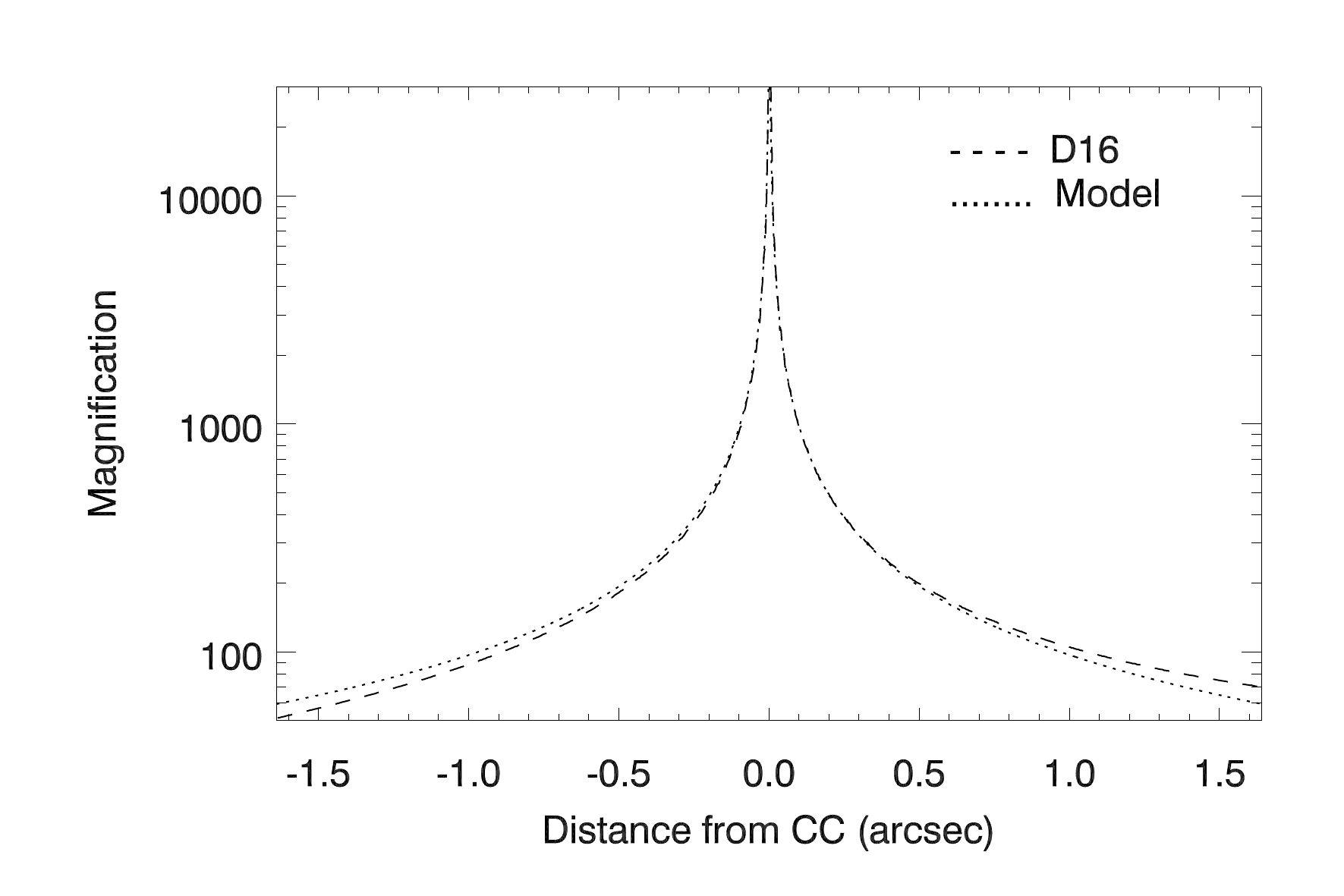}}
  \caption{Magnification along a direction perpendicular to the CC at the position of the Icarus event. The dashed line corresponds to the model 
           of Diego et al. (2016). The dotted line is an analytical model following Eq.~\ref{Eq_mu_theta}. The left side of the curve corresponds to the inner part of the CC 
          (or negative parity, $a_1<0$; see text) where the magnification falls faster than the simple analytical model. The right side of the curve is for the region where the 
          parity is positive, $a_1>0$. } 
   \label{Fig_MuModel}  
\end{figure}  

A gravitational lens is characterized by the lens equation
\begin{equation}
\beta=\theta-\alpha(M,\theta),
\label{Eq_lens}
\end{equation}
where $\beta$ is the position of the background source, $\theta$ is the observed position in the sky of the lensed image, and $\alpha(M,\theta)$ 
is the deflection angle produced by the lens with mass $M$. The dependence of $\alpha(M,\theta)$ on the position $\theta$ results in Eq.~\ref{Eq_lens} 
being nonlinear. Consequently, for a given position $\beta$, it is sometimes possible to find multiple solutions to Eq.~\ref{Eq_lens} with each solution representing 
a different counterimage. Each counterimage is magnified by a factor $\mu$. 
Since lensing preserves the surface brightness of the background source, 
$\mu$ can be defined as the ratio between the observed size (i.e., area) of the counterimage, $d\Omega_{\theta}$, and the intrinsic size of the 
background source, $d\Omega_{\beta}$. 
For a given lens model the deflection field $\alpha(M,\theta)$ is known, and the magnification can be computed in a given position from the derivatives 
of the deflection field. The inverse of the magnification is defined as 
\begin{equation}
\mu^{-1} = a_1 a_2 = (1-\kappa-\gamma)(1-\kappa+\gamma) = (1-\kappa)^2-\gamma^2,
\label{Eq_mu}
\end{equation}
where $\kappa$ and $\gamma$ are the convergence and shear (respectively), and are combinations of the derivatives of the deflection field. We introduce 
the inverse of the magnifications, $a_1$ and $a_2$, that will be used later in this work. 
At a tangential CC, $a_1=0$. On each side of the CC, $a_1$ 
takes positive and negative values (parity). 
The sign of $a_1$ gives the parity of the image, so images with negative parity have $a_1<0$ and images with positive parity have $a_1>0$.\footnote{In the Appendix, the distinction between $a_1>0$ and $a_1<0$ becomes more evident.} 

Counterimages that form near a CC can be magnified by very large factors. 
At the CC, the magnification diverges and $d\beta/d\theta=0$. We can take advantage of this property to Taylor expand the lens equation around the CC,
\begin{equation}
\beta=\beta_o +  \frac{d\beta}{d\theta}(\theta-\theta_o) + \frac{1}{2}\frac{d^2\beta}{d\theta^2}(\theta-\theta_o)^2  + ...
\label{Eq_lens2}
\end{equation}
We choose $\beta_o$ and $\theta_o$ as the origins of the reference systems in the source and image plane, 
respectively, and redefine $\beta = \beta-\beta_o$ and $\theta = \theta-\theta_o$. 
The second term cancels out at the position of the CC, leaving to second order
\begin{equation}
\beta = \theta^2/\Theta,
\label{Eq_beta_theta}
\end{equation}
where we have defined the constant $\Theta^{-1}= (1/2) d^2\beta/d\theta^2$. At the position of the CC ($\theta=0$) we satisfy the condition $\mu^{-1}=0$, and to first order  $\mu^{-1} = d\beta/d\theta \propto \theta$. 
Hence, in the image plane we obtain for the total magnification (i.e., the magnification of the two images on each side of the CC)
\begin{equation}
\mu = \frac{\mu_o}{\theta}
\label{Eq_mu_theta}
\end{equation}
near the CC, where $\mu_o$ is a constant that depends on the lens mass and geometry. This condition is satisfied for most lenses up to $\theta \approx 1''$ (see Fig.~\ref{Fig_MuModel}). The asymptotic behavior when $\theta \gg 0$ is $\mu=1$ in the external side of the CC ($a_1>0$) and $\mu=0$ at the position of the lens for a point-source lens.

The magnification in Eq.~\ref{Eq_mu_theta} is expressed in the image plane. 
In terms of the position in the source plane, we can use Eq.~\ref{Eq_beta_theta} to obtain
\begin{equation}
\mu = \frac{\mu_o/\sqrt{\Theta}}{\sqrt{\beta}}.
\label{Eq_mu_beta}
\end{equation}
The maximum magnification is obtained when the source touches the caustic --- that is, when the distance from the center of the source to the caustic equals the radius
of the source, $R$. Then we obtain
\begin{equation}
\mu_{\rm max} = \frac{\mu_o/\sqrt{\Theta}}{2\sqrt{R}}.
\end{equation}

\begin{figure}  
 \centerline{ \includegraphics[width=9cm]{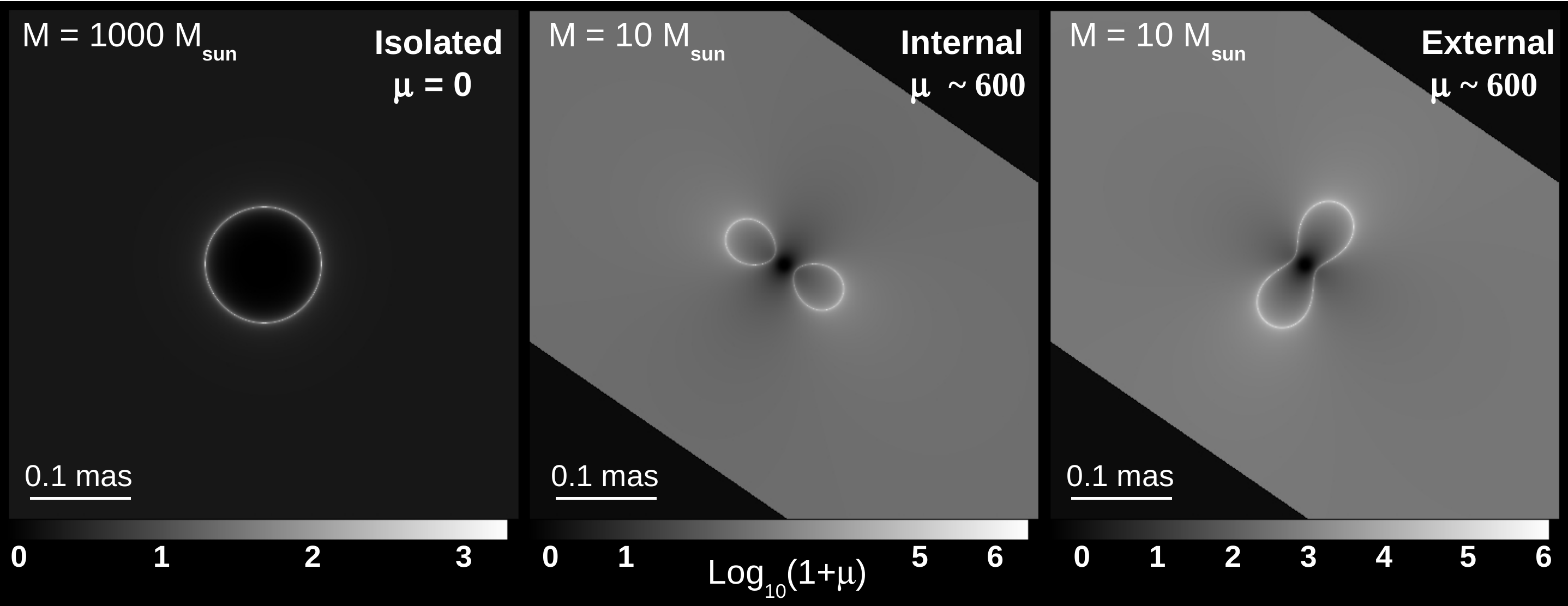}}
  \caption{The left panel shows magnification for a source at $z=1.49$ around a star with $M=1000$\,M$_{\odot}$ at $z=0.55$. The Einstein ring can be clearly seen as a circle. The middle and right panels indicate the 
           magnification (for a source at  $z=1.49$) caused by two stars with much smaller masses ($M=10$\,M$_{\odot}$) at $z=0.55$, but that are close to a 
           CC of a galaxy cluster also at $z=0.55$. The main CC (not shown) runs perpendicular to the gray band. The middle panel is for the side of the CC where $a_1<0$ and the right panel for 
           the side where $a_1>0$. The circular configuration of the Einstein ring transforms into a figure-eight pattern.    
   } 
   \label{Fig5}  
\end{figure}  

Eqs.~\ref{Eq_beta_theta},~\ref{Eq_mu_theta}, and~\ref{Eq_mu_beta} are very useful for characterizing the  properties of the counterimages near a CC.
The values of $\mu_o$ and $\Theta$ can be obtained for a given lens and at a given position after fitting several positions near the CC.
For a cluster like MACS1149 at the position of the Icarus event (and a background source at $z=1.49$), $\mu_o\approx 150''$ and $\Theta\approx 68''$ for the model of \citet[][hereafter D16]{Diego2016}. These values may change by as much as a factor of $\sim 2$ for alternative models that still predict the CC in the correct position depending on the slope of the potential at the position of the CC, but we will adopt them below as realistic examples (see Section~\ref{Sect_Uncertainties}). For these values of $\mu_o$ and $\Theta$, if the background star is a giant star with radius $R=100\,{\rm R}_{\odot}$, the maximum magnification can be as high as $\mu_{\rm max}\approx 10^6$ at the CC near the 
Icarus position. If a background source is moving in the source plane with a constant velocity $v_p=d\beta/dt$ in a direction perpendicular to the caustic, 
the apparent observed velocity  of the counterimages in the image plane is 
\begin{equation}
v_{\rm obs}=\frac{d\theta}{dt}=\frac{\Theta}{2\mu_o}v_p\mu,
\label{Eq_v}
\end{equation}
where we have used Eq.~\ref{Eq_beta_theta} to relate $\theta$ with $\beta$ and replaced $\theta$ with $\mu$ using Eq.~\ref{Eq_mu_theta} after doing the derivative. 
Hence, at total magnifications $\mu \ga 1000$, the counterimages would appear to move at superluminal speeds (for this particular configuration). This has an interesting implication, since counterimages that move with larger apparent velocities can cover a larger region in the lens plane, probing more substructure as the source moves across the lens plane. Conversely, this can be seen as the microcaustics being more cramped in the source plane as we approach the main caustic.

If DM is clumped (like in some wave dark matter models), or if a significant fraction of DM is made of compact objects such as PBHs, the probability that in a fixed period of time a given counterimage passes behind a clump of DM or a PBH will be higher near a CC, where counterimages probe the lens plane at a faster rate. 
The most interesting scenario to constrain the fraction of DM in compact objects is near tangential CCs. 
Radial CCs are normally close to the center of the cluster, where the intracluster light or the stars from the brightest cluster galaxy (BCG) can overwhelm the possible signature from compact DM.

In this work, we will focus on the case of tangential CCs, where the exploitation 
of crossing events may be most fruitful \citep[see, however,][where a microlensing event near a radial CC in the same cluster, MACS1149, is used to unveil a supermassive black hole $\sim 10$\,kpc from the center of the BCG]{Chen2017}.  

Tangential CCs form when $1-\kappa-\gamma = 0$ (while $|1-\kappa+\gamma|>0$). 
If the lens plane is populated by small microlenses, they will contribute to the convergence (or surface mass density) with a small factor ${\bar{\kappa}}_*(\theta) \ll 1$, 
where $\bar{\kappa}_*(\theta)$ is the mean surface density of a point-like star with mass $M$ within a radius $\theta$ --- that is, $\bar{\kappa}_*(\theta)\Sigma_{\rm crit}=M/(\pi\theta^2)$.
Near the CC, the condition for diverging magnification becomes $1-\kappa-\gamma = \mu_t^{-1}=\bar{\kappa}_*(\theta)$, where $\mu=\mu_t\mu_r=(a_1a_2)^{-1}$. 
Hence, we can conclude that
\begin{equation}
\theta_E = \sqrt{ \frac{M\mu_t}{\pi\Sigma_{\rm crit}} }.
\label{Eq_thetaE}
\end{equation}

\begin{figure}  
 \centerline{ \includegraphics[width=9cm]{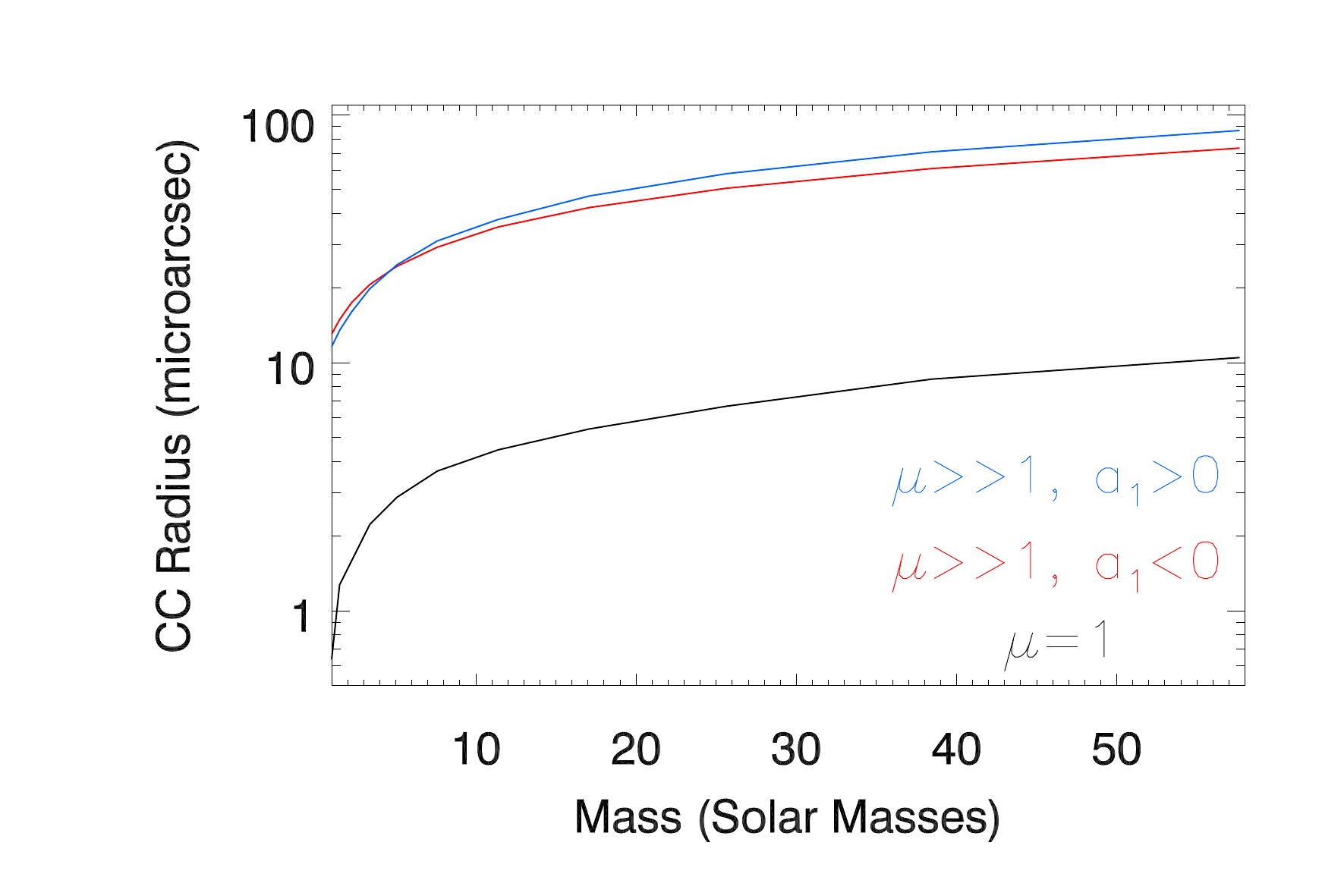}}
  \caption{Change in size of the micro-CC as a function of mass. The black solid line is for an isolated star (not in a strong-lensing deflection field) and gives the standard Einstein radius. 
           The red and blue curves correspond to the cases where the star is located $\sim 0.13''$ from the CC on either side of the CC (see Section \ref{sect_S2} for the definition of $a_1$). 
           The CC radius is defined as the perimeter of the CC divided by 2$\pi$. The red and blue curves are roughly a factor $\sqrt{\mu_t} \approx \sqrt{100}=10$ times higher than the black curve.} 
   \label{Fig_SizeCC}  
\end{figure}  

The above result has profound implications. A microlens at a position near a CC, where the magnification is $\mu_t\approx 1000$, will behave (to first order) 
like an isolated microlens, but a thousand times more massive (see Fig.~\ref{Fig5}). As shown in Section~\ref{subsect_Jupiter}, in the last moments before a star crosses the cluster caustic, the magnification can become of order $10^6$, allowing the detection of substructures with masses comparable to a Jupiter mass \citep[see also Eq.~23 in][where an expression similar to Eq.~\ref{Eq_thetaE} is introduced as the dimensionless radius]{Paczynski1986}. 

Despite being based on some approximations, like neglecting higher-order terms, the expressions above seem to match remarkably well the results derived from detailed 
numerical calculations. Fig.~\ref{Fig_SizeCC} shows the change in effective radius (defined as the perimeter divided by $2\pi$) for a microlens that is isolated 
(no external field; bottom curve) and for a microlens that is embedded in a lensing potential with $\mu_t\approx100$ (top curves). All curves grow with radius 
as $\sqrt{M}$, but in the case of the microlens in a lensing potential, the amplitude is increased by a factor $\sim\sqrt{\mu_t}$ as predicted 
by Eq.~\ref{Eq_thetaE}. 

The magnification around a microlens in a field with external shear and convergence has previously been studied in detail \citep[see, e.g.,][]{Schechter2002}. In the Appendix, we present a brief 
and simplified description of a single microlens in an external field at high magnification.

\section{Numerical results}\label{sect_S4}
The results presented in the previous section (see also the Appendix) give us useful insights into the behavior of the magnification around a microlens near a CC. 
However, in most realistic scenarios the CC region will be populated by a number of microlenses having, in general, different masses. In order to explore 
this more realistic regime, we resort to numerical simulations where the magnification is computed from simulated data.\footnote{The reader can find movies based on these simulations showing the formation and destruction of micro-images as a function of time (movies 1 through 4) at this site: https://cosmicspectator.org/2017/06/30/dark-matter-under-the-microscope/ and also in section~\ref{sect_animations} in the Appendix.} 

We will assume the background source is a luminous giant star, which have radii ranging between $\sim 100$\,R$_{\odot}$ and  $\sim 1000$\,R$_{\odot}$. We adopt the value $R_{\rm star}=1000$\,R$_{\odot}$. 
The results presented in this paper are virtually the same for smaller stars, except for the 
maximum magnification reached when a microcaustic is being crossed (after the star touches a given microcaustic). 
In this case, the maximum peak magnification would grow by a factor $\sqrt{1000/R_{\rm star}}$ (see Eq.~\ref{Eq_mu_beta}) if the microlens is at sufficiently large distances from nearby microlenses. 
The deflection field in the simulations contains a smooth component from the large-scale cluster potential and small-scale fluctuations from the microlenses. 
Normally, extremely luminous (background) stars can be found in star-forming regions where the density of stars may be relatively high. In this work we ignore the effect of neighboring stars and consider only the effect over one of those stars. If several background stars are moving on the web of caustics, each star would produce a series of peaks as they cross microcaustics. In realistic scenarios, only the brightest stars will produce peaks that can be measured, with the remaining stars contributing to a stochastic background of small fluctuations in the light curve. For instance, K17 argue that in order to explain the Icarus event, the background star needs to be extremely luminous and hence very rare.

For the smooth-scale potential we assume 
a realistic lens model, in particular the lens model of D16 for the cluster MACS1149 in the region of the Icarus and Iapyx events; see K17, where the same model was also used to interpret the observations. The model produces a CC (for a background source at $z=1.55$) that falls in between the positions of Icarus and Iapyx. Following K17, in this work we assume that the CC is exactly between Icarus and Iapyx as predicted by various models \citep{Richard2014,Oguri2015,Kawamata2016,Diego2016}. However, the reader should note that other interpretations are also possible where, for instance, the CC is closer to Icarus than to Iapyx, in which case the two events would be produced by two different background stars (see K17 for a more detailed discussion of this and other alternative interpretations). The hypothesis that there are multiple bright stars in the background moving between microcaustics would also be supported if the third event of K17 is confirmed as an additional microlensing event. In this case, since this new image does not appear aligned with the direction where counterimages of the same star are expected to form, a second background star would be needed.

Alternatively, when a fast computation is needed over a large area (for instance, in Fig.\ref{Fig_MuModel}), we use the analytical model from \citet{Blandford1987}
for the smooth component, where we tune the lens parameters to produce a magnification pattern similar to the model above. 
In Section~\ref{SectSourceplane}, some of the results computed in the source plane come from a small region at very high resolution. 
For this particular case, we use a simplified model for the macromodel that is given by just two parameters --- the surface mass density ($\kappa_{\rm smooth}$) and shear ($\gamma_{\rm smooth}$). 
These two parameters are considered constant, which is a valid assumption given the small area being simulated. Tuning $\kappa_{\rm smooth}$ and $\gamma_{\rm smooth}$ to the desired values allows us to quickly produce simulations with a variety of magnifications from the macrolens model.

\begin{figure*}  
 \centerline{ \includegraphics[width=18cm]{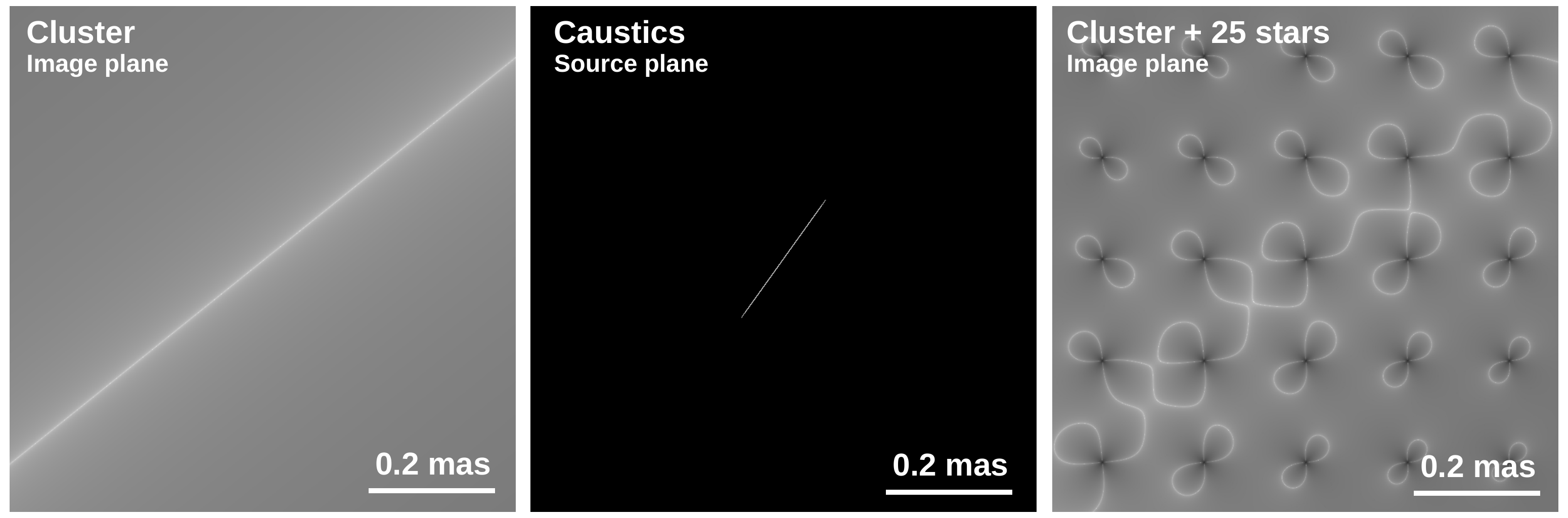}}
  \caption{The left panel shows a close-up region ($0.8 \times 0.8$ milliarcsec$^2$) around the main CC for a galaxy cluster (MACS1149). 
           The middle panel shows the corresponding caustic region with the same scale. The right panel shows the disrupted CC when 
           25 microlenses are added in the image plane. The mass of each microlens is 0.01\, ${\rm M}_{\odot}$. Note how the microlenses increase their 
           associated micro-CCs as they approach the cluster main CC. The orientation is determined by the sign of the quantity 
           $\mu_t^{-1}=1-\kappa-\gamma$. By definition,  $\mu_t^{-1}=0$ at the cluster main CC. } 
   \label{Fig_SmallMicro}  
\end{figure*}  
The microlenses are assumed to be point masses and 
are randomly distributed. Unless otherwise noted, the masses of the microlenses are drawn from a 
\cite{Spera2015} initial-final mass function where the only surviving stars in the intracluster medium are less massive than 1.5\,M$_{\odot}$ (above this mass, the remnants of more massive stars are also included in the simulation). The same model is also discussed by K17 together with other alternative models (see K17 for details). The mass function is normalized to match the inferred stellar surface mass density at the Icarus position \citep[as estimated by][]{Morishita2016}.

We place stars in a region (or extended region, hereafter) which is slightly bigger than the final simulation region (or target region, hereafter). This is done in order to minimize edge effects. The extended region contains the target region plus buffer zones around it, of 0.2 milliarcsec (mas) width each extending in the vertical direction. (The left and right edges of the simulation are not used for the computation of the light curves, so we do not add an extra buffer on these two edges.) This buffer zone is sufficiently large to account for the individual effect of the largest microlenses that could be found beyond (but near) the edge of the target simulated region. The target region is a band of width 1 mas and length 10 mas, aligned in the direction where counterimages form and it is contained in the extended region of width 1.4 mas and length 10 mas. The total number of microlenses included in this extended region is $18,686$, and they are placed randomly within the extended region. The total mass of the microlenses in the extended region is $\sim 4000$ M$_{\odot}$.
When PBHs are included in the simulation, we use the same distribution of stars and add the effect of randomly placed PBHs. The number of PBHs is determined by the fraction of total mass that is in the form of PBHs. This number scales as $N_{\rm PBH} \approx 30\,F_{\rm PBH}$ per milliarcsec$^2$, where $F_{\rm PBH}$ is the fraction (in percent) of mass in the form of 30 M$_{\odot}$ PBHs. 
This results  in $420 \,F_{\rm PBH}$ PBHs (with  30 M$_{\odot}$ each) in the extended region. 
Finally, we subtract from this extended region the contribution to the deflection field from a smooth mass distribution with the same surface mass density as the microlenses (stars plus remnants, or stars plus remnants plus PBHs), so the total surface mass density remains constant.

The simulations are made at a resolution of 1\,$\mu$arcsecond in the image plane. As mentioned earlier, the target region is a band of width 1000 pixels and length 10,000 pixels in the direction where counterimages are expected to form (i.e., at an angle $\alpha_c\approx-40^{\circ}$). 
The length of the simulated box ($\sim 10$ mas/cos($\alpha_c)$) maps into a corresponding length in the source plane of $\sim 100\,\mu$arcsecond. This is enough to follow a moving background source at $z\approx 1.5$ with $v\approx 1000$ km/,s$^{-1}$ during $\sim 1000$~yr.
This resolution is sufficient to resolve the elongated arcs that form when a background star crosses a microcaustic, if the background star is at least a few tens of solar radii.

We assume that the source is moving perpendicular to the main caustic of the cluster. The simulated light curves have a weak dependence on this angle, since the microcaustics are stretched by a large factor in the direction of the main caustic of the cluster. Only if the background sources are moving in a direction very close to the main axis of the microcaustics would the simulated light curves be significantly different --- but this is unlikely since the probability of moving in this narrow range of angles is small. If the source is not moving perpendicular to the main axis of the caustics but at a different angle, the simulated light curves would still be very similar to the ones presented in this work, except that the time it takes for a source to cross a microcaustic would be stretched by a factor cos($\alpha_{\rm sc})^{-1}$, where $\alpha_{\rm sc}$ is the angle between the main axis of the microcaustic and the direction of motion of the source. The reader can find videos at https://cosmicspectator.org/2017/06/30/dark-matter-under-the-microscope/ (movies 5 and 6) extracted from the simulations and showing the effect of the motion of a source as it travels through a web of caustics that is moving parallel, or at an angle with the main axis of the microcaustics. A source moving parallel to the caustics may be interpreted as a source moving in a region with a small surface density of microlenses (see also Section~\ref{SectSourceplane} below). Also, if the velocity is very small, it may be erroneously interpreted in a similar way.

When the lensed images form farther from a micro-CC, the dimension of the micro-images is typically 
smaller than the pixel size of the simulation. In this case (but also when the micro-image is resolved), the real dimension of the micro-image is computed 
at the subpixel level (making use of approximations that allow resolving scales much smaller than the simulation pixel). This is achieved by interpolating the deflection field so any position in the source plane can be mapped into the corresponding interpolated position in the image plane, effectively achieving infinite resolution. Simple, fast interpolations are sufficient because the deflection field is extremely smooth. The smoothness of the deflection field is guaranteed since it is simply the superposition of the deflection field from the cluster and the deflection field from the microlenses. The former is orders of magnitude larger than the latter, so the small perturbations from the microlenses do not break the smooth condition needed for the simple interpolations. The only place where the simple interpolation may break down is when one is looking for counterimages very close to the microlens, since in this case the deflection field diverges. Luckily, these positions correspond to the lowest magnification regions, so those counterimages can never be observed.

The magnification is computed as the ratio of the total area in the image plane divided by the area of the background star in the source plane (i.e., $\pi R_{\rm star}^2$), and we neglect limb-darkening effects (this would add a small correction during a caustic crossing event that is most important in the last moments of the event). We also neglect interference effects, since both the background stars and the microlenses are sufficiently large.

\subsection{Multiple Microlenses with the Same Mass}

The large magnifying power of a galaxy cluster near its CC can allow for detailed study of both the background objects and the substructure in the lens plane itself.
A point-like microlens with a mass $M$ (like a star or a BH) in the lens plane will behave like a lens with effective 
mass $\mu M$ (see Eq.~\ref{Eq_thetaE}). 
In the simple scenario where the microlens is isolated (i.e., no other microlenses nearby), the magnification ($\mu$) of a cluster at a distance less than $0.1''$ from the CC can easily reach values 
above $\mu=1000$ for a point-like background source. At this magnification, a background compact bright object such as a giant star will be boosted by $\sim 7.5$\,mag. This boost factor 
may be sufficient to make luminous stars at $z>1$ detectable with deep observations. In this situation, a microlens with mass $M=10^{-2}$\,M$_{\odot}$ in or near the line of sight to the background 
star and close to the cluster CC will behave (in terms of its lensing effect) like a microlens with mass $M>1\,{\rm M}_{\odot}$. This makes it possible to detect the microlens in the light curve of the background source. 

If no microlenses are present in the lens plane, on small scales (less than $0.1''$) the cluster CC can be approximated by a straight line (see left panel in Fig.~\ref{Fig_SmallMicro}), 
and the magnification grows as the inverse of the distance to the CC (Fig.~\ref{Fig_MuModel}). 
The corresponding caustic is equally well described by a straight line, but the magnification grows as the inverse of the square root of the distance to the 
caustic. Hence, a large magnification of several thousand requires an incredibly small separation between the background star and the caustic, making this type of configuration very rare, and observing 
a caustic crossing very unlikely (see middle panel in Fig.~\ref{Fig_SmallMicro}, where we show the incredibly narrow region in the source plane that maps on the image plane in the left panel). 

A cluster caustic crossing event is expected to be very short lived (several hours or a few days, depending on relative velocity and star radius)
and involves very large magnifications when the lens plane contains no microlenses. The dependence on the square root of the separation 
between the star and the caustic means that the precise moment of the caustic crossing event can be predicted, since the observed flux evolves as $1/\sqrt{t-t_{\rm o}}$, where $t$ is time and $t_{\rm o}$ 
is the time of crossing. When microlenses are included in the lens plane, the situation can be very different since microlenses can significantly disrupt the CC (see Fig.~\ref{Fig_SmothCC_vs_MicroCC}). The disruption is most prominent near the CC and decreases with 
the distance to the CC (see Fig.~\ref{Fig_SmallMicro}). 
In the source plane, the corresponding caustic region gets expanded by a factor that depends on the number of microlenses and their masses. 
A larger caustic region means that observing a microlensing event becomes more likely, since a moving background star may intersect multiple microcaustics in a given period of time 
(as opposed to intersecting just the main CC of the cluster). One of these microcaustics can also be intersected many years before the source crosses the position of the 
main caustic. This translates into a dramatic increase in the probability of seeing caustic events before (but also after) crossing the position of the main caustic. 
\begin{figure}  
 \centerline{ \includegraphics[width=9cm]{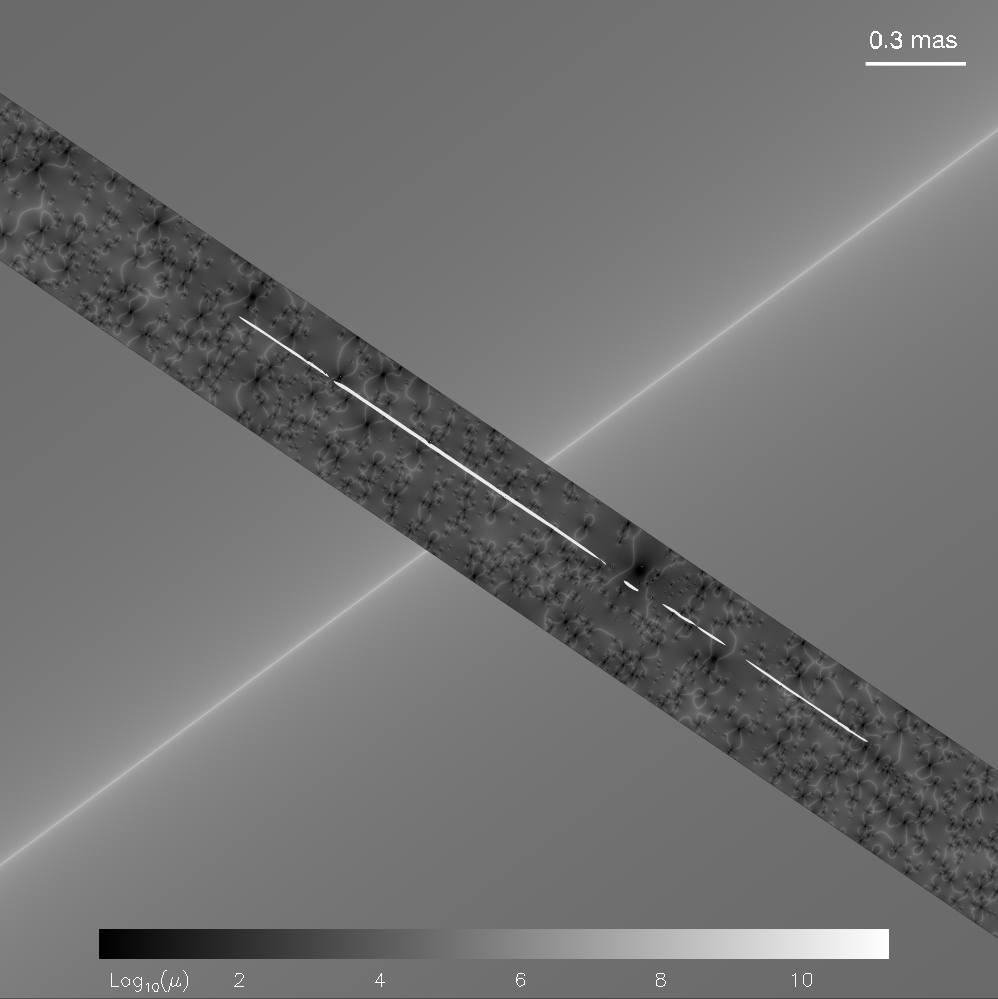}}
  \caption{The diagonal band at $\sim -40^{\circ}$ shows the magnification pattern when microlenses are added in the lens plane around the position of the main 
           CC. The light-gray broken line close to the middle of the band is the lensed image of a background source (or train of micro-images). 
           The rest of the image shows the magnification of the smooth lens model. For illustration purposes, the background source ($R \approx 0.01$\,pc) 
           is much larger than a typical giant star ($R \approx 10^{-5}$\,pc).}
   \label{Fig_SmothCC_vs_MicroCC}
\end{figure}  
\subsection{Source Plane Interpretation}\label{SectSourceplane}
%
\begin{figure*}  
 \centerline{
 \includegraphics[width=9cm]{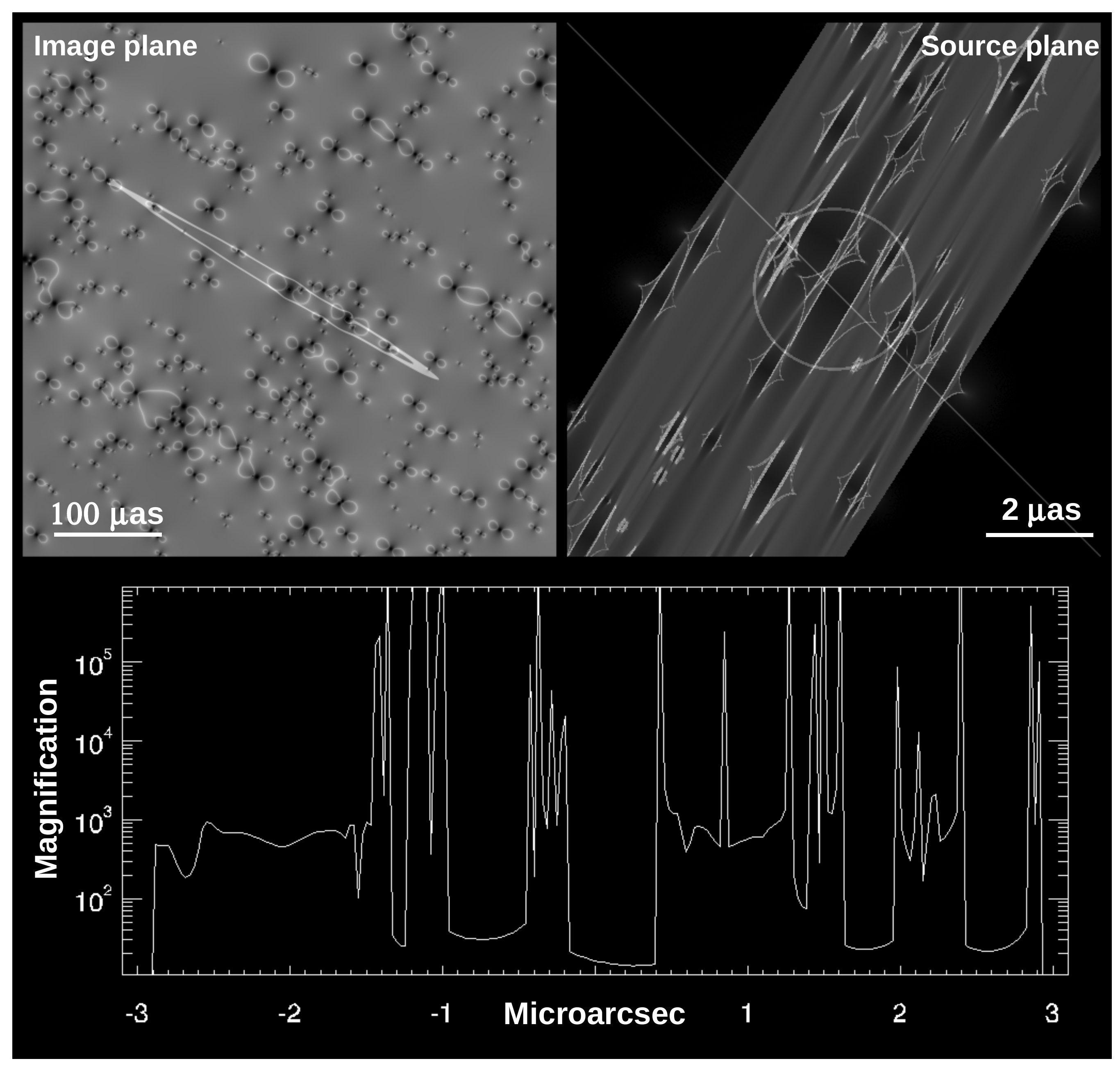}
 \includegraphics[width=9cm]{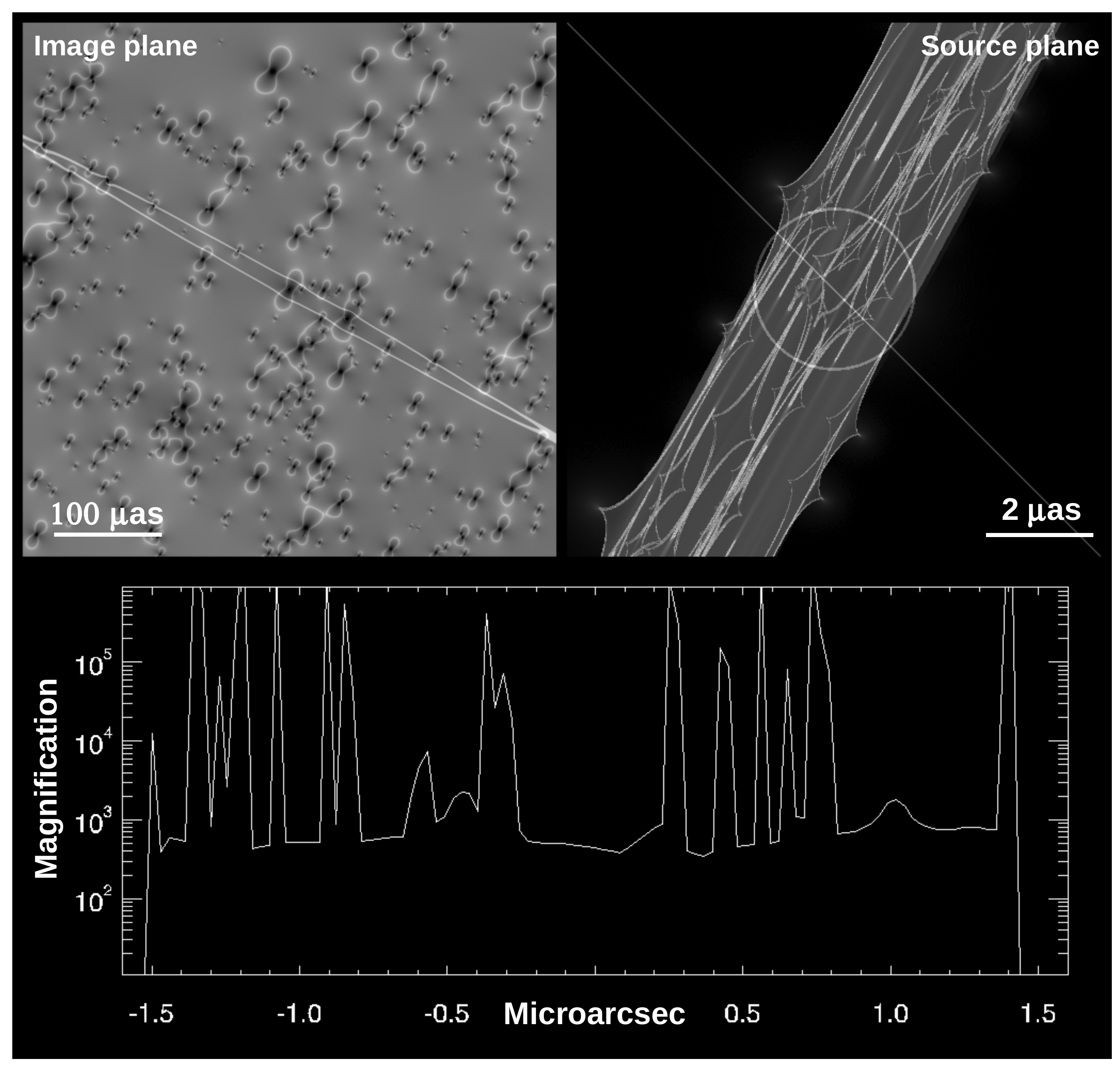}}
  \caption{Image and source plane for a small region. The left panel corresponds to an area on the side with negative parity, while the right panel is for an area on the side with positive parity. In each panel, the top-left region shows the image plane and the top-right a zoomed region in the image plane. A source with a geometry similar to the circle in the source plane would map into the marked  elongated ellipse in the image plane. The diagonal line in the source plane marks the track shown in the bottom part of each panel where the maximum magnification per pixel is represented.} 
   \label{Fig_SourcePlane}  
\end{figure*}  
The plots in the previous section show the magnification pattern in the image plane. 
However, the magnification in the image plane for individual micro-images is normally not observed (unless the number density of microlenses is very small). To better understand how the magnification of the multiple images work, it is more useful to represent the magnification in the source plane. The mapping between the image and lens planes is done through the lens equation. 

In this subsection we present a small portion of the source plane computed at higher resolution than the simulations used in the bulk of this work. The higher resolution is attained by interpolating the deflection field from the main simulation to smaller scales. Figure~\ref{Fig_SourcePlane} shows the image plane, source plane, and a cross-section of the magnification in the source plane for both sides with negative and positive parities. The image planes display the characteristic hourglass shapes discussed in previous sections. The source plane shows the familiar overlapping and stretched diamond shapes. The side having negative parity clearly reveals the gap with low magnification in the central region of the diamond-shape caustics. This is a familiar result found in earlier work \citep[see, for instance,][]{Chang1984,Schechter2002}. This gap of low magnification results in periods of low flux in the observed light curves, on the side with negative parity. 
The bottom plot in each panel shows the cross section along the diagonal line in the source plane. 

The circles in the source plane represent a source with a radius of $\sim 1.5$ microarcsec. That source would simultaneously see the caustics from the negative and positive parity sides, but we have separated them here for clarity purposes. The corresponding amplified image is shown as an ellipse (with small distortions) in the image plane. A source with this size would produce only one counterimage on the side with negative parity, and another one on the side with positive parity (i.e., only two macro-images and no micro-images) since the size of the source is significantly larger than the characteristic scale of the microcaustics. Also, the counterimage on the side with negative parity could not be hidden by microlenses with masses similar to those in the simulation, since portions of the source would always overlap with regions of high magnification in the source plane.
 
Note how a source that is significantly larger than the width of the microcaustics produces counterimages that are notably less magnified on the side with negative parity. This phenomenon would not take place if there where no microlenses, since in that case the magnification within the circular region would be very similar for both parities. This can be understood if we integrate the total magnification within the circle in the source plane. The gap between the caustics results in a smaller total magnification in the enclosed area. Consequently, relatively small regions in the source plane, of angular size a few times the typical size of the caustics (like the circular regions in Fig.~\ref{Fig_SourcePlane} or $R \approx 1.5$ microarcsec $\approx 0.01$ parsecs at $z\approx 1.5$), could show a ratio in the flux between the two counterimages (positive/negative parity) of $\sim 1.3$ (although the exact value depends on multiple factors like source size, mass of microlenses, distance to the macro-CC, etc.). A similar property has been exploited in the context of QSO microlensing, as for instance by \citet{Mediavilla2017}. 
An explanation of this phenomenon is given by \cite{Schechter2002}, where the authors demonstrate with a simple toy model how macrominima and macrosaddle points can have their magnifications affected significantly in the presence of microlenses.

Finally, movies 7 and 8 at https://cosmicspectator.org/2017/06/30/dark-matter-under-the-microscope/ show how the observed magnification depends on the point where the trajectory of the background star intersects the microcaustics. 

In order to get a better view of the source plane in the different scenarios, we make an ensemble of alternative simulations at even higher resolution where we vary both the magnification of the macromodel and the surface mass density of microlenses. We adopt the same redshift for the cluster lens and background source as in the Icarus and Iapyx events. For this particular set of simulations, we adopt a simplified model where the surface mass density from the macromodel ($\kappa_m$) and the shear ($\gamma_m$) are fixed, instead of adopting the model from \citet{Diego2016} used in the main simulations. This is valid approximation since the simulated region is very small. By varying  $\kappa_m$ and $\gamma_m$, we can easily simulate a given region of the lens plane with the desired  magnification from the macromodel. For simplicity, we also assume that the component of the shear in the vertical direction is zero (that is, $\gamma_m=\sqrt{\gamma_1^2 + \gamma_2^2}=\gamma_1$), so the deflection field has its main component in the horizontal direction. Since both $\kappa_m$ and $\gamma_m$ can be expressed in terms of the derivatives of the deflection field ($\alpha$), these relations can be reversed, and we can also express the derivatives of the potential as a function of  $\kappa_m$ and $\gamma_m$,
\begin{equation}
  \alpha_x^x = \frac{1}{2}(\kappa_m + \gamma_m),
\end{equation}
\begin{equation}
  \alpha_y^y = \frac{1}{2}(\kappa_m - \gamma_m),
\end{equation}
\begin{equation}
  \alpha_x^y = \alpha_y^x = \gamma_2 = 0,
\end{equation}
where $\alpha_i^j$ is the derivative of the $i$ component of $\alpha$ with respect the coordinate $j$. 

With these equations we can describe the deflection field (except for an irrelevant constant) of the macromodel. The deflection field from a population of microlenses is added linearly. When considering microlenses, instead of $\kappa_m$ one needs to use $\kappa_m - \kappa_{*}$, with $\kappa_{*}$ the convergence from the surface mass density of microlenses. This guarantees that the total convergence (cluster plus microlenses) is equal to the target $\kappa_m$. 

The specific values of $\kappa_m$ and $\gamma_m$ are determined by the value of the magnification to be simulated, $\mu_m=\mu_t\times\mu_r$. One can easily find that for the side with negative parity $\gamma_m = (\mu_r^{-1}+\mu_t^{-1})/2$ and  $\kappa_m = 1-\gamma_m+\mu_r^{-1}$, while for the side with positive parity we have  the condition $\gamma_m = (\mu_t^{-1}-\mu_r^{-1})/2$ and $\kappa_m = 1-\gamma_m-\mu_r^{-1}$. We vary $\mu_r$, $\mu_t$, and $\kappa_{*}$, producing a set of simulations for the cases with positive and negative parities. The simulations consider a large circle of radius 0.465 milliarcsec where we place the microlenses randomly. The pixel scale is 31 nanoarcseconds and we compute the total deflection field in a narrow horizontal band of width 744 $\mu$arcseconds and height 93 $\mu$arcseconds By construction, this area maps into an area a factor $\mu=\mu_r\times \mu_t$ times smaller in the source plane. Without loss of generality, we fix $\mu_t=1.5$, so in the source plane and to first order, the simulated region maps into an area of dimension $93/1.5\times744\mu/1.5$ arcsecond$^2$. As discussed below, at high optical depth, the effective magnification is smaller than the one for the macromodel, so the simulated region can be larger than the one inferred from the values above. 

By inverse ray tracing, we compute the magnification in the source plane after interpolating the original deflection field to achieve effective resolutions of $\sim 3$ nanoarcseconds per pixel (or $\sim 1000$ solar radii at $z=1.5$). 
\begin{figure*}  
 \centerline{ \includegraphics[width=18cm]{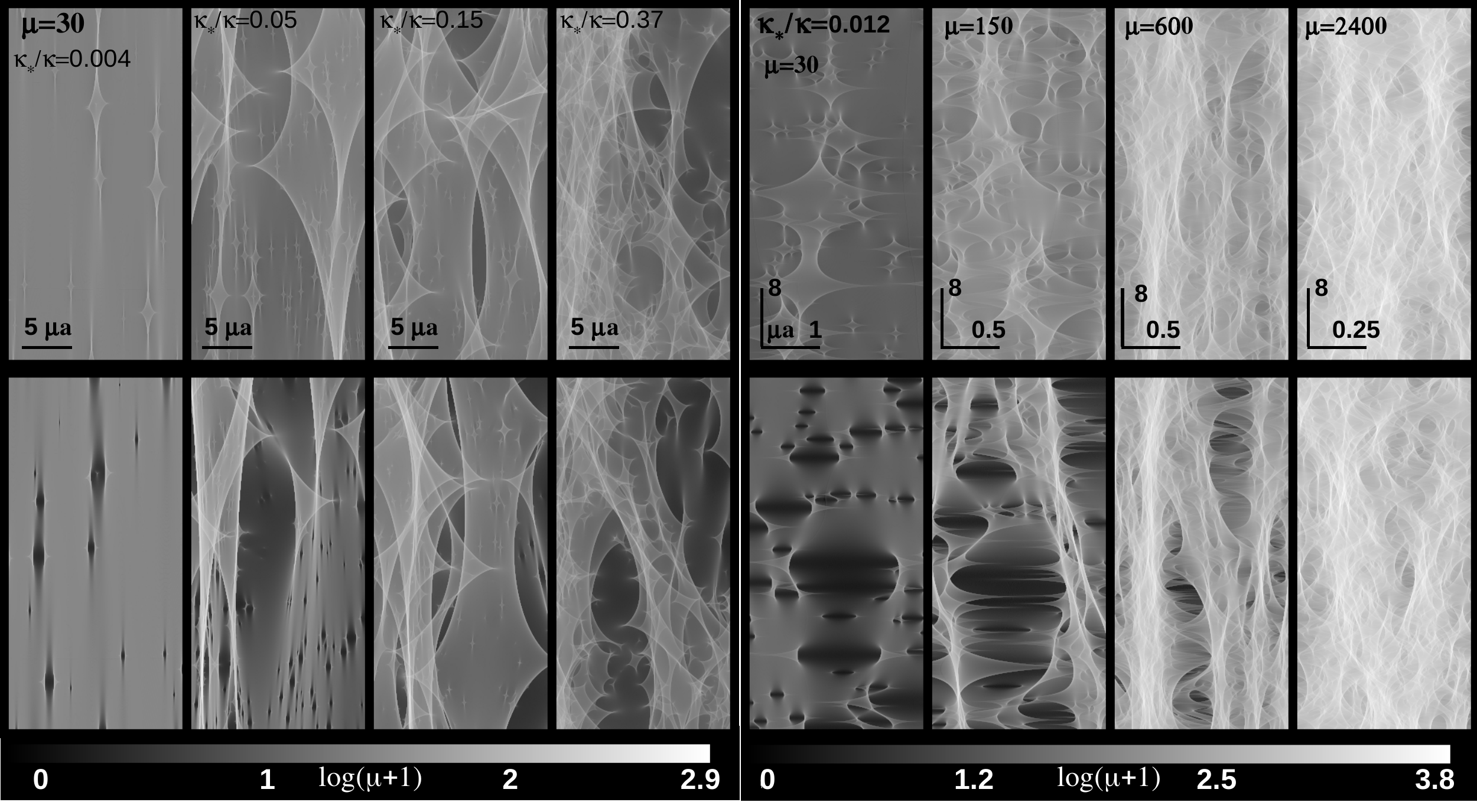}}
  \caption{Zoom-in of the source plane at high resolution. The left block (with 8 panels) shows a small region in the source plane at constant magnification from the macromodel and varying surface mass density of microlenses. The upper row shows the plane with positive parity and the bottom row shows the corresponding plane with negative parity. Both planes overlap in the source plane but are displayed separate here for clarity purposes. The right  block with 8 panels shows the source plane at constant surface mass density of microlenses and varying magnification from the macromodel. The upper and bottom rows correspond to the planes with positive and negative parity, respectively. Note how the source plane has been compressed in the $y$ direction by factors ranging from 8 at $\mu=30$ to 32 at  $\mu=2400$. At very high magnification, both planes with negative and positive parity resemble each other. A source at $z=1.5$ traveling at $1000$ km s$^{-1}$ with respect to the caustics would move $\sim 1 \mu$arcsec every ten years.
          } 
   \label{Fig_Microcaustics}  
\end{figure*}  
A small area of the source plane is shown in Fig.~\ref{Fig_Microcaustics} for each simulation. The results in this plot are divided into two groups. On the right side we show the source plane at fixed $\kappa_{*}$ but varying magnification. On the left we show the simulated source plane at fixed magnification but varying $\kappa_{*}$. The magnification considered for this example is moderate ($\mu=30$), but it serves our purposes as it shows better the structure of the caustics in the source plane. For larger magnifications the behavior would be qualitatively similar to what is shown on the left block of  Fig.~\ref{Fig_Microcaustics}. Since we have the sides with opposite parity projecting back into the same region in the source plane, at a given pixel in the source plane one would get a bundle of rays (from the inverse ray tracing method) coming from the side with positive parity and a different bundle coming from the side with negative parity. 
The mapping between the image plane and the source plane around a critical curve can be visualized as a sheet of paper being folded by its middle point (critical curve). In the source plane, the fold represents the caustic, with the two halves of the sheet forming two overlapping planes. A source will project into these two planes; when unfolded (i.e., the image plane), the sheet of paper will show two images which are symmetric with respect to the folding line.

To better illustrate the differences between the sides with positive and negative parities, we show the source plane for each of the two planes in the source plane described at the end of the previous paragraph and also compute the statistics of each plane separately. This makes sense when comparing with observations, since the statistics of the observed lensed image depends on the parity as we show later and has been demonstrated in earlier work \citep[e.g.,][]{Schechter2002}. The top row shows the plane with positive parity while the bottom row shows the plane with negative parity, where the characteristic microsaddle points with low magnification can be appreciated clearly. The two columns with 
 $\kappa_{*}/\kappa=0.05$ and  $\kappa_{*}/\kappa=0.15$ in the left block contain a stellar component consistent with the upper limit in K17 (that is, with 19 M$_{\odot}$\,pc$^{-2}$ or  $\kappa_{*}/\kappa=0.012$ and a Salpeter spectrum in the low-mass regime) plus microlenses with  30 M$_{\odot}$ mimicking a monochromatic population of PBHs. As mentioned above, the value 19 M$_{\odot}$\,pc$^{-2}$ is motivated by the updated estimate in K17. This is almost 3 times more mass than the value of $\Sigma_o=7 {\rm M}_{\odot}$\,pc$^{-2}$ used in the rest of this work, but the careful reader will notice that $\kappa_{*}/\kappa$ is instead a factor 4 larger than the value of 3\% corresponding to $\Sigma_o$. This is due to the fact that for this set of high-resolution simulations with 19 M$_{\odot}$\,pc$^{-2}$, the value of $\kappa$ considered is 0.66 instead of the value from the model in D16 ($\kappa=0.9$) used in the rest of this work.  $\kappa \approx 0.66$ is the value required by $\mu$ once $\mu_t$ is fixed to 1.5 and $\mu_r>>1$ as described above.
The left panel with $\kappa_{*}/\kappa=0.004$ has approximately four times fewer microlenses than the model with  $\kappa_{*}/\kappa=0.012$ and no PBHs (of 30 M$_{\odot}$). The last column with  $\kappa_{*}/\kappa=0.37$ represents a population of PBHs but with a power-law spectrum (for the mass function) similar to the one used to simulate the stellar component from the ICL. 

On the right side of Fig.~\ref{Fig_Microcaustics} we show similar plots, but this time the surface mass density of microlenses is fixed (to a value consistent with the upper limit on  $\kappa_{*}$ in \citealt{Kelly2017}, $\kappa_{*}/\kappa=0.012$ and no PBHs) and we vary the magnification. The case with $\mu=30$ can be compared directly with the cases presented in the left block. 

\begin{figure*}  
 \centerline{\includegraphics[width=9.5cm]{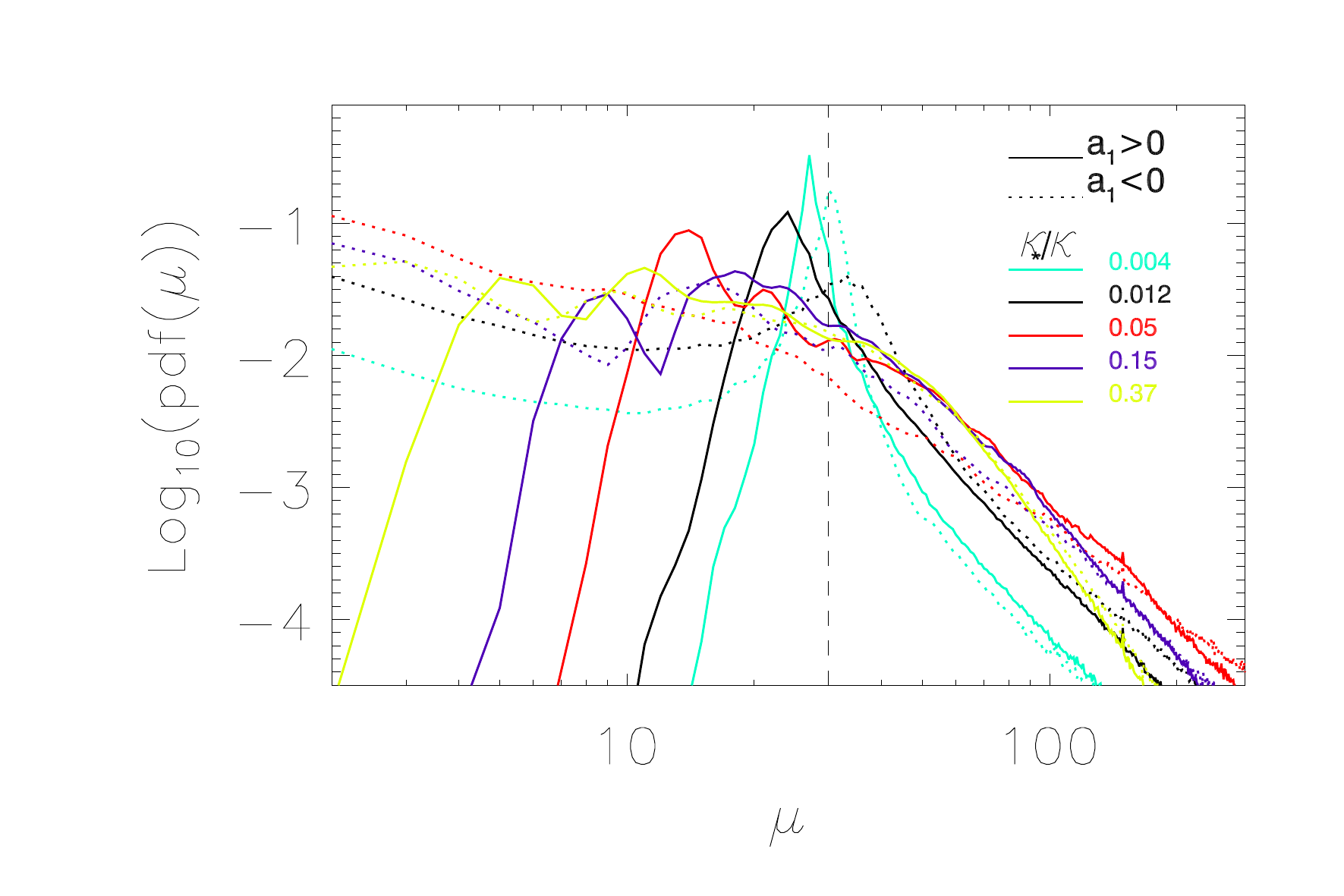}
             \includegraphics[width=9cm]{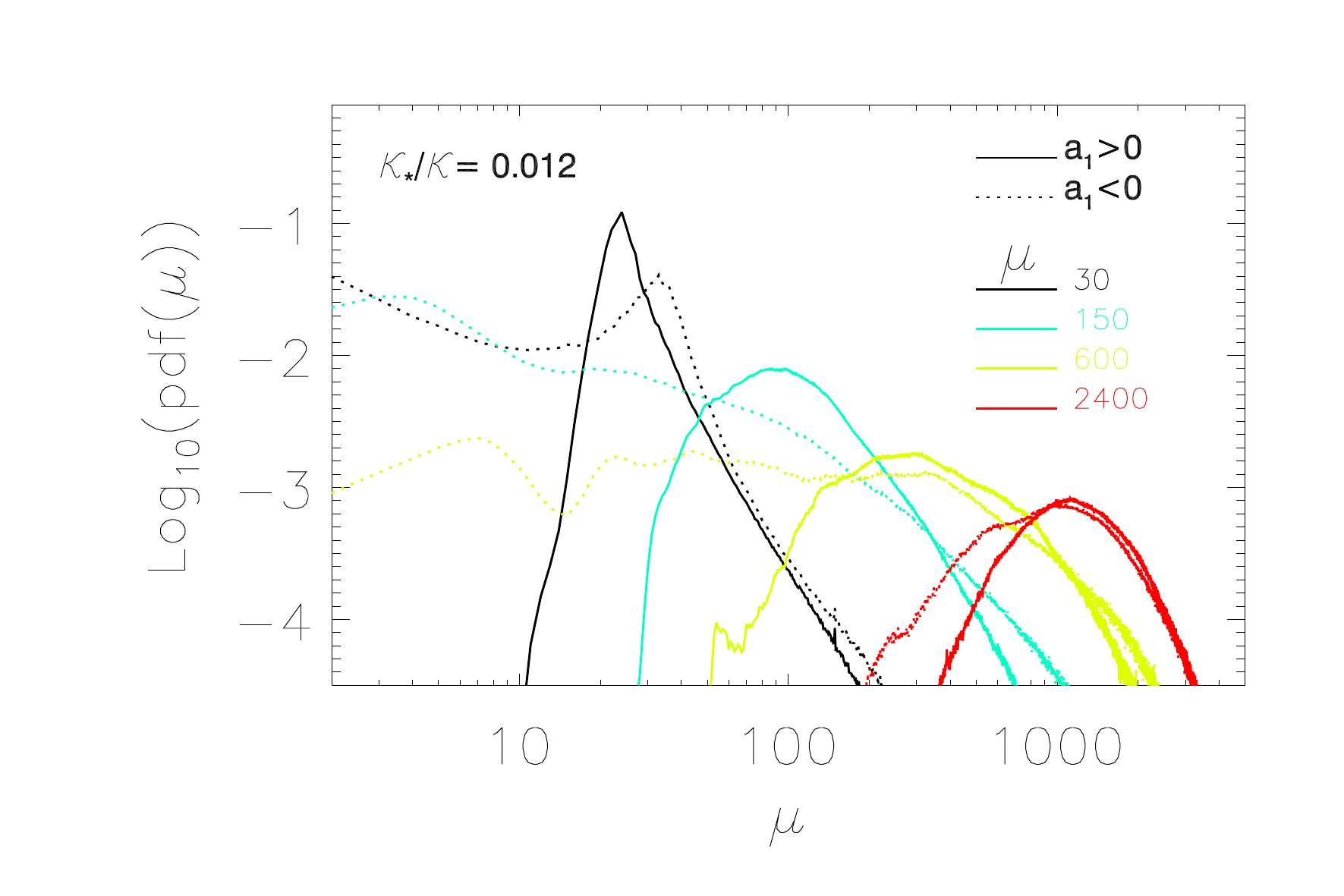}}
  \caption{{\em Left panel} Probability of magnification extracted from the simulations shown in Fig.~\ref{Fig_Microcaustics} (left block). All models have the same magnification from the macromodel ($\mu=30$) marked with a vertical dashed line. For each curve we vary the surface mass density of microlenses and the parity. Positive parity is shown with solid lines and negative parity with dotted lines. At low optical depths, the probability of high (and low) magnification grows as the surface mass density of microlenses. The tail at high magnification falls as the expected $\mu^{-3}$. At high optical depth of microlenses this scaling breaks down and reverses at the saturation regime for the tail at high magnification. {\em Right panel} Similar to the left panel but the surface mass density of microlenses is fixed and we vary the magnification of the macromodel. Note how at high magnification the probability converges towards a low-normal and starts to look similar for both planes with positive and negative parities. Also, the side with negative parity has a higher probability of having extreme magnifications.
          } 
   \label{Fig_HistogramsMuSource}  
\end{figure*}  

Fig.~\ref{Fig_Microcaustics} makes the degeneracy between $\kappa_{*}$ and $\mu$ evident. When  $\kappa_{*}$  is sufficiently high, the source plane saturates with overlapping caustics. The same is also true for moderate values of $\kappa_{*}$ but at large magnifications. In this case the overlapping of the caustics is produced by the high magnification. In order to quantify the differences, we compute the probability distribution function (PDF) of the magnification in the source plane from these simulations.  

The result is shown in Fig.~\ref{Fig_HistogramsMuSource}. The left plot shows the PDF for the case with fixed magnification and varying $\kappa_{*}$. For small values of $\kappa_{*}$, the PDF of the magnification shows a clear peak near the magnification of the macromodel, $\mu_m$. The side with positive parity peaks at slightly smaller values than $\mu_m$ while the side with negative parity peaks at values very close to $\mu_m$. At high magnification, both sides behave very similarly, with the tail of the PDF falling like $\mu^{-3}$, typical of isolated lenses. At low magnifications, the differences are significant between the two parities. When $\kappa_{*}$ is increased, we observe that the differences between the two parities grow as well. The peaks of the PDF separate more, with the one from positive parity clearly below and the one from the negative parity clearly above $\mu_m$. At high magnifications, there seems to be an excess of probability on the side with negative  parity with respect to the side with positive parity. Also, the probabilities of having high and low magnification increase as $\kappa_{*}$. 

As we increase $\kappa_{*}$, the peaks in the PDF disappear, and for sufficiently large values of $\kappa_{*}$, the PDFs of both parity sides start to resemble each other. This is the saturation regime at which the notion of sides with positive and negative parity loses its meaning (the sign of $a_1=1-\kappa-\gamma$ can adopt positive and negative values on both sides of the main CC). 
 The right plot in Fig.~\ref{Fig_HistogramsMuSource} shows the case of fixed $\kappa_{*}$ and varying magnification. The model with $\mu_m=30$ (black curves) is the same as the 
black curve model in the left plot. We observe a similar trend, but now the excess of probability at high magnification in the side with negative parity is more evident (specially at $\mu=150$ and $\mu=600$). Also, 
for $\mu_m=2400$, the PDF of the sides with negative and positive parity are almost similar and deviate from $\mu^{-3}$ at high magnification. Instead, the PDF resembles a log-normal distributions, which typically appear in multiplicative processes. This is attained by the combined effect of multiple overlapping caustics and the high magnification of the macromodel. Finally, we note that similarly to what happens in the left panel, when we increase the magnification of the macromodel, the average magnification in the simulated region deviates from the one we would have obtained from the macromodel (i.e., with no microlenses). In particular, we find that in the presence of microlenses with $\kappa_{*}/\kappa=0.012$ and in the side with positive (negative) parity, the averages of the magnifications are  29.9 (30.2), 150.4 (144.5), 564.8 (566.3), and 1455.7 (1345.3) for macromodel magnifications of 30, 150, 600, and 2400, respectively.

From results like those shown in Fig.~\ref{Fig_HistogramsMuSource}, one can extract important properties of the magification, but they do not contain all the information. 
The magnification is highly non-Gaussian as shown in this figure, and hence the PDF alone gives an incomplete picture of the problem. For instance, the PDF plots shown in  Fig.~\ref{Fig_HistogramsMuSource} do not account fro the correlations that are evident in  Fig.~\ref{Fig_Microcaustics}. 
Higher-order statistics like the correlation function or power spectrum are useful discriminators in this type of situation. 
%

\section{Disruption of the Cluster CC by Microlenses}\label{sect_S5}
\begin{figure*}  
 \centerline{ \includegraphics[width=18cm]{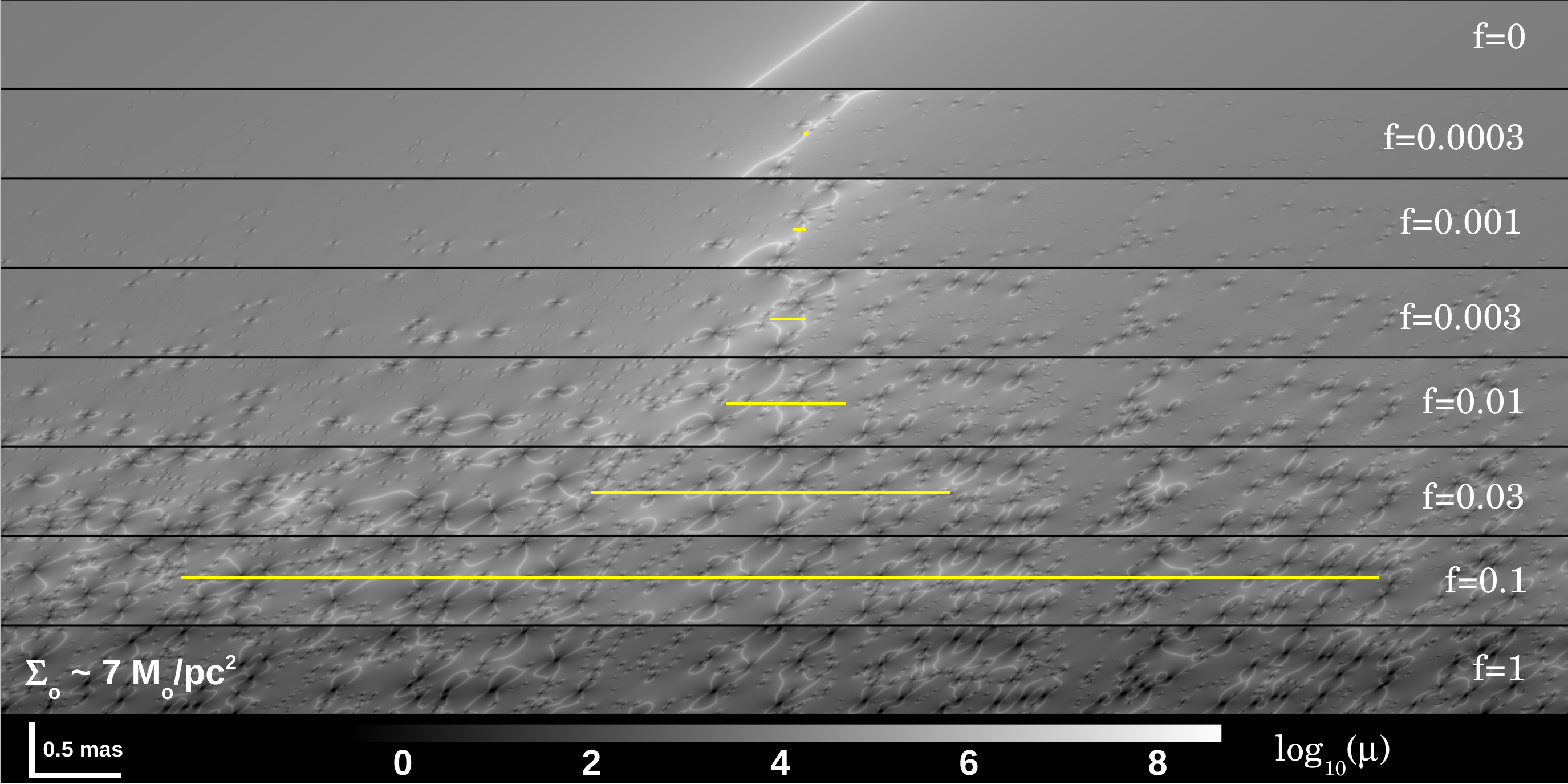}}
  \caption{Disruption of the CC as a function of microlens surface mass density. Each panel shows the CC region when a population 
           of microlenses with $\Sigma = f\Sigma_o$ is present. The case $f=1$ corresponds to the model of Spera (2015) at the position of 
           Icarus. The yellow lines show the approximation in Eq.~\ref{Eq_tau2}. The last panel at bottom does not 
           show the yellow line since it extends beyond the boundaries of the plot. The total surface mass density (i.e., smooth plus microlens) is the same in all panels.} 
   \label{Fig_SmoothCC_vs_FICL}  
\end{figure*}  

In this section we study in more detail the effects of microlenses at the position of, or very close to, the CC. 
For the microlenses we adopt as a reference the Spera (2015) model normalized to $\Sigma_o \approx 7\, {\rm M}_{\odot}$\,pc$^{-2}$ (similar to the surface 
mass density in surviving stars inferred at the position of Icarus). 

When microlenses are present in the vicinity of the CC, the infinitesimally narrow CC widens, with overlapping critical lines that form a complex network (see, e.g., Fig.~\ref{Fig_SmothCC_vs_MicroCC}). 
This network extends up to a maximum range that depends on the total number and masses of the microlenses. In the case of Fig.~\ref{Fig_SmothCC_vs_MicroCC}, this network 
extends well beyond the displayed field of view. The change in the network when the amount of microlenses is varied is made more evident in Fig.~\ref{Fig_SmoothCC_vs_FICL}. 
The magnification pattern gets shifted around, with regions of high magnification becoming regions of low magnification, and vice versa. 
When the amount of microlenses is small (i.e., small fraction $f=\Sigma/\Sigma_o$), the main CC becomes sharper by trading high magnification 
by lower magnification with the surrounding area in the lens plane. As more microlenses are added, the disruption becomes more serious, and at some point between $f=0.003$ and $f=0.01$ in 
Fig.~\ref{Fig_SmoothCC_vs_FICL} the CC itself transforms into a network of micro-CCs. 
The extension of this network around the main CC marks the region where microlensing events are more likely to be observed.
An interesting consequence is that the typical magnification one would expect 
is changed by the addition of microlenses. In Fig.~\ref{Fig_Median} we show the median of the distribution of magnifications (of individual micro-images) computed at different distances from 
the position of the main CC. For each distance, the distribution of magnifications and its median are computed in an area of $1.9 \times 0.5$\,mas$^2$. Adding 
microlenses thus results in a reduction of the typical magnification of micro-images that one would have obtained without them. This median magnification, however, cannot be normally observed, since the lensed image forms (typically) an unresolved train of micro-images and what we observe is the total flux of all micro-images (an exception being at low optical depth, where the the total flux is usually given by one micro-image). We show later, however, that if the lens plane is populated by massive microlenses (a few tens of solar masses), the separation between micro-images can reach a few milliarcseconds, opening the door to future high-resolution observations of the individual micro-images.

This change in the magnification is also evident in Fig.~\ref{Fig_SmothCC_vs_MicroCC}, where we display a small region  around the 
main CC in the case of the smooth lens compared with the magnification pattern when microlenses are added (diagonal band). The figure also shows 
a lensed background object (with $\sim 0.01$\,pc radius), or train of micro-images, at the moment of maximum magnification. The lensed image breaks up into multiple smaller components. 
For smaller background sources (such as a large star), the lensed image would break up into even more smaller pieces.    

When $f$ is sufficiently small, the effect of the microlenses is small and the magnification 
behaves like in the smooth lens model case, except when we approach the CC. At short distances from the CC, even small microlenses can have a significant impact on 
the magnification pattern. As $f$ grows, the range at which the CC gets disrupted grows as well. For values of $f\approx 0.001$ the disruption is still significant up to scales 
of a few milliarcsec. In this situation, if macro-images are being formed on both sides of the main CC at a distance of a few milliarcsec, a telescope like the {\it Hubble Space Telescope (HST)} observing the unresolved macro-images would start to see not only a change in flux over time but also a change in the observed position of the peak, since the observed images would appear to be jumping back and forth between the two sides with opposite parity. 
For values of $f \approx 1$ it is impossible to determine the exact location of the main CC and the magnification pattern is completely disrupted over a scale of 
hundreds of milliarcseconds. If $f \gg 1$ the disruption can extend to scales of an arcsec and these types of microlensing events would be much more common. The fact that no similar microlens event 
has ever been reported before {\it Icarus/Iapyx} is a simple indirect indication that the optical depth of microlenses cannot be much higher than that from the stellar component (with $\kappa \approx 10^{-3}$--$10^{-2}$). 
\begin{figure}  
 \centerline{ \includegraphics[width=9cm]{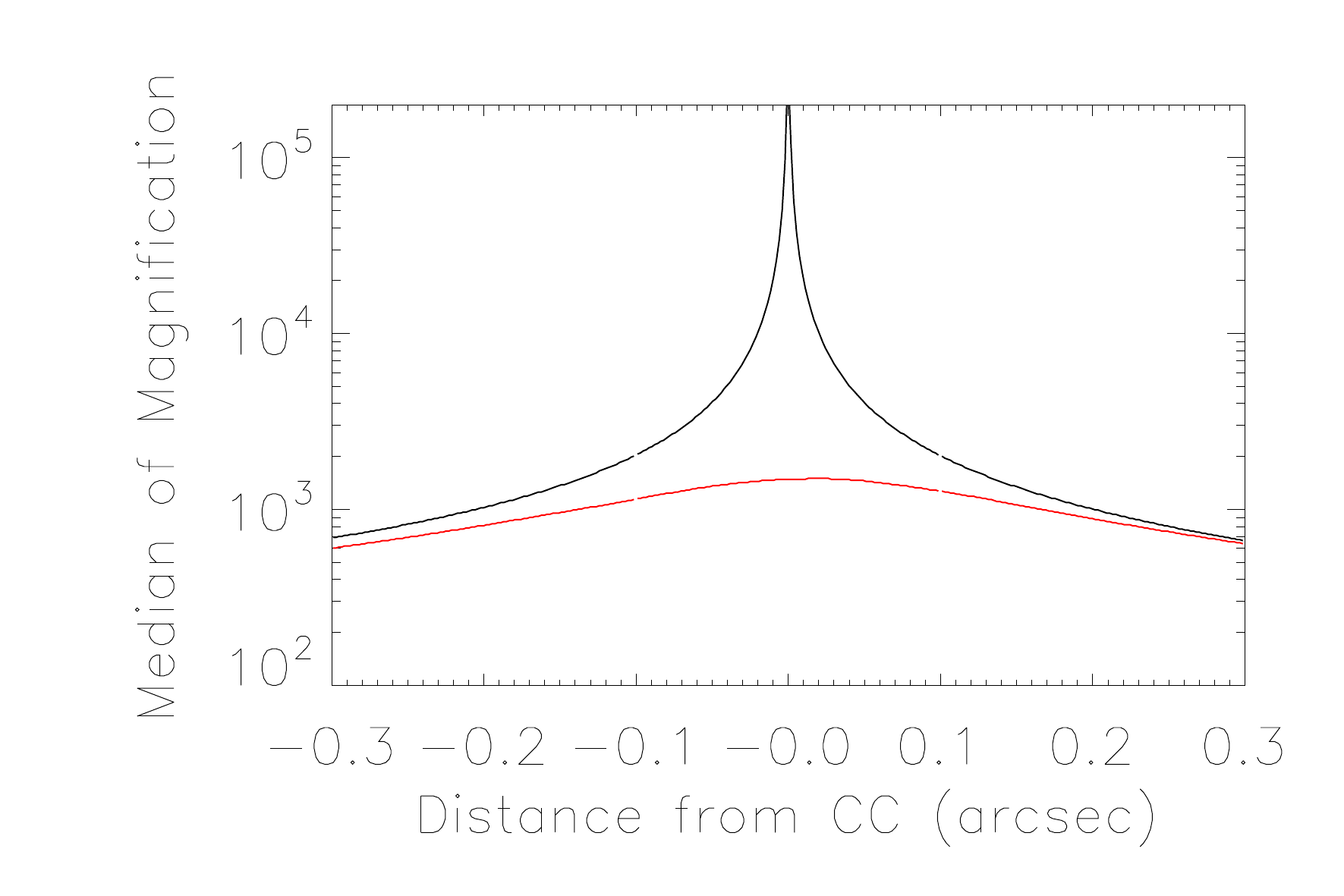}}
  \caption{Median of the magnification (of micro-images in the lens plane) as a function of distance from the main CC 
           (negative distances mean they are measured toward the left of the CC and positive toward the right).
           The black line corresponds to the smooth model (no microlenses) and the 
           red line is for the case when microlenses are added (with $\Sigma \approx 7 \,{\rm M}_{\odot}$\,pc$^{-2}$).}           
   \label{Fig_Median}  
\end{figure}  
%

\section{Light Curves}\label{sect_S6}
%
\begin{figure*}  
 \centerline{\includegraphics[width=9cm]{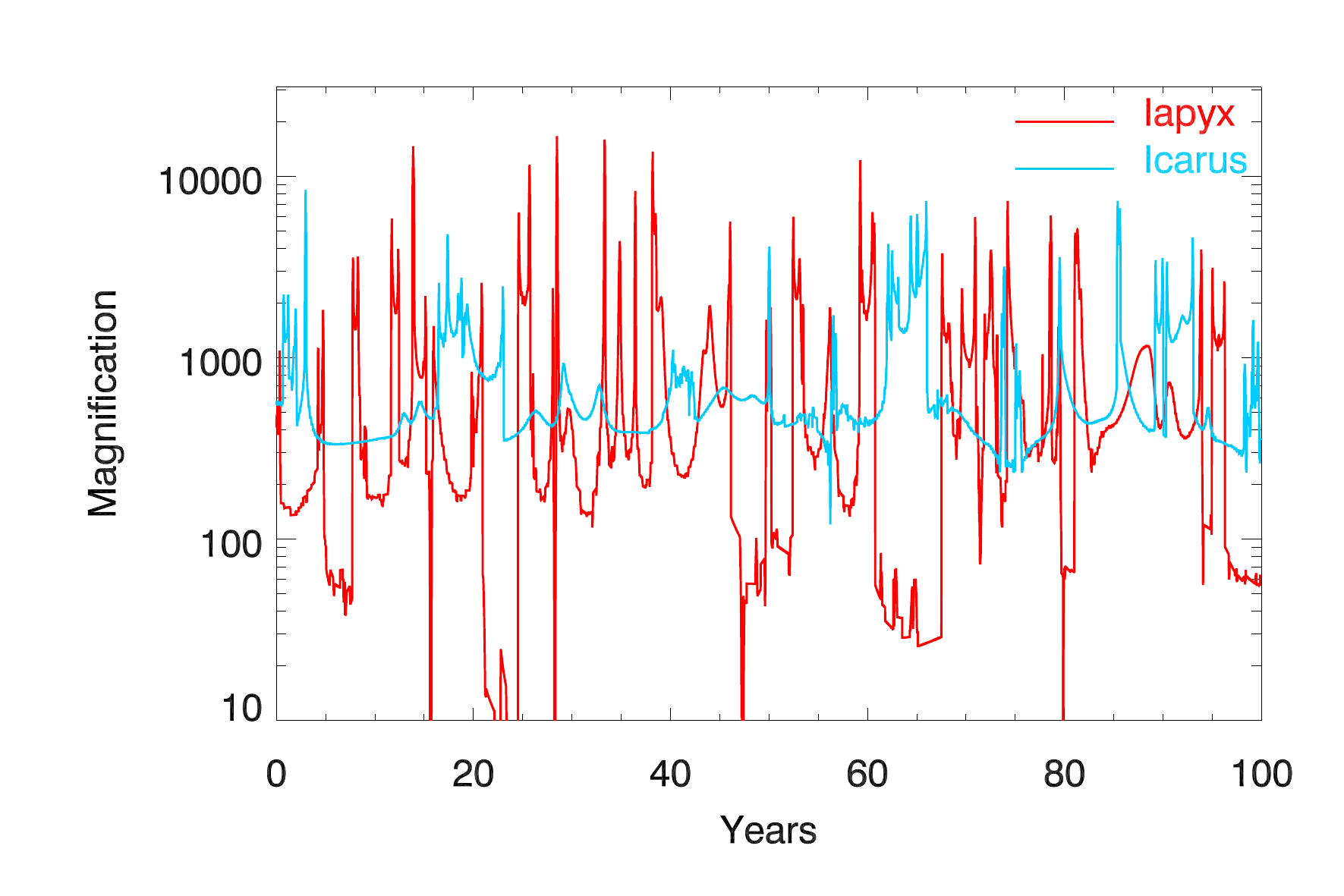}
             \includegraphics[width=9cm]{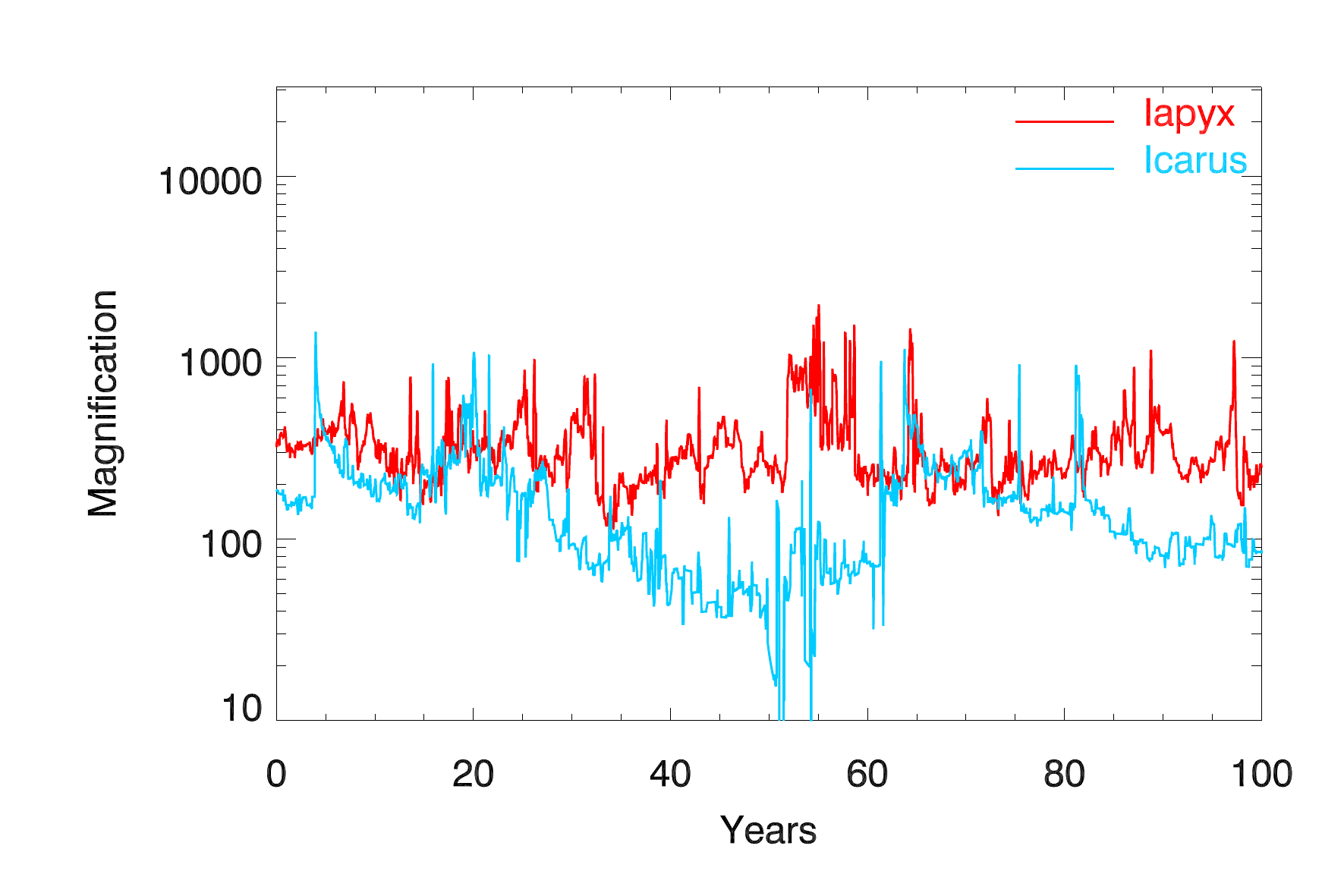}}
  \caption{{\em Left panel}. Fragments of the simulated light curves for Icarus ($a_1>0$) and Iapyx ($a_1<0$) based on the Spera model for the ICL. 
           {\em Right panel}. Similar to the left panel, but when 10\% of DM is substituted by PBHs with 30\,M$_{\odot}$ each. The right panel also includes 
           the ICL microlenses, and the total surface mass density is the same in both cases. }
   \label{Fig_LCSpera400yr}  
\end{figure*}

The results presented in the previous section show examples of the magnification pattern near the main CC and in a narrow region of the source plane when microlenses are present. However, 
observations can only sample the magnification in the source plane (unless micro-images are resolved).
Through observations we can measure the total flux as a function of time --- that is, the light curve. 

To simulate the light curves, we place a background star in the source plane moving with a relative velocity of 1000\,km\,s$^{-1}$ toward the main caustic. The results presented in this section can easily be rescaled (stretched or compressed) to any velocity ($v$) by the factor $v/1000$. A relative velocity of 1000\,km\,s$^{-1}$ is a reasonable assumption 
given the redshifts of the lens and source.  
Figure~\ref{Fig_LCSpera400yr} shows a small segment of the simulated light curves for Icarus (blue) and Iapyx (red). We assume the Spera model (left panel) and the Spera+PBH(10\%) 
model (right panel). 
In the case when microlenses are only ICL stars (left panel), the light curves for Icarus and  Iapyx can be very different when microlenses populate both sides of the main CC. 
Macro-images on the Iapyx side ($a_1<0$) can {\it disappear} for periods of 10\,yr or more.

This is a consequence of the low-magnification regions 
present between the semidiamond-shape caustics discussed in Section~\ref{SectSourceplane}, on this side of the main CC. 

As shown by \cite{Schechter2002}, macrosaddle points (like the ones on the side with $a_1<0$) can be fainter because they can lack microminima, while macrominima must have at least one microminimum.

Another interesting difference is the amplitude of the peaks, which can be higher on the Iapyx side (see Section~\ref{SectSourceplane}, where the same effect is observed in the PDF). This tradeoff between low-magnification periods and higher peaks conserves the total integrated flux when integrating over long periods of time. 
We find that for surface mass densities of microlenses comparable to $\Sigma_o$, the average of a light curve converges (both on the side with positive and negative parities) toward the value of the macromodel when averaging the light curve over a few hundred years.
The right panel of Fig.~\ref{Fig_LCSpera400yr} shows the corresponding light curve when 10\% of the mass in the lens plane is substituted by PBHs with a mass of 30\,M$_{\odot}$ each (this case also includes the microlenses from the ICL).
In this case, the light curves on both sides of the main CC are more similar, and we do not observe periods of low magnification on the side with $a_1<0$. We also note that when the fraction of PBHs is high, the clustering of PBHs introduces large-scale temporal and spatial correlations in the magnification pattern that can result in long periods of relatively low or high magnification, as shown in the right panel of Fig.~\ref{Fig_LCSpera400yr}.
As we show below in Section \ref{sect_S8}, when the optical depth of microlenses is sufficiently high (for instance, when 10\% of all mass is in microlenses), we are in the saturation regime and 
the properties of the light curves must be similar. We also notice that the peaks on both sides are also smaller as a consequence of the reduction in Einstein radius. 
The associated Einstein radius of the microlenses no longer scales like $(M\mu_t)^{1/2}$; more precisely, the effective $\mu_t$ is smaller when we reach the saturation regime. This is an 
interesting result since it implies that an event like {\it Icarus} would require, on average, a brighter star if a significant fraction of the DM in the lens plane is made of PBHs. At the same 
time, hiding {\it Iapyx} for at least 10\,yr seems unlikely in this scenario and would require the presence of a second (and fainter) star in the source plane to explain {\it Iapyx}. 

This argument is in agreement with the probabilities estimated by K17, which show that a scenario where a sizable fraction (more than a few percent) of the DM is in the form of compact lenses is less agreeable with the data than a scenario where the microlenses are just the ICL stars. Also, K17 found that simulations suggest that the existing data have sensitivity to the mass function of stars and remnants (given the assumptions made about the magnification and stellar mass density). 
However, the observed light curve presented by K17 does not have as many data points as one would desire to achieve good constraining power. A clear discrimination between different models (other than a preference for models with a low fraction of PBHs or microlenses forming binary systems as discussed by K17) is not yet possible. More data are needed (more peaks, better cadence, smaller photometric errors) in order to clearly distinguish between the different possible scenarios. We hope that regular monitoring of this cluster will provide such data in the near future.

If we imagine we can monitor Icarus and Iapyx for $\sim 10,000$\,yr, we
would witness the moment where both events merge and disappear as the background star crosses the last microcaustic (see Section \ref{sect_S7}). 
If the deflection field were perfectly smooth (without microlenses), the light curves of Icarus and Iapyx would be featureless and the flux would increase as $1/\sqrt{t-t_o}$, as the background star approaches the main caustic and would reach a magnification of $\sim 10^6$ in the last moments before disappearing. 
Conversely, if the lens plane near the CC is populated with microlenses from the ICL stars, PBHs, or both, the light curves will be rich in features like the ones 
shown in Fig.~\ref{Fig_LCSpera400yr}. In this scenario, we will observe hundreds or thousands of peaks, each having a magnification between $\sim 10^3$ and  $\sim 10^4$. 
The total flux integrated over the 10,000\,yr will be the same, but the light curves will be very different.  

It is also interesting to track the fluxes of the individual micro-images. 
In Fig.~\ref{Fig_LC_MaxPixel} we compare a fragment of the simulated light curve (for Icarus) with the magnification in the lens plane at the position where the 
brightest micro-image is formed. During the valley periods, both magnifications coincide because there is only one dominant micro-image (there are other counterimages but 
with significantly smaller fluxes), so the total flux of the macro-image is basically given by the flux of the dominant micro-image. 
During a peak, new bright micro-images form around the micro-CC. The total flux is then typically larger by a factor 2 or 3 than the flux of the brightest micro-image. The result shown in Fig.~\ref{Fig_LC_MaxPixel} demonstrates how, at moderate optical depth, during most of the time the train of micro-images 
is very compact in size (only one dominant micro-image during the more frequent valley periods). 
However, during a peak event, it may be possible that the separation between different portions of the train 
may be large enough so the overall size of the train of micro-images could be distinguished from that of a point source. This will be explored in more detail later in this paper. 

\subsection{Estimating Rates}

%
\begin{figure}  
 \centerline{\includegraphics[width=9cm]{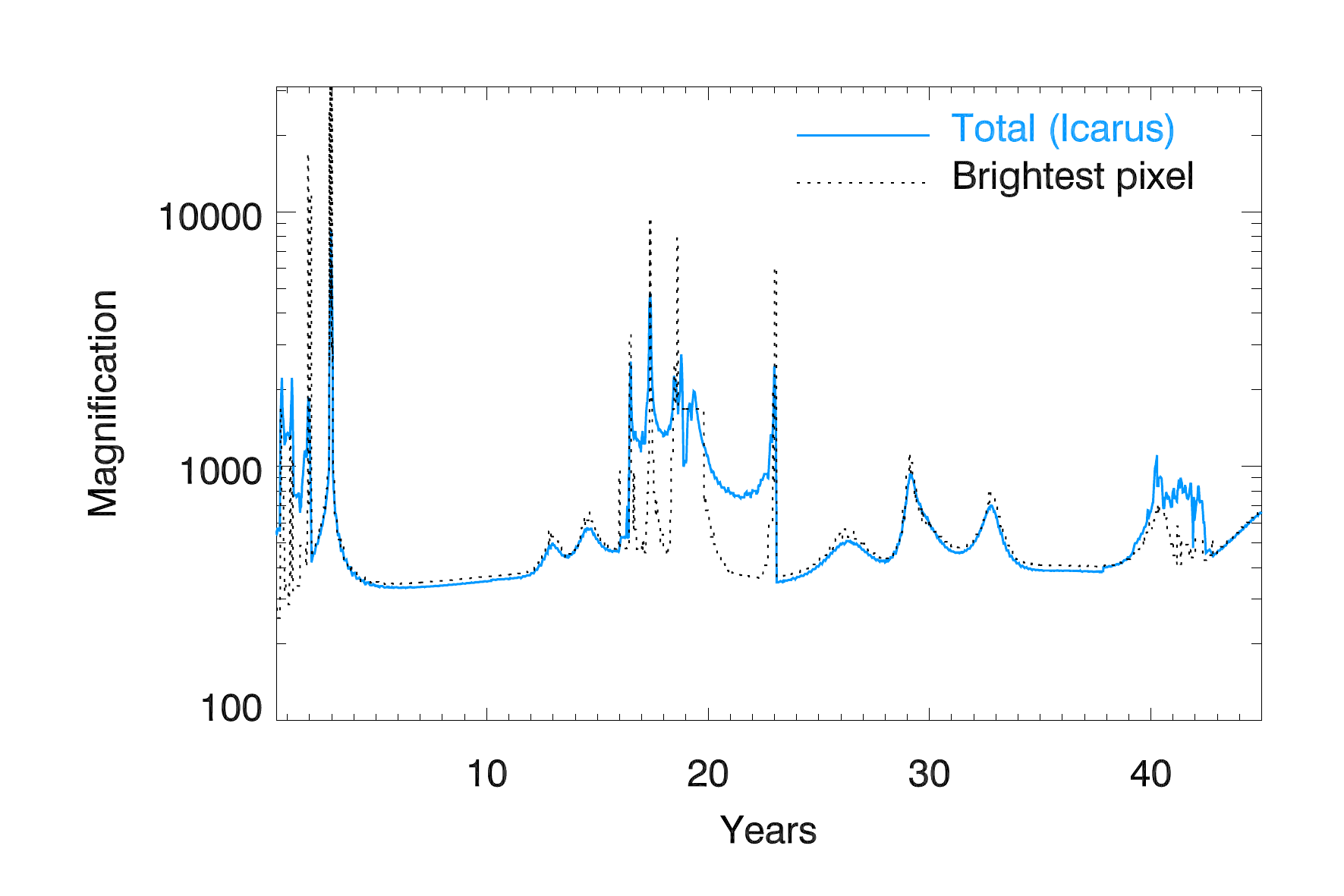}}
  \caption{Fragment of the simulated light curve for Icarus (blue line) compared with the underlying magnification in the pixel of the simulation 
            that contains the largest magnification (dotted line). 
            The dotted line is generally above the solid line because it does not account for the extended nature of the micro-image (which smooths out the observed magnification).} 
   \label{Fig_LC_MaxPixel}  
\end{figure}  
The simulated light curves provide a detailed picture of the number of peaks expected over a time period, but they are computationally expensive to obtain. The rate of peaks can be 
estimated after doing some simple approximations and taking advantage of the scalings presented in this work. 
First we assume that most of the events are produced by microlenses in the range between 0.5\,${\rm M}_{\odot}$ and 2.0\,${\rm M}_{\odot}$. 
This assumption is reasonable since more-massive microlenses are rare (especially for realistic models where the heaviest stars are short lived, leaving remnants 
with low to moderate masses) and less-massive microlenses have a smaller impact parameter (or Einstein radius) for strong lensing (we should also add that the abundance of low-mass stars 
is not as well constrained, either).

We assume the Spera model with the properties of the lens model of D16 and the surface mass density estimated at the Icarus position (just for convenience, but the arguments given below extend to any surface mass density). For this configuration we expect 
3.8 stars per projected pc$^2$ in the assumed mass range. The typical separation between these stars is then 0.5\,pc or $d\theta=0.08$ milliarcseconds. Taking advantage of 
Eq.~\ref{Eq_beta_theta}, we can compute the corresponding typical separation in the source plane between microcaustics corresponding to microlenses at the position of Icarus 
($\theta=0.13''$), 
$d\beta\approx 2\theta d\theta/\Theta=3.1\times10^{-7}$
arcsec, where we have ignored the term  $d\theta^2/\Theta$ since $d\theta \le \Theta$. 
A background star traveling with a velocity of $v=1000$ km s$^{-1}$ would take $\sim 2.6$\,yr to cover the angular distance $d\beta$, so we should expect 1 microlensing event every 2.6\,yr on average. This number is in reasonable agreement with both the rate estimated from the light curves and the observed rate estimated for Icarus (see K17). 

\begin{figure*}  
 \centerline{ \includegraphics[width=8cm]{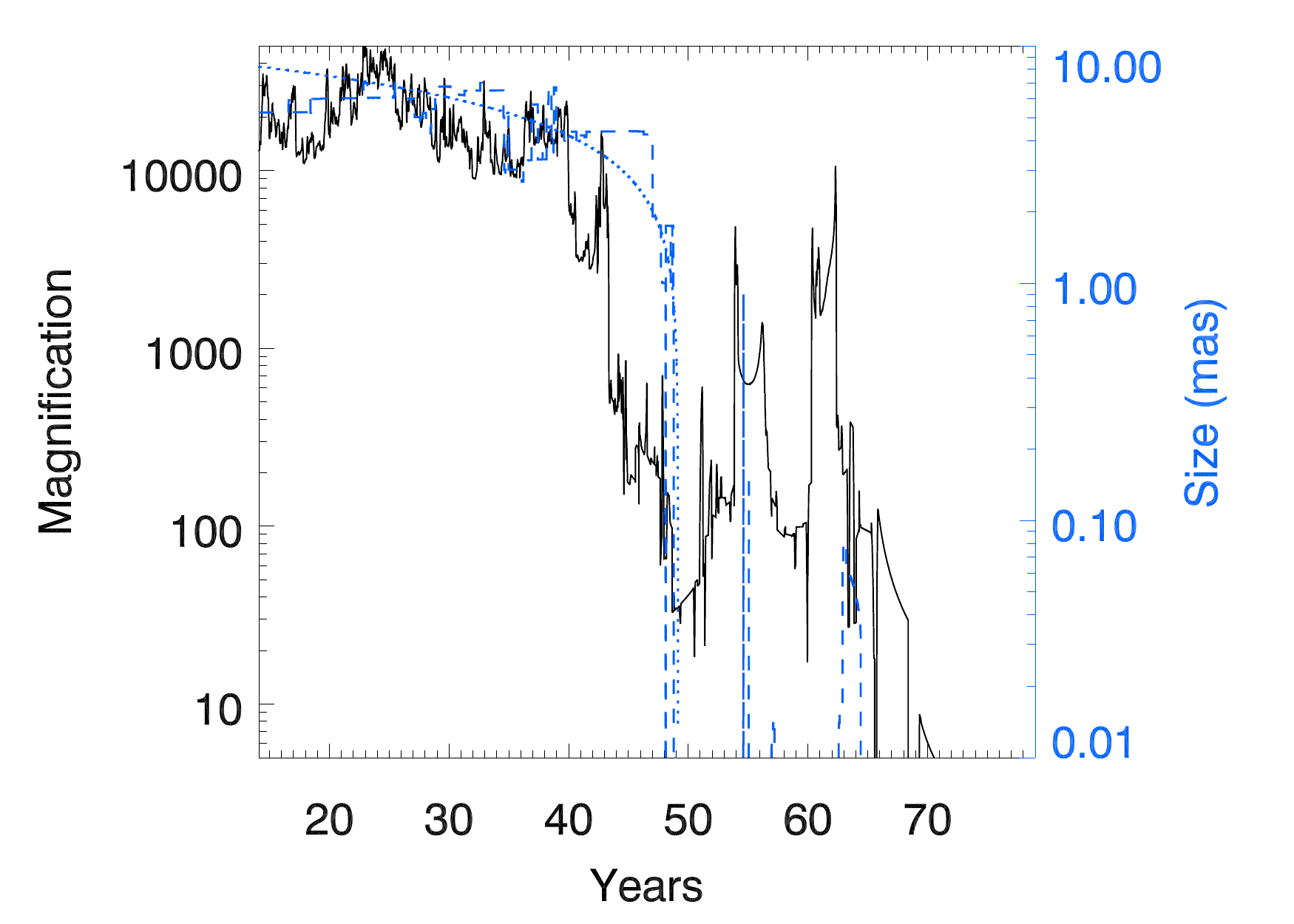}
              \includegraphics[width=8cm]{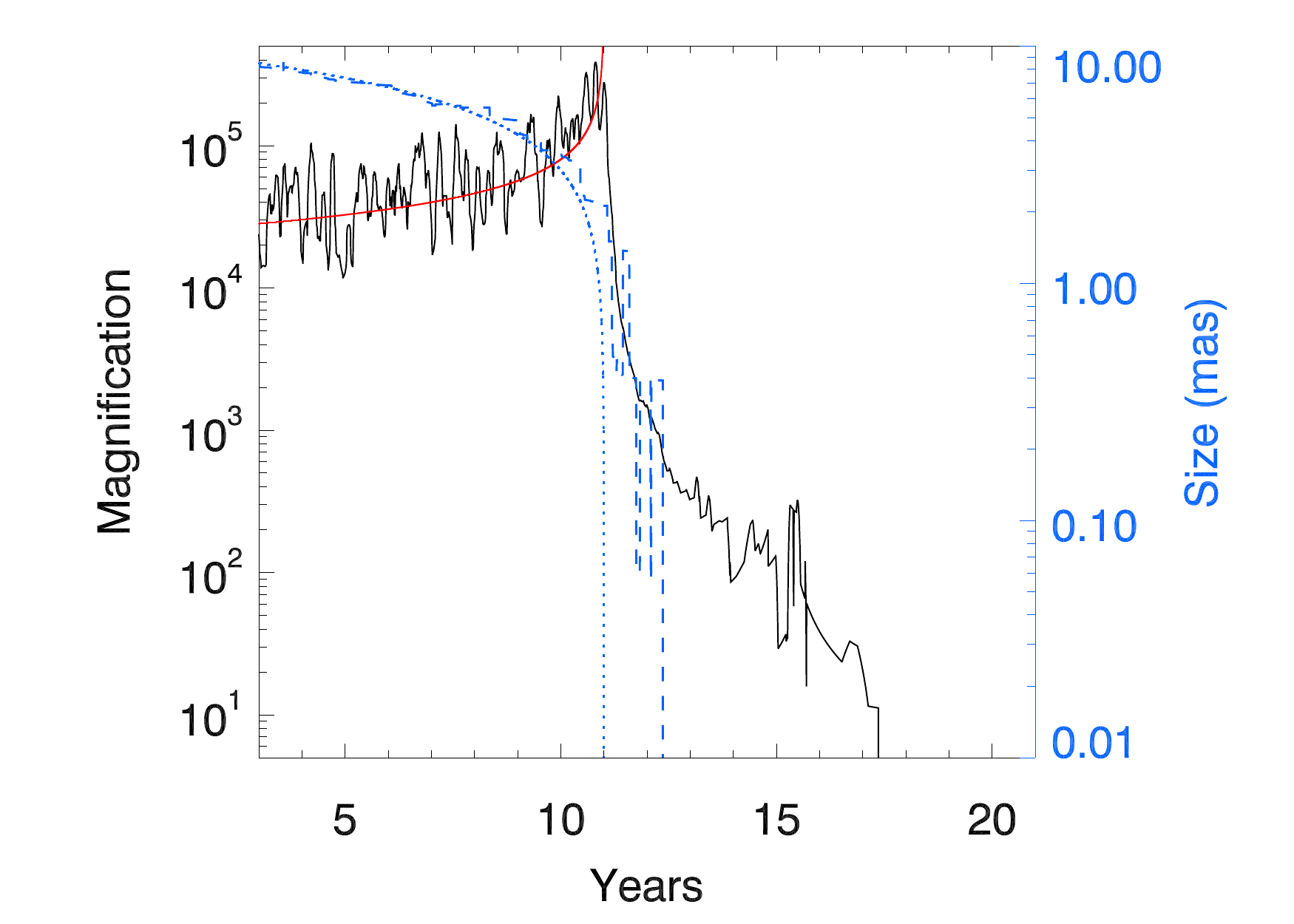}
            }
  \caption{{\em Left panel:} Magnification during the last years before the disappearance of Icarus. The dashed line shows the maximum separation between all micro-images on 
          both sides of the CC (or the size of the combined train of micro-images). 
         This separation can be seen as the maximum extension of the set of micro-images that could be resolved by a telescope with microarcsecond resolution. 
          The dotted line shows the expected theoretical behavior for the change in size in the smooth lens model, 
          ${\rm Size} \propto \sqrt{|t-t_o|}$, where $t_o$ is the caustic crossing time. The exact crossing time is undetermined since there is not a single caustic, 
          but a reasonable choice would be the point where the magnification is maximum.
          The microlenses are consistent with the ICL constraints at the position of Icarus and assuming a Spera model ($f=1$ in Fig.~\ref{Fig_SmoothCC_vs_FICL}). 
          {\em Right panel:} Similar to the left panel, but with a smaller contribution from microlenses (30 times smaller, 
           or $f=0.03$ in Fig.~\ref{Fig_SmoothCC_vs_FICL}). 
           Note how owing to the increase in magnification the separation between the two trains of micro-images evolves more quickly. 
           The red smooth line is the expected magnification for a smooth model ($\propto 1/\sqrt{|t-t_o|})$.
           } 
   \label{Fig_IcarusDeath1}  
\end{figure*}  

This estimate does not take into account the size of the microcaustics, which would grow as we approach the main caustic or would be smaller as we move away from it. 
A more accurate estimate of the rate should take into account this filling factor, or optical depth. Since the optical depth of microlenses depends on the distance to the main CC, 
at large distances (smaller optical depths) the rate of events should decrease significantly as the separation between microcaustics gets larger. 
For the microlenses assumed at the beginning of this subsection, the Einstein radius ranges between $1.3\ \mu$arcsec and $2.5\ \mu$arcsec before we account for the magnification 
effect of the cluster. 

Once the magnifying power of the cluster is taken into account, we have seen how these Einstein radii can be enlarged by 1 order of magnitude when the distance to the main CC is $0.13''$ ($\mu_T \approx 100$; see Eq.~\ref{Eq_thetaE}), bringing the Einstein radii to $\sim 20$\,$\mu$arcsec. 
Since the typical separation between microlenses in the lens plane is $\sim 80$\,$\mu$arcsec (see above in this subsection), it means that an average of 3 out of 4 microlenses will be missed 
by the moving macro-image. Thus, we should expect a ratio of 1 event every $\sim 10$\,yr 
instead of 1 event every 2.6\,yr 
for microlenses with masses between 
$0.5\times{\rm M}_{\odot}$ and $2.0\times{\rm M}_{\odot}$. This is in excellent agreement with the observed light curve for the Icarus event, where the observed light curve in Fig. 3 of K17 suggests one major microlensing event over a period of $\sim 12$\,yr (although we should note that the observed light curve has large gaps). The arguments presented above are based on the simplifying assumption that all microlenses have a mass in a relatively narrow range. More realistic scenarios, like the ones presented in the simulations, will introduce additional variability and even long-scale temporal correlations (not considered above) in the presence of clustered massive microlenses.

\section{The death of Icarus}\label{sect_S7}

Based on our current understanding of the Icarus/Iapyx events, we should expect to see more peaks in the future at the position of these events. 
Assuming the source is heading toward the main caustic, for transverse velocities on the order of 1000\,km\,s$^{-1}$, 
it might take $\sim 10,000$\,yr for the source to cross the main caustic and become unobservable. Conversely, if the background source
is moving away from the main caustic, we are witnessing the lensed source  $\sim 10,000$\,yr after it first crossed the position of the main caustic. 
Based on our model for the underlying microlenses and the cluster potential, we can predict how the last moments (or first moments) 
of the multiply lensed images will take (or took) place. For simplicity we assume that the source is heading toward the main caustic from now on. 
The results presented in this section can be reversed to describe the opposite situation where the background source is moving in the 
opposite direction --- that is, away from the main caustic.  

Figure ~\ref{Fig_IcarusDeath1} shows the light curve moments before and after the background source crosses the main caustic. 
When approaching the last moments, the macro-images are broken into multiple groups or trains. Each train comprises a bright micro-image surrounded by smaller ones, usually aligned in the direction of the cluster deflection field.  
The last moments in the light curve show how the individual trains of micro-images disappear at different moments. When one of the trains vanishes the total flux drops by a large fraction. The last surviving micro-images have small fluxes and roam close to the largest microlenses (that is, in regions with low magnification). In principle, every single 
microlens (assuming they are truly point sources) would have a small micro-image near its center since the deflection angle of a point source diverges at the position of the microlens 
\cite[there is a nice visual demonstration of this effect by][]{Lewis1993,Witt1993}. These small micro-images can survive a much longer time than the trains. 
However, these low-magnification images have such small fluxes that they can be neglected in terms of their contribution to the observed flux. 

Figure ~\ref{Fig_IcarusDeath1} also shows the maximum extension, or size, of the set formed by all micro-images. We compute this size based on the micro-images that have magnification larger than 50. At time $\sim 50$\,yr in the left plot, there is only one surviving micro-image with magnification above 50 and the apparent size of the train drops to zero (a second micro-image appears and vanishes quickly at times $\sim58$ and $\sim65$\,yr).     
The size evolves as $\sqrt{|t-t_o|}$, which is inversely proportional to the apparent velocity of the macro-images. Although not shown in the plot, there is a long tail past 100\,yr where micro-images with small magnification survive for a long period close to the largest microlenses (with $\mu\approx 1$ or below). This tail and the evolution of the size can be better appreciated when the contribution from microlenses is smaller.
 
In the right panel of Fig.~\ref{Fig_IcarusDeath1} we show a similar plot, but this time the masses of all the microlenses are divided by a factor 30. 
Even though this may not represent a realistic scenario, it does mimic a situation where the optical depth of the microlenses is a factor 30 times smaller than in the left panel of  Fig.~\ref{Fig_IcarusDeath1}. (The deflection field scales as the mass; hence, rescaling the masses is a much faster way of producing multiple realizations with different lensing optical depths.) 
A more realistic simulation corresponding to a factor 30 times smaller surface mass density would contain fewer (but heavier) stars and the right panel in Fig.~\ref{Fig_IcarusDeath1} would have fewer (but more pronounced) peaks. 

At low optical depth (right panel of Fig.~\ref{Fig_IcarusDeath1}), the light curve in the last moments starts to resemble the expected behavior of the {\it classic} caustic crossing event with a smooth model \citep[see, for instance,][]{Jordi1991}, with the typical $1/\sqrt{|t-t_o|}$ change in flux (red solid line). 
In this case the relative separation between the two trains of micro-images follows closely the $\sqrt{|t-t_o|}$ law and could be extrapolated to earlier times to predict the crossing time from real observations or constrain the relative velocity between the background source and the caustic. 
A background source close to the time of caustic crossing will produce two macro-images separated by a distance that is proportional to $\sqrt{|t-t_o|}$. The proportionality constant involves the relative velocity between the source and the caustic. Future high-resolution observations (for instance, with the Extremely Large Telescope and a maximum expected resolution of $\sim 1$ mas) may be able to discriminate between point sources and extended sources with sizes larger than a few mas. Monitoring the change in the size of the unresolved pair of macro-images can be used to constrain the relative velocity between the source and caustic. Similarly, the rise in observed flux in the last moments exhibits a similar dependency with the relative velocity between the source and the caustic. From Fig~\ref{Fig_IcarusDeath1}, the period of flux decline after the peak is sensitive to the number and type of microlenses. Nevertheless, in the presence of microlenses, a caustic-crossing event needs to be monitored regularly for several decades to produce light curves that capture the rise and decline of the flux as the caustic network is being crossed. This will require an effort similar to the production of light curves for QSOs and other variable objects. 

\section{Prospects for Constraints on the Fraction of Compact DM}\label{sect_S8}

A popular candidate for DM is PBHs \citep[see, e.g.,][for a recent discussion]{Carr2016}. 
They are formed in the first moments of the Universe and, other than through gravitational interactions,  
they do not play any significant role in nucleosynthesis or baryonic physics after inflation (with the exception of very small PBHs that can evaporate quickly and inject 
energy into the Universe, or the very massive ones that can transform mass into energy through their associated accretion disks). 
Other than in these extremely low- and high-mass regimes, PBHs would be very elusive, interacting only through gravity with the surrounding matter. 
A wide range of masses has been excluded for PBHs as a significant contributor to the DM. 
PBHs below a certain mass ($\sim 10^{11}$\,kg) would have evaporated by now \citep{Hawking1974}. PBHs with slightly higher masses 
would be strong sources of $\gamma$-rays. 

Current $\gamma$-ray observations have ruled out this mass range \citep{Barnacka2012,Carr2016b}. 
At the other extreme, a large number of massive PBHs ($M>100\,{\rm M}_{\odot}$) 
would disrupt globular clusters or impact the baryonic physics in the early universe. At intermediate masses, there are constraints of varying strengths from pulsar timing or microlensing 
\citep{Kashiyama2012,Niikura2017}. A window where the constraints on the fraction of PBHs still needs to be improved is $M \approx 30\,{\rm M}_{\odot}$. \cite{Bird2016} made the interesting 
suggestion that the first LIGO event (that involved two BHs with masses in this range) could have been the coalescence of two PBHs in this mass regime \citep[see also][]{Sasaki2016,Clesse2017}. If true, this would offer 
a natural explanation for the extreme BH masses measured by LIGO \citep{LIGO2017a,LIGO2017b}.

In this section we explore the possibility of constraining the fraction of PBHs with a caustic crossing event.  
If PBHs make up a significant fraction of the DM, they would disrupt the caustic. As seen in Fig.~\ref{Fig_SmoothCC_vs_FICL}, 
the disruption is proportional to the optical depth of microlenses. This scaling can be easily derived from 
basic principles (for simplicity, we assume all mirolenses have similar masses). The optical depth, $\tau$, can be estimated as the number of microlenses, $N$, times the area of their associated Einstein ring ($\pi\theta_E^2$) per unit area, $A$, as
\begin{equation} 
\tau = \frac{N\pi\theta_E^2}{A}.
\end{equation}
If we introduce the surface mass density of microlenses with mass $M$, $\Sigma=NM/A$, and make use of Eqs. (\ref{Eq_thetaE}) and (\ref{Eq_mu_theta}), we get
\begin{equation} 
\tau=\frac{\Sigma}{M}\frac{4\pi GM}{c^2}\frac{D_{ds}}{D_dD_s}\mu_t= (4.2\times10^{-4})\,\Sigma({\rm M}_{\odot}/{\rm pc}^2)\frac{a_2\mu_o}{\theta},
\label{Eq_tau}
\end{equation}
where we remind the reader that $a_2$ is the inverse of the magnification in the direction parallel to the main CC.

As expected, $\tau$ scales with $\Sigma$ and is independent of the masses of the microlenses. If enough PBHs are present near the main CC, at some point their CCs start to overlap (or, similarly, the microcaustics overlap in the source plane). 
This corresponds to $\tau=1$. From Eq.~\ref{Eq_tau}, we can infer the width of the saturation region around the main CC. Since  Eq.~\ref{Eq_mu_theta} 
gives the total magnification and each train of micro-images carries half the magnification, after accounting for this factor of two the width of the saturation region is simply
\begin{equation} 
\Delta\Theta= (4.2\times10^{-4})\,\Sigma({\rm M}_{\odot}/{\rm pc}^2)a_2\mu_o.
\label{Eq_tau2}
\end{equation}
This simple approximation works remarkably well when tested with the simulated data (see Fig.~\ref{Fig_SmoothCC_vs_FICL}, where the length of the yellow lines 
is derived from  Eq.~\ref{Eq_tau2}  for  the model of D16, $a_2\mu_o \approx 0.2\times150''=30''$). Despite the assumption of similar masses made to derive 
Eq.~\ref{Eq_tau}, we find that when the optical depth is computed exactly for a realistic distribution of masses (by integrating the areas within the Einstein ring for each microlens), 
the disagreement is approximately only a factor of 2 or 3. In particular, Eq.~\ref{Eq_tau} predicts $\tau=0.7$ at the position of Icarus (assuming the model of D16 with $a_2\mu_o \approx 30''$) and for $\Sigma \approx 7\, {\rm M}_{\odot}/{\rm pc}^2$, while the integration of a realization of the Spera model between 
$M=0.01\,{\rm M}_{\odot}$ and the maximum mass in the realization ($M \approx 70\, {\rm M}_{\odot}$) renders $\tau=0.3$. For alternative models like those of \citet{Woosley2002} and \citet{Fryer2012}, we obtain similar values of 0.26 and 0.28, respectively.
From Eq.~\ref{Eq_tau2} and for the model of D16, we obtain $\Delta\Theta \approx 0.1''$. 


Interestingly, this distance is smaller than (but comparable to) the separation found between Icarus and Iapyx ($\Delta\Theta \approx 0.26''$). 
This suggests that the adopted $\Sigma$ may actually be close to its real value. Much higher values of $\Sigma$ would result in a larger disruption of the main CC and would 
make it very difficult to observe a nearly constant flux for a period of $\sim 10$\,yr at the position of Icarus. 
Conversely, significantly smaller values of $\Sigma$ would translate into very small probabilities of observing a peak like that witnessed for Icarus in May 2016. 

We can define the effective surface mass density of microlenses\  $\Sigma_{\rm eff} = \Sigma a_2\mu_o/\theta$ (see Eq.~\ref{Eq_tau}), which is inversely proportional to the distance to the CC. 
Hence, at large separations from the CC, both $\Sigma_{\rm eff}$ and $\tau$ tend to zero and the observed light curve should be featureless (i.e., no microlensing events and just a slowly varying flux).
On the other hand, a given area in the image plane at $\Delta\Theta/2$ from the main CC maps into a larger area in the source plane than the same area in the image plane would if it were at a smaller distance from the main CC (i.e., smaller $\theta$). Thus, one would naively expect the probability of seeing microlensing events to be maximized at the point where we reach the saturation regime. 
Future observations of similar events should be found at comparable distances, or that scale with the estimated surface mass density of microlenses as shown in Eq.~\ref{Eq_tau2}. 

This argument can be used to indirectly infer that the fraction of DM in the form of compact objects (like PBHs) must be low (independently of their masses). 
The value of $\Sigma\approx7\, {\rm M}_{\odot}$\,pc$^{-2}$ corresponds to a fraction of the total mass of $\sim 0.3$\%. The relatively smooth behavior of the Icarus 
light curve (1--3 minor peaks in the last $\sim 10$\,yr) suggests that the fraction of DM in microlenses cannot be significantly higher than 3\% (or the mass of the PBH is orders of magnitude smaller than 1~M$_\odot$). 
If the fraction of DM is much higher, the width of the saturation region would extend to farther distances. According to Eq.~\ref{Eq_tau}, a  3\% fraction would 
saturate a region of $\sim 1''$ around the CC. Current observatories have the resolution and sensitivity to witness such events but none has been reported prior 
to {\it Icarus/Iapyx}. If we assume that the spiral arm extending all the way to the CC in MACS1149 contains a nearly 
constant surface density of stars (as suggested by its nearly constant surface brightness), during the last $\sim 10$\,yr we would have expected to see additional 
peaks farther away from the CC. The fact that both Icarus and Iapyx (but also {\it Perdix}, also called LS1 /Lev~2017A); see K17 for more details) appear at distances $\sim 0.1''$ from the CC points to a low 
fraction of DM ($< 1\%$) in the form of microlenses. This conclusion is, in principle, independent of the microlens mass. However, this is not entirely true at 
sufficiently low microlens masses since in this case the microlenses start to behave as a smooth distribution of DM (their associated Einstein rings would be too 
small compared with the dimension of the macro-images). 
Even if individual events cannot be resolved, deep observations around the CC can reveal fluctuations that are not observed at larger distances from the CC. The strength and extension of these fluctuations can be used to constrain $\Sigma_{\rm eff}$ and hence the fraction of compact objects.

In the low-mass microlens regime, the sensitivity to the mass of the microlens depends on two variables:  
(i) the surface mass density of massive microlenses, $\Sigma$, and (ii) the underlying magnification. If massive microlenses (like, for instance, stars and remnants from the ICL) 
have a large $\Sigma$, small microlenses play a negligible role (the magnification pattern is entirely dominated by the massive ones). Only when the contribution from 
massive microlenses is small, can the signature of small microlenses (like PBHs with planet-like masses) be detected. As shown in Section 2, the Einstein ring associated with a small microlens can be highly amplified when it lies very close to a cluster CC. In the source plane, larger microcaustics from massive microlenses would overlap on top of the smaller microcaustics overwhelming them. A clear illustration of this effect was shown in the right panel of Fig.~\ref{Fig_SmallMicro}, where 
$M=0.01\, {\rm M}_{\odot}$ microlenses can disrupt the main CC in a way that could be quantified during a caustic crossing event. The ideal scenario for constraining compact 
DM in the low-mass regime is to monitor a background galaxy that lies behind a caustic that is itself far away from any member galaxy or with minimal ICL. 
This will minimize the effect of more massive microlenses (like stars) offering a clean view of the structure of DM in the small-mass regime. The same argument can be applied to PBHs with higher masses. If a caustic crossing event is found in a region where the contribution from the ICL is known to be negligible and the event shows signatures of microlensing, PBHs could be constrained more easily than in a region where the ICL produces microlensing events.

Giant arcs are ideal candidates, since the exact location of the CC can be estimated with great accuracy owing to symmetry principles. Knowing the exact location of the 
CC is very important to break the degeneracy between the microlens mass and the magnification (see Eq.~\ref{Eq_thetaE}). 
In particular, elongated thin blue arcs tend to form in regions where $\kappa\approx\gamma\approx0.5$, since then $a_1=1-\kappa-\gamma\approx0$ and $a_2=1-\kappa+\gamma\approx1$. 
The first ($a_1\approx0$) condition produces arcs with very large tangential magnification while the second condition ($a_2\approx1$) is needed to produce very thin arcs 
(small radial magnification). Thin giant arcs are normally found in elliptical potentials in regions relatively far away from the central BCG and where the intracluster light is moderate or 
small, reducing the negative impact of microlenses from the ICL that could be mistaken for PBHs. Blue arcs are also more likely to contain younger giant stars that can be very luminous and make ideal background sources. 

\begin{figure}  
 \centerline{ \includegraphics[width=9cm]{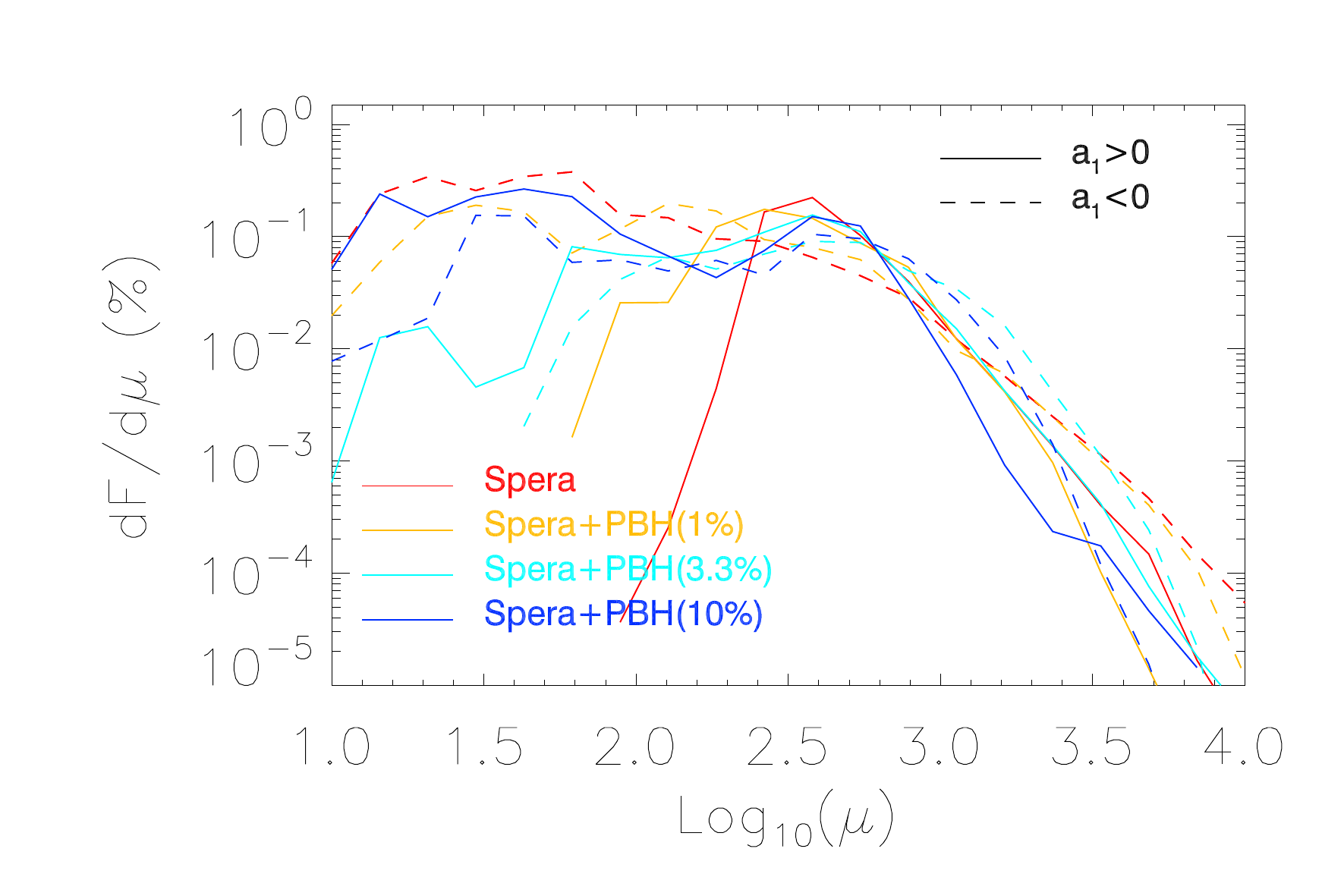}}
  \caption{Histogram of the magnification computed from the simulated light curves (at $0.13''$ distance from the main CC). 
          Similarly to what is observed in Fig.~\ref{Fig_HistogramsMuSource}, the distinction between different models is not obvious 
          at high magnification. 
          Note how for the case with 10\% DM microlenses, both sides of the main CC ($a_1>0$ and $a_1<0$) behave 
          similarly, as this case has already reached the saturation level ($\tau>1$). Finally, note also how the probability still scales as $\mu^{-3}$ at high magnifications.}
   \label{Fig_HistogramMu3}  
\end{figure}  

From the observational point of view, having data with a high cadence (one data point every week or two weeks and daily when the flux aproaches the maximum) is important for constraining the shape of the peaks. The width of the peak depends most strongly on the ratio $R/v$, where $R$ is the radius of the background source and $v$ the relative velocity between the source and the caustic.  Deep observations of the event are also useful for improving the photometry and ruling out (or confirming) smaller peaks that could be hidden between the most prominent ones. Also, if the background source consists of a binary system, different events would have a similar pattern with two consecutive peaks.

\subsection{PDF of the Light Curves.}\label{subsect_stat}

The PDF of the observed magnification (extracted from light curves spanning $\sim 400$~yr) is shown in Fig.~\ref{Fig_HistogramMu3}. 
Clearly, a compensating effect is taking place when $F$ is higher. The PDFs look remarkably similar independent of the value of $F$. 
The situation is very similar to the result presented in Fig.~\ref{Fig_HistogramsMuSource}. When $F$ is higher, the PDF shown in Fig.~\ref{Fig_HistogramMu3} shifts toward lower magnifications. However, in the observed light curve, moderate magnifications ($\mu \approx 10^3$) have similar probabilities. This is a consequence of the trains breaking up into smaller {\it bits} as $F$ grows. Perhaps the most interesting difference is the trend observed at very low magnifications ($\mu \approx 10$) and very high magnifications ($\mu \approx 10^4$). When $F$ is high, it is more difficult to {\it hide} an entire macro-image from the observer (i.e., the total flux of the macro-image is very low) on the side where $a_1<0$. This is a consequence of the overlapping caustics in the source plane, as we have seen earlier. For high magnifications, Fig.~\ref{Fig_HistogramMu3} also shows a deficit of bright peaks when $F$ is higher. When $F$ is higher, the side with negative parity ($a_1<0$) and the side with positive parity ($a_1>0$) behave similarly, as shown in Fig.~\ref{Fig_HistogramMu3}. The fact that the observed light curves from Icarus and Iapyx look very distinctive suggests indirectly that  $F$ must be $\lesssim 1\%$.
 
\subsection{Power Spectrum}

\begin{figure*}  
 \centerline{ \includegraphics[width=9cm]{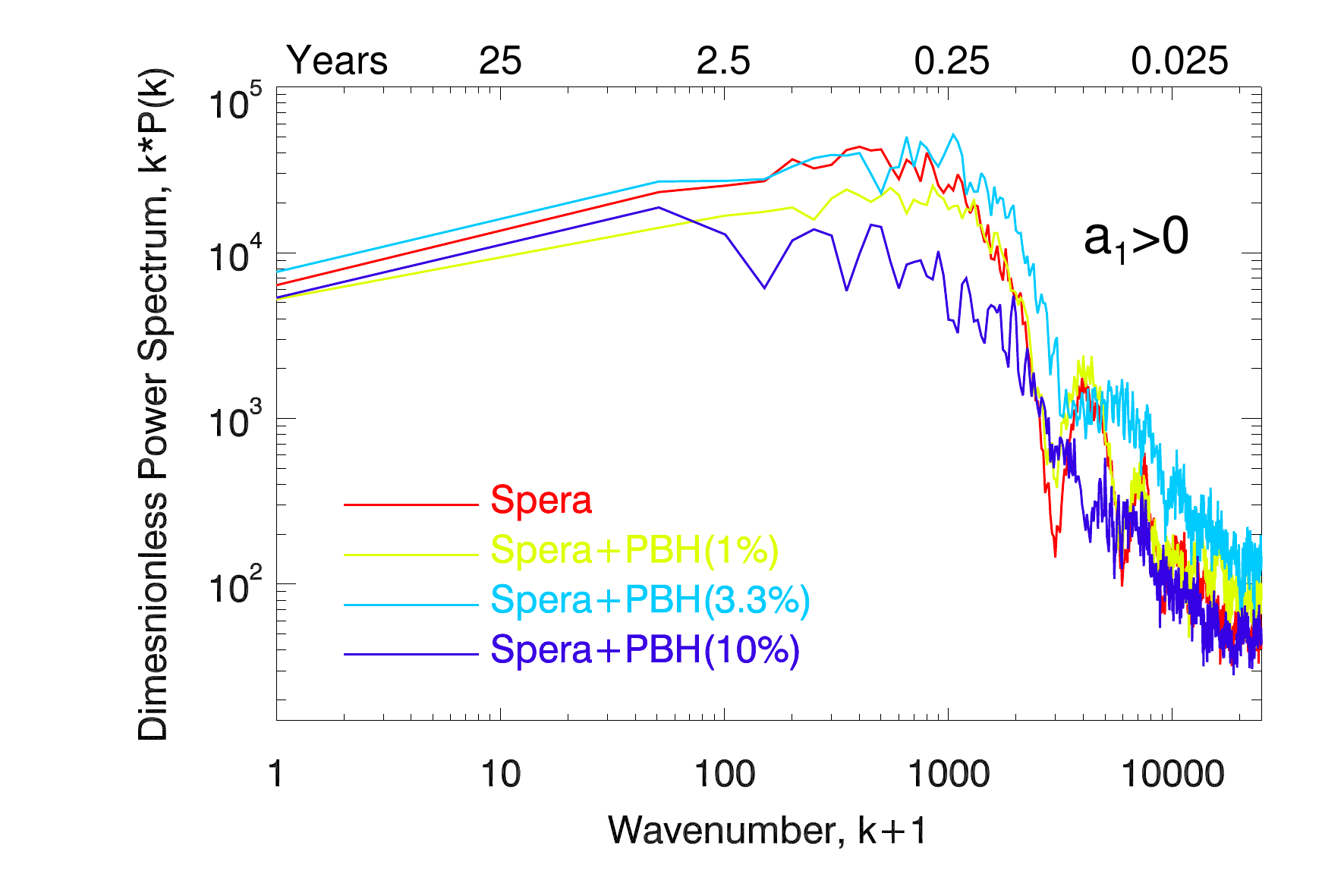}
              \includegraphics[width=9cm]{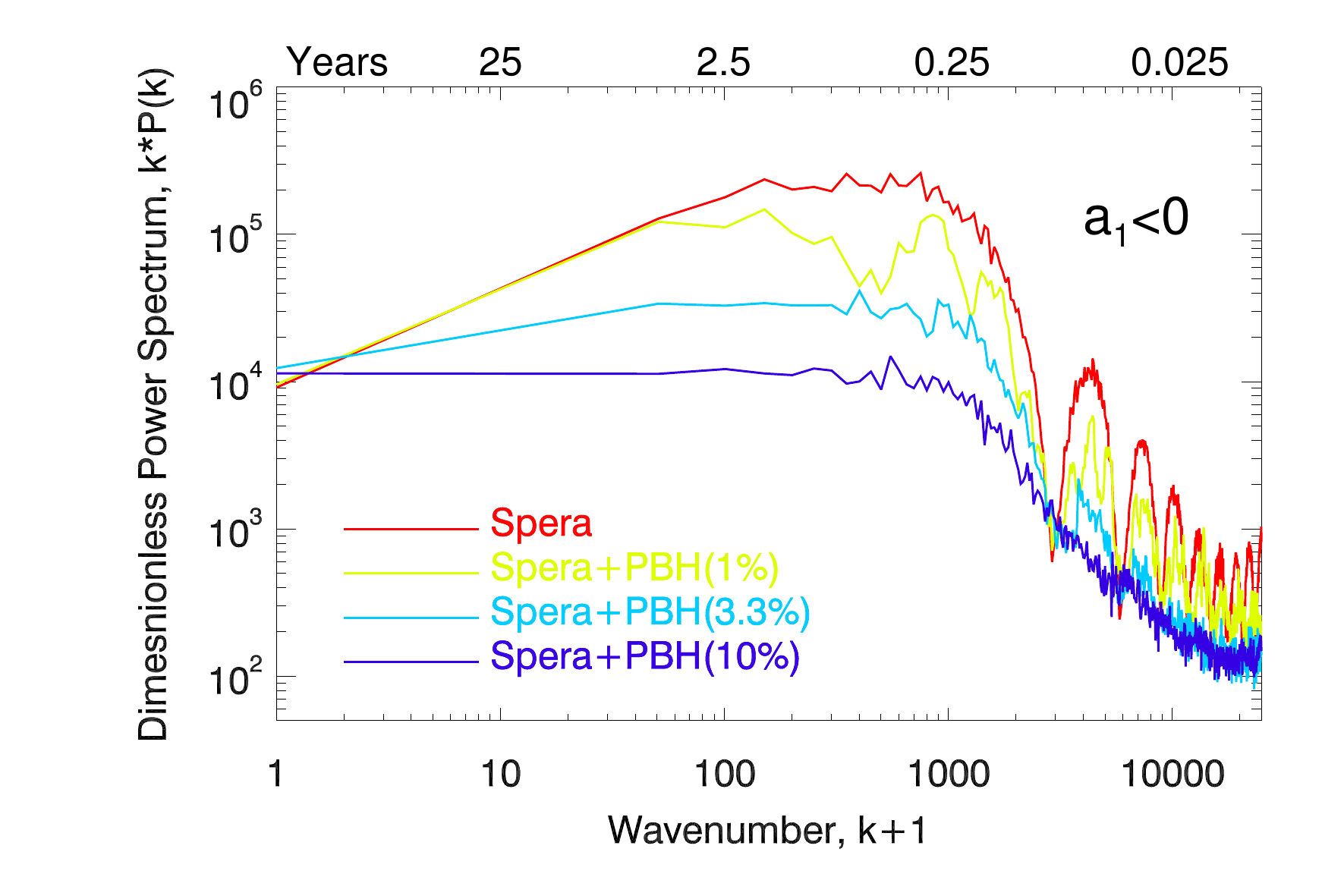} }
  \caption{Power spectrum of simulated light curves. The prominent oscillations or peaks on small scales in the red curve for the $a_1<0$ case in the right panel 
            are due to the low-magnification periods (the reader can think of the Fourier transform of a top-hat). Similar, but more modest, oscillations can be 
            appreciated in other cases as well.}
   \label{Fig_PowerSpectrum}  
\end{figure*}  

The distribution of peaks and valleys depends on the optical depth of the microlenses. 
A correlation function, or power spectrum, of the light curves can highlight hidden features in the light curve such as correlations that are not easily identified in the PDF. 
Correlations between peaks are expected, for instance, when a single microlens produces a double peak. The separation between the peaks is, in this case, generally proportional 
to the square root of the mass of the microlens. We compute the power spectrum of the light curves and show the result in Fig.~\ref{Fig_PowerSpectrum} for the models with only ICL stars (Spera) and for models where we add to the Spera model a fraction of the total mass in the form of compact DM (or PBHs).

\begin{figure}  
 \centerline{ \includegraphics[width=9cm]{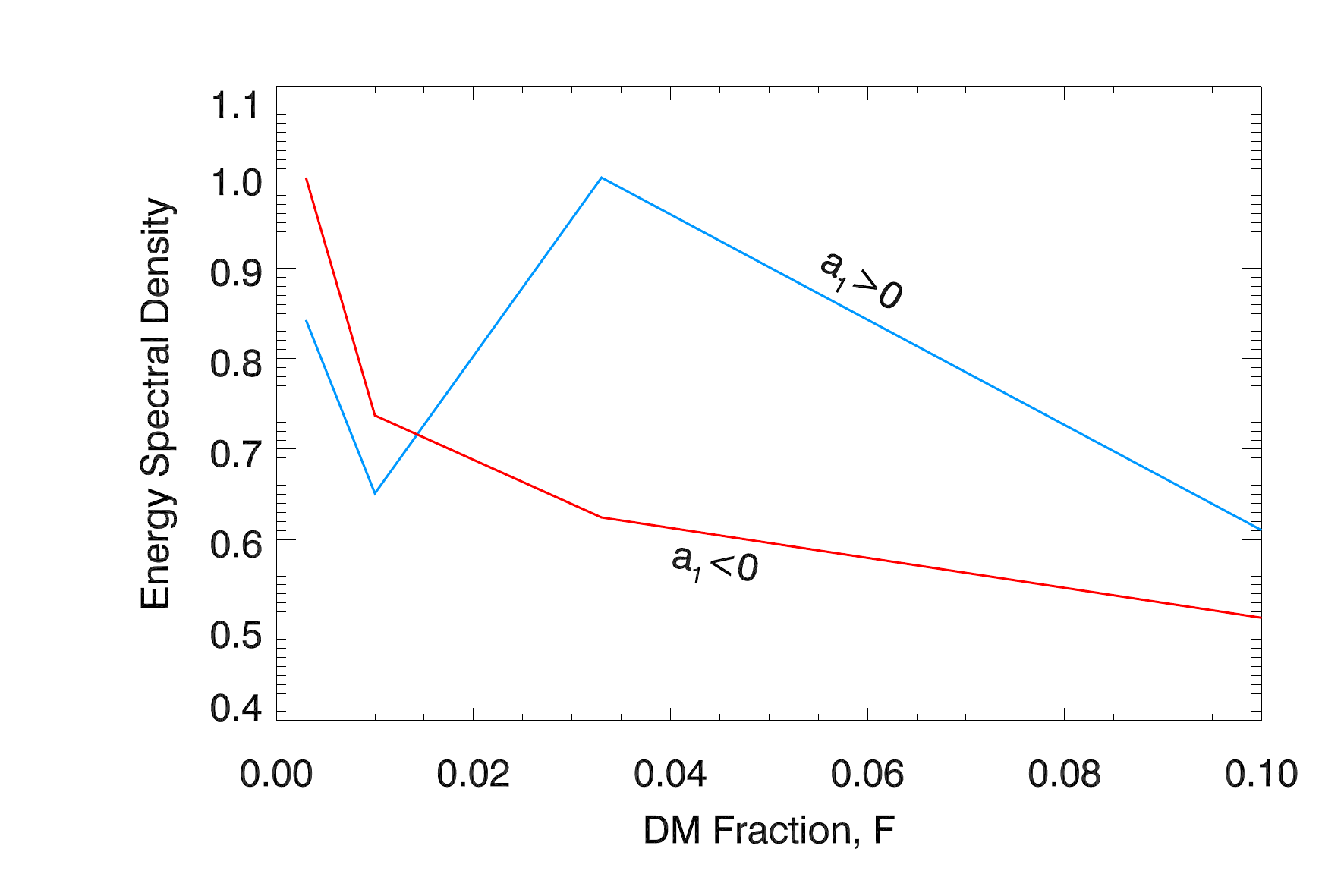}}
            \caption{Energy spectral density of the light curves as a function of the fraction of mass in the form of microlenses, $F$. 
                     In both cases, the curves are normalized to 1 to better show their relative change with $F$. Note how the side with $a_1<0$ shows a more consistent trend.}
   \label{Fig_ESD}  
\end{figure}  

The power spectrum seems to be more sensitive to the optical depth of microlenses on the side where $a_1<0$. This behavior is also found in the PDF of the light curves. Similarly to \cite{Schechter2002}, we find that fluctuations are accentuated when the fraction of microlenses relative to the smooth component is smaller.  
Interestingly, on the side with $a_1<0$ and for the cases with higher $F$ (PBH = 3.3\% and PBH = 10\%), the power spectrum behaves like pink noise (or $1/f$ noise) on large scales, 
a behavior which is found in many situations in nature. 
One of these situations is relevant for this work and refers to {\it self-organized criticality}\footnote{Note, however, that the original claim between the $1/f$ noise and the self-organized criticality is not formally correct, as discussed by \cite{Jensen1998}.} systems introduced by \cite{Bak1987} \citep[see][for more detailed descriptions of these systems]{Jensen1998,Christensen2005}. 

The $1/f$ nature of the power spectrum on large timescales (more than  a few months) indicates that there is no characteristic timescale on the side where $a_1<0$. 

A similar argument could have been used for the PDF of the magnification or the light curves, which also exhibit power-law behaviors typical of systems where self-organization is involved. In this  case, it is the magnification itself that is redistributed across the image plane, owing to the constraint that the total magnification (or observed flux) remains constant when integrating over long periods of time independent of the substructure in the lens plane.  
When $F$ is large (PBH = 10\%), the properties of the power spectrum resemble those of time series that fall in the categories of stationary 
fractals \citep{Rodriguez2014}, which can be linked to renewal processes (a generalization of a Poissonian process) 
in which the time intervals between events are not correlated. This is the expected behavior for the separation between events when the distribution of projected microlenses is random.

The energy spectral density (ESD) can be used to measure the total amount of signal contained in the power spectrum,
\begin{equation}
E = \int{|\mu(t)|^2 dt},
\end{equation}
which by Parseval's theorem is equivalent to the integral of the square of the power spectrum. Fig.~\ref{Fig_ESD} shows the ESD for the Icarus and Iapyx simulated light 
curves and for the models shown in Fig.~\ref{Fig_PowerSpectrum}. A clearer trend is observed on the side where $a_1<0$, suggesting that ESD of the fluctuations on this side 
of the main CC may be more capable of discriminating between different optical depths of microlenses.

To produce accurate power spectra that can discriminate among different models it will be necessary to monitor these events for decades, similarly to what is being done for QSOs and other variable objects. With smaller samples, the two-point correlation function offers an interesting alternative.

\section{Discussion}\label{sect_S9}

Earlier work suggests that from a statistical point of view, it is very difficult to distinguish the mass distribution of the microlenses for a fixed 
surface mass density and shear \citep{Wyithe2001}. On the contrary, \cite{Schechter2004} show how the PDF of magnifications is sensitive 
to the mass distribution of microlenses. Our results partially support this. Some differences exist for similar optical depths, but the differences 
are relatively small. Interestingly, these differences seem to be more accentuated on the inner part of the main CC ($a_1<0$). 

One of the most striking features in the 
light curves, and that is also sensitive to the optical depth of microlenses, is the ``hiding'' periods when the flux of a macro-image falls below the detection limit. 
These low-magnification periods observed in our simulated light curves are similar to those found by \citet{Kayser1986} or \citet{Paczynski1986} when the dimensionless 
surface mass density of stars, $\sigma=\Sigma_s/(1-\Sigma_c)$, is negative (here $\Sigma_s$ is the surface mass density of stars and $\Sigma_c$ is the surface 
mass density of the smooth distribution of matter). The counterintuitive results of \cite{Schechter2002} can be also interpreted in light of our own results. 
They find that ``contrary to naive expectation, diluting the stellar component of the lensing galaxy in a highly magnified system with smoothly distributed dark matter increases rather than decreases the microlensing fluctuations caused by the remaining stars.'' Our simulations show that increasing the fraction of mass in compact objects results in lowering the typical magnification and consequently the fluctuations. Also, overlapping caustics smooth out large fluctuations.

Although in our simulations, $\sigma$ is always positive ($\Sigma_s>0$ and $\Sigma_c<1$), making the masses of 
the microlenses negative in the simulation changes the magnification pattern of the $a_1>0$ side into a pattern similar to the $a_1<0$ side and vice versa. That is, 
it would be equivalent to the cases discussed by  \citet{Kayser1986} and \citet{Paczynski1986} with  $\sigma<0$. 
When comparing the histograms of intensities (or magnifications) in \cite{Paczynski1986} with our results, we find a similar trend. 
Negative values of $\sigma$ (or negative optical depth following Paczy\'{n}ski's argument) reproduce our results for $a_1<0$.

The work of \cite{Paczynski1986} also presents another interesting result relative to the extension of the train of micro-images, or the {\it halo} in Paczy\'{n}ski's terms. 
As  $\sigma$ increases, the size of the halo of micro-images grows as well. In our case, we observe the same phenomenon with the trains extending over larger distances along 
the direction where micro-images form (i.e., in the direction of the deflection field). When $\Sigma_s$ is sufficiently large, micro-images start to form in a more complex pattern, breaking the straight line found when  $\Sigma_s$ is low and adopting a configuration that more resembles the halo described by \cite{Paczynski1986}. 

Perhaps the most interesting practical application of caustic crossing events is their ability to constrain the fraction of compact DM. 
For a given surface mass density of microlenses, $\Sigma$, the optical depth scales linearly with $\Sigma$. One can use this fact to 
set limits on $\Sigma$ by observing the slow rise (or decline) in flux of a magnified background star as it approaches (or departs from) the cluster caustic. If $\Sigma$ is sufficiently 
large and the position of the caustic or CC can be determined with relative precision, microlensing events should become ubiquitous thousands of years before (or after) 
the star crosses the cluster caustic. This can be visualized easily in the source plane for the case of high $\Sigma$. 

The rate of observed events like Icarus at the position of background arcs known to cross caustics gives us indirect information about the 
level of disruption of cluster caustics and/or about the luminosity function of the background object. This is easier to see in the ideal case where the caustic is not disrupted.  
If the mass-luminosity relation follows analytical models ($L \propto M^3$ in normal stars where gas pressure and gravity are balanced, and $L \propto M$ in heavy stars where 
radiation pressure overwhelms gas pressure), we should expect a classic IMF that falls like $dN/dM \propto M^{-2.3}$ to translate into luminosity functions $dN/dL \propto L^\alpha$ with 
$-2.3<\alpha<-1.1$ depending on the mass (or luminosity) of the star. As discussed by K17, owing to the $1/\mu^2$ probability of being magnified by a factor larger than $\mu$, 
if $dN/dL \propto L^{-2}$, the smaller probability of being magnified by a larger factor $\mu$ gets exactly compensated by the larger abundance of lower-luminosity stars making all unlensed 
luminosities have the same probability of being observed.  
A stellar luminosity function, $dN/dL$, that is steeper than $L^{-2}$ results in lower-luminosity stars being more likely to be observed above a certain flux limit in caustic crossing 
events than higher-luminosity stars. On the contrary, if $dN/dL$ is shallower than $L^{-2}$, brighter stars will be more likely to be observed than the less luminous ones. When the cluster caustic is disrupted by microlenses, the argument above remains the same, since the probability of being magnified by a factor larger than $\mu$ still retains its fundamental form at high magnifications, $\mu^{-2}$ (see Fig.~\ref{Fig_HistogramMu3}), except that now the normalization of the probability is smaller and very rare extremely luminous stars may be required to produce the observed flux.\\

\subsection{Uncertainties in the Lens Model}\label{Sect_Uncertainties}

Despite the high quality and excellent agreement between the different lens models that have been published for clusters like MACS1149, there are still significant uncertainties in the lens models that limit the capability of using CC crossing events as probes of DM (or substructure in general). For the work presented here, one of the most relevant systematics is the uncertainty in the magnification near the CC. \cite{Meneghetti2017,Priewe2017} show how discrepancies of order 50\% in the magnification are typically found between state-of-the-art lens models that are considered otherwise {\it accurate}. 
Here we compare the predictions made by two of these models derived under very different assumptions. 

The first one is the free-form model 
from D16 derived using the WSLAP+ free-form code \citep{Diego2005,Diego2007}. The second one is from \citet{Kawamata2016} derived using the parametric code GLAFIC \citep{Oguri2010}. The two models correctly predict the position of the CC between Icarus and Iapyx, and 
have successfully anticipated the position and time of reappearance of SN Refsdal. 
However, as shown in Fig.~\ref{Fig_Masamune_vs_Chema}, the two models 
disagree by a factor $\sim 2$ in the predicted magnification at Icarus and Iapyx.
(A third model, by \citealt{Richard2014}, predicts a magnification similar to the D16 model 
with $\mu_o=150\pm10$.) 
The reason for this disagreement can be found in the small differences in the deflection field near the critical region. The condition for a CC to occur is 
that  a certain combination of derivatives of the deflection field cancel out. In particular, the inverse of the magnification is given by
\begin{equation}
\mu^{-1} = 1 - \alpha_x^x - \alpha_y^y + \alpha_x^x\alpha_y^y - \alpha_x^y\alpha_y^x,
\label{Eq_mu2}
\end{equation}
where $\alpha_x$ and $\alpha_y$ are the deflection field in the $x$ and $y$ directions and $\alpha_i^j$ is the partial derivative of  $\alpha_i$ with respect to the 
coordinate $j$ ($i,j=x,y$).
From the expression above, it is clear that small changes in the deflection field near the point where the inverse of the magnification is zero can have a significant impact on 
the delicate balance between the derivatives in Eq.~\ref{Eq_mu2}. The magnification is then more sensitive to the small differences in $\alpha$ near the CCs, where in relative terms these differences are larger. 

\begin{figure}  
 \centerline{ \includegraphics[width=9cm]{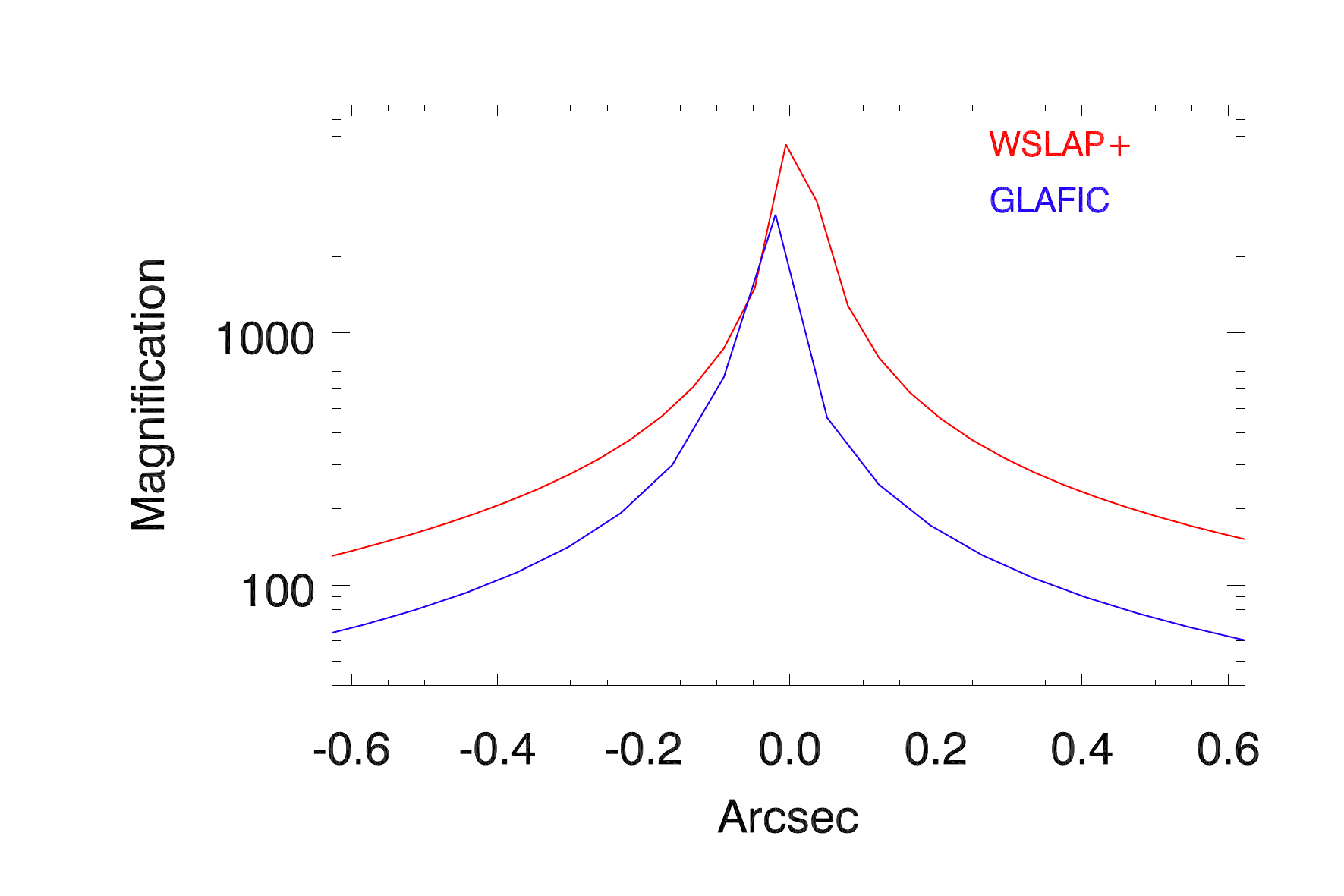}}
  \caption{Comparison of the predicted magnification between the models of D16 (red) and \citet{Kawamata2016} (blue) along a line that crosses the CC in an orthogonal direction and at the position between Icarus and Iapyx.} 
   \label{Fig_Masamune_vs_Chema}  
\end{figure}  

Small differences in the inferred deflection field between different lens models are expected.  
For instance, each method assumes different profiles for the halos around member galaxies or for the cluster halo(s), which results in potentials that are more or less shallow.\footnote{The deflection field is related to the lensing potential through partial derivatives.} 
Also, the inferred positions of the background sources used to constrain the lens models differ from model to model. 
These {\it unknown} positions are treated as variables that need to be determined together with the mass distribution (or potential) 
of the galaxy cluster. By comparing the models of D16 and \citet{Kawamata2016}, we find that CC positions, $\kappa$, and $\gamma$ generally agree to within percent levels (except in the surroundings of small member galaxies considered in one model but not in the other). However, at the position of Icarus 
and Iapyx, the deflection angle disagrees by a factor of $\sim 40\%$. In absolute terms, the difference between deflection angles is small (only $\sim 3''$), but since both  
deflection angles are relatively small in this part of the lens plane, it translates into a large relative difference. 
The $\sim 3''$ difference in the deflection field originates in the different predicted positions of the background source used to constrain this part of the lens 
plane. The small differences in the derivatives of the deflection field around the CC translate into a factor of $\sim 2$ in the magnification. 

The uncertainty in the magnification has direct implications on our ability to constrain the masses of the microlenses. As shown in Section \ref{sect_S2}, the effective lensing 
mass of the microlens scales as $\sqrt{\mu}$. A factor of 2 uncertainty in the magnification then translates into a factor of $\sqrt{2}$ uncertainty in the mass of the microlenses. 
Future observations may need to rely on independent calibrators of the magnification. One such calibrator could be a SN~Ia in a background galaxy and that happens 
to be near the CC. For instance, in \cite{Rodney2015}, SN HFF14Tom is used to compare the observed magnification with the one predicted by several lens models, finding in general 
good consistency but also a small systematic bias in many of the lens models that tend to overpredict the observed magnification by $\sim 10$--20\%. 

\begin{figure}  
 \centerline{ \includegraphics[width=9cm]{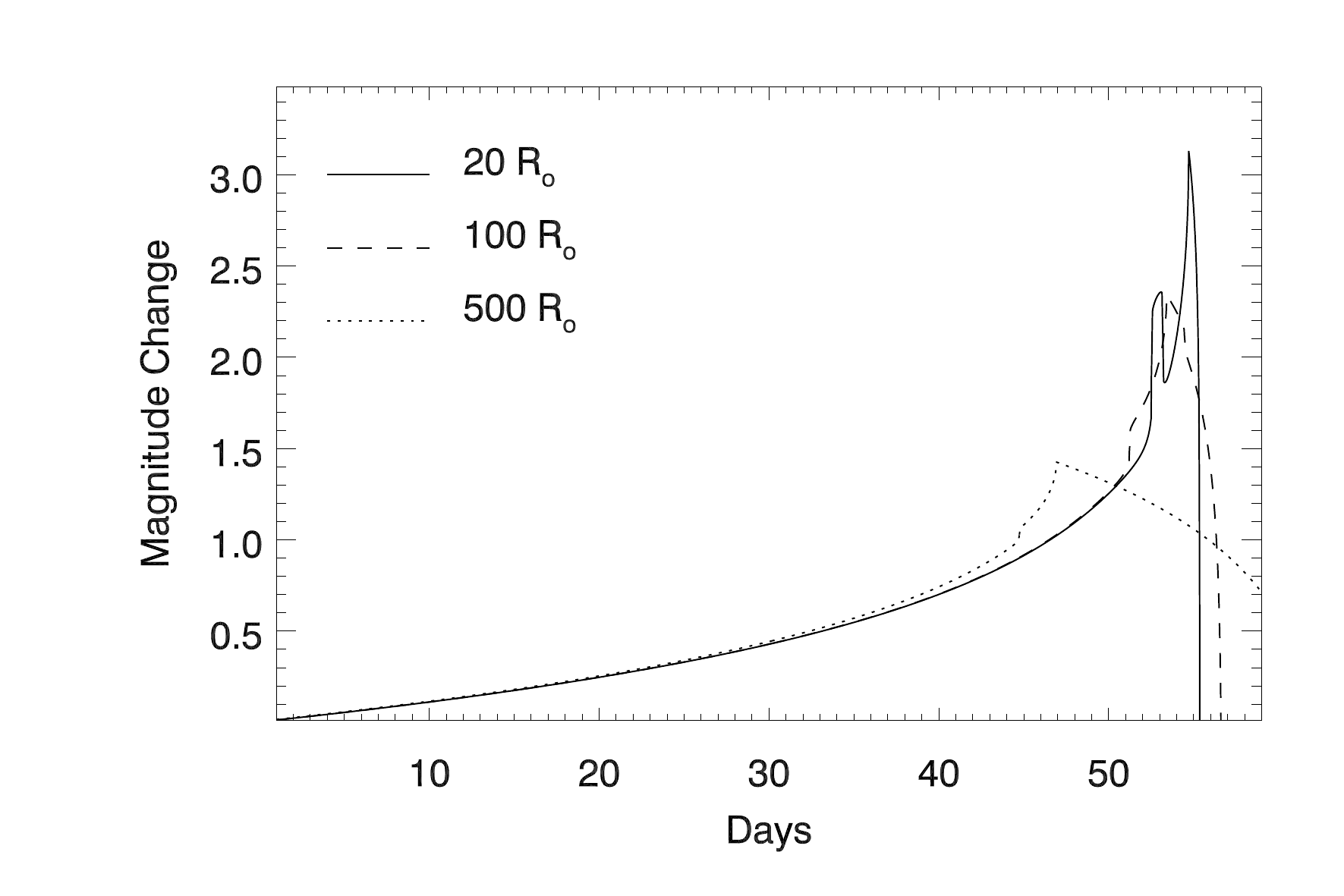}}
  \caption{Light curve for the ideal situation of a smooth model plus a small microlens near the CC. 
           The microlens has one Jupiter mass and the background star is moving at $v=1000$\,km\,s$^{-1}$ relative to the caustic and has a 
           radius of 20\,${\rm R}_{\odot}$ (solid line). The dashed and dotted lines show the light curve for 
           background stars with radii  of 100\,${\rm R}_{\odot}$ and 500\,${\rm R}_{\odot}$, respectively. 
           The ordinate shows the relative change in flux expressed in magnitudes.} 
   \label{Fig_Jupiter}  
\end{figure}  

\subsection{Exploring the Small Mass Compact DM Regime}\label{subsect_Jupiter}

The best constraints on the compact DM when $\Sigma$ is small should be obtained in a portion of the CC where the contribution from ICL stars (and remnants) is negligible. 
If ICL stars are populating the lens plane at the position of the CC, they will limit the ability to constrain the amount of compact DM. 
The ideal situation is to monitor a blue arc (to maximize the number of luminous stars) known to cross a CC and located far from the ICL or member galaxies. 
Ideally, such an arc contains a very bright star that is, on average, visible when magnified with the macromodel magnification in order 
to construct its light curve to probe DM (on the contrary, it will be difficult to distinguish if different events correspond to different stars). 
In such situations, the possibilities of detecting small-size microlenses increase significantly. Moreover, as discussed earlier in the paper, monitoring arcs behind a galaxy cluster CC has the added advantage of probing a wide area in the lens plane for a fixed amount of time. This is a consequence of the increase in density of microcaustics 
at high magnifications. The reader can find an interesting discussion on this subject in \cite{Oguri2017}.
 
If the contribution from ICL stars is very small, the light curve should contain very few peaks and the main caustic would be minimally disrupted, allowing the approach to the caustic to have maximal magnification. As shown earlier in Fig.~\ref{Fig_IcarusDeath1}, at optical depths of microlensing which are a factor of 30 smaller than the value found at Icarus/Iapyx (i.e., $\Sigma=\Sigma_o/30$), the fluctuations in the light curve may still be significant, and the microlenses from the ICL may overwhelm the signature of an unknown population of microlenses with low mass (for instance, PBHs with masses smaller than 1\,M$_\odot$). However, the results shown in the right panel of Fig.~\ref{Fig_IcarusDeath1} may not be the most realistic representation of a population of ICL microlenses with a small optical depth from the ICL. Estimating the precise amount of ICL that minimally disturbs the CC (needed to explore the low-mass regime of a hidden population of microlenses) is not trivial, but we can make some approximations. 

First, and for the sake of simplicity, we assume that all the stars in the ICL have 1\,M$_\odot$. (This is not a bad approximation, since we want to estimate the number density of objects which, for most IMF models, is dominated by stars with masses close to 1\,M$_\odot$.) Then, we make use of the scaling for the effective Einstein radius of the microlens, $\theta_e\sqrt{\mu_t}$ ($Q=1$) (see Appendix). Since from Eq. 5, $\mu_t = \mu/\mu_r = \mu_o/(\mu_r\theta)$, we can find the value of $\theta$ for which the effective Einstein radius of the microlens is equal to the distance to the CC, $\theta$. We can require that the separation between microlenses must be at least twice this angle in order for the CC not to be completely disrupted. 

For the values in the D16 model, we then infer that the surface mass density of stars must be $\sim 10^6 {\rm M}_{\odot}/{\rm arcsec}^2$, or roughly two orders of magnitude smaller than the value found at the position of Icarus and Iapyx. Even though this may seem like a big difference, we have to realize that Icarus and Iapyx are found relatively close to the cluster center ($\sim 7''$ from the BCG), where the contribution form the ICL is still important. A simple fit to the total observed light around the BCG of MACS1149 (in the F160W band) shows how the unresolved flux around the BCG falls as $1/d^2$ (up to $\sim 30''$ from the BCG), where $d$ is the distance to the BCG. This is also the expected behavior if the ICL trace the DM, which from simulations (but also observations) falls as $\rho(r) \propto r^{-3}$ in the outer regions of a cluster (with $\rho(r)$ the 3-dimensional mass density). Hence, in order for this unresolved flux to fall two orders of magnitude with respect to the ICL found at $7''$, one should move a factor of $\sim 10$ (that is, $\sim 1'$) from the BCG (less if the ICL falls faster than the assumed $1/d^2$ at larger radii, since there are fewer galaxies that can lose stars to the ICL). The CC of clusters at $z\approx1.5$ does not reach these distances, but at higher redshifts ($z>3$) the CC can extend up to $\sim 1'$ along the main axis of the cluster. However, at these redshifts, the flux of a background star is a factor $\sim 5.5$ times smaller than at $z=1.5$ (or $\sim 1.8$\,mag fainter for $z=3$). One would need extremely luminous stars at high redshift and a powerful telescope capable of observing their redshifted emission. Luckily, such combinations, will exist in the near future. The {\it James Webb Space Telescope (JWST)} may be able to observe the first Population III stars through caustic crossing events \citep[see][where this scenario is studied in detail]{Windhorst2018}. Population III stars crossing a caustic far from the projected center of the cluster (to avoid the contamination from ICL microlenses) can then be used to constrain the fraction of compact DM to unprecedented levels.

This scenario offers the possibility of constraining PBHs in the less explored low-mass regime ($M< 0.1\, {\rm M}_{\odot}$). 
If we consider the ideal situation of a smooth deflection field with negligible contribution from the ICL stars, a microlens with small mass, $M$, close enough to the CC will behave as a larger microlens with mass $\mu M$, and will produce a peak before the maximum flux is reached. If the background star 
is large, this peak may merge with the primary peak, making the identification of the microlens harder (see Fig.~\ref{Fig_Jupiter}). However, if the background star is sufficiently small (few tens of solar radii), the peaks may be distinguished and the relative height of the peak due to the microlens can be used to constrain the mass of the microlens.  Fig.~\ref{Fig_Jupiter} shows how a microlens with a Jupiter mass at $z=0.55$ can produce a change in flux of almost half a magnitude for a period of $\sim 1$ day.

Finally, as discussed in the previous subsection, the ideal target area would contain a {\it magnification calibrator} 
such as a Type Ia SN near a CC, which can be used to directly estimate the magnification in that part of the lens plane and reduce the uncertainty in the lens 
model magnification. SN Refsdal is relatively close to Icarus, but it is also too close to a medium-size member galaxy which overwhelms the cluster contribution. 

\section{Conclusions}\label{sect_S10}

This work studies the particular case of microlenses very close to a cluster CC. 
As one approaches the CC, the magnification changes rapidly, affecting the way microlenses disrupt the magnification pattern. 
The main results of this work are summarized below. 

\noindent
$\bullet$ {\em Superluminal substructure probe}. We have shown how the observed velocity of the macro-images is proportional to the magnification. Consequently, the apparent transverse motion of the observed macro-images becomes superluminal when the distance to the CC is sufficiently small and the magnification is sufficiently large. Under these circumstances, moving macro-images probe more substructure in the lens plane than a classic microlensing event. \\
\noindent
$\bullet$ {\em Distinctive light curves.}
As found in previous work, the light curves on each side of the CC look quite different (at low optical depths and magnifications smaller than 1000), with periods of very low flux on the side where $a_1<0$ not present on the $a_1>0$ side. This deficit in flux on the side with  $a_1<0$ is compensated by brighter peaks when a micro-CC is intersected by the background star. At high optical depth ($\tau>1$) or extreme magnifications (from the macromodel) of several thousand, there is little distinction between the positive and negative parity sides.\\ 
\noindent
$\bullet$ {\em Reduced maximum magnification.} 
Macro-images of a small background source, like a bright star, crossing a cluster CC can be magnified by factors of $\sim 10^6$,  
provided there is no substructure that disrupts the CC. When microlenses populate the lens plane, the maximum magnification 
is reduced, but it can still reach values of several thousands. \\
\noindent
$\bullet$ {\em Flux borrowing.} 
Microlenses can {\em borrow} photons from the main caustic thousands of years before the background star crosses the position of the main caustic, making the observation of these events much more likely than previously thought. 
This produces peaks in the light curve with magnification factors of several thousands. If microlenses are ubiquitous, multiple peaks are expected hundreds or thousands of years before 
the background star crosses the main caustic. The number of peaks (and their intensity) depends on the density of microlenses and their masses. 
The constraint on flux conservation implies that during the crossing of the main caustic, the magnification is orders of magnitude smaller than for the smooth 
DM model when microlenses are disrupting the main caustic. Moderate disruption of cluster caustics increases the probability of detecting microlensing events of fainter background stars, while significant disruption produces microlensing events much farther away from the main CC that can be observed only when the background star is extremely luminous.\\
\noindent
$\bullet$ {\em Saturation and $1/f$ noise.}
When microlenses populate the lens plane, there is a region around the main CC where the optical depth exceeds the critical value of 1. The width of this region depends on the surface 
mass density of microlenses and the properties of the cluster deflection field. Within this region, the microcaustics overlap in the source plane and the magnification pattern is changed substantially. 
The power spectrum of the light curves shows features on larger timescales that resemble the ubiquitous $1/f$ noise present in self-organized criticality or renewal processes studied in 
other fields. \\
\noindent
$\bullet$ {\em Constraints on PBHs.}
The conditions required for the interior macro-image ($a_1<0$) to disappear during an extended period is that there exist low-magnification areas in the source plane. This effect was also predicted in earlier work, and it was shown to be important for low contributions from microlenses \citep[see, e.g.,][]{Schechter2002}. When enough massive microlenses are added in the lens plane, for instance if PBHs make up a significant fraction of the DM (larger than a few percent), it becomes harder to satisfy this condition since the microcaustics start to overlap on top of the low-magnification regions. 
Thus, the lack of extended periods where one of the macro-images {\it vanishes} can be used to set a limit on the fraction of DM in the form of microlenses (like PBHs). 
So far, the data provided by {\it Icarus} seem too sparse to place strong constraints on the fraction of DM that can be in a compact form (although they appear to favor a low fraction of DM as PBHs). Continuous monitoring of both {\it Icarus} and {\it Iapyx} in the near future will allow us to derive such strong constraints after the identification of additional peaks and the precise modeling of these peaks together with the extended periods between peaks. Interestingly, K17 report a possible nearby third event ({\it Perdix}, or LS1 / Lev~2017A) that could correspond to a different background star being lensed by the same web of caustics.
\\

\noindent
$\bullet$ {\em Complementary to QSO microlensing}. 
An obvious advantage of this type of observation when compared with QSO microlensing is that owing to the smaller intrinsic size of the background source (a few solar radii as opposed to $\sim0.01$\,pc ($\sim0.001$\,pc) for optical (X-ray) accretion disk) and the larger apparent motion of the macro-images, the timescale of a microlensing event is much shorter (by 3 or 4 orders of magnitude) and the magnifications during maxima are larger (by 1.5 or 2 orders of magnitude). 
Light curves spanning several years can intersect several microlenses and can be used to determine a census of microlenses. 
The constraints derived this way on the population of microlenses are then affected mostly by local substructure and less by projection effects 
acting on larger scales. 
\\
\noindent
$\bullet$ {\em Cosmic microscopes.} When the optical depth of microlenses is very small, a caustic crossing event can be used to probe very small masses near the CC. 
A microlens at cosmic distance with a mass similar to that of Jupiter can produce features in the light curve that can be observed with current technology provided the radius of the background star is sufficiently small. This opens the exciting possibility of probing compact DM in the low-mass regime, unreachable by other means, where at extreme magnifications of order $10^6$, a Jupiter-like microlens would behave as a microlens with a mass $\sim 1000$\,M$_{\odot}$. Monitoring (with {\it JWST}) of high-redhsift galaxies rich in luminous Pop~III stars that are intersecting a critical curve far away from the ICL are ideal targets for this kind of study \citep{Windhorst2018}.

\section{Acknowledgments}  

The authors wish to thank 
Juan M. Lopez, Miguel Angel Rodriguez, Gary Bernstein, Evencio Mediavilla, and the anonymous referee for inspiring discussions and useful comments. 
J.M.D. acknowledges the support of projects AYA2015-64508-P (MINECO/FEDER, UE), AYA2012-39475-C02-01, and the consolider project CSD2010-00064 funded by the Ministerio de Economia y Competitividad. 
J.M.D. acknowledges the hospitality of the Physics Department at the University of Pennsylvania for hosting him during the preparation of this work. T.J.B. gratefully acknowledges the Visiting Research Professor Scheme at the University of Hong Kong. M.O. is supported in part by World Premier International Research Center Initiative (WPI Initiative), MEXT, Japan, and JSPS KAKENHI Grant Numbers 26800093 and 15H05892. M.J. acknowledges support by the Science and Technology Facilities Council (grant number ST/L00075X/1). Support for S.R. through {\it HST} programs GO-13386, GO-14199, and GO-14208 was provided by NASA through grants from the Space Telescope Science Institute (STScI), which is operated by the Association of Universities for Research in Astronomy, Inc., under NASA contract NAS 5-26555.
A.V.F. and P.L.K. were supported by NASA/{\it HST} grants GO-14041, GO-14199, GO-14208, GO-14528, GO-14872, and GO-14922 from STScI; they are also grateful for financial assistance from the Christopher R. Redlich Fund, the TABASGO Foundation, and the Miller Institute for Basic Research in Science (U.C. Berkeley).

%

\appendix\label{sect_Appendix}
This Appendix presents the basic formalism for a single microlens near a CC. We consider the simple scenario where the deflection field from the cluster is oriented in the vertical direction ($y$ axis in the equations below). 
In this configuration, the CC from the cluster will be a horizontal line, the tangential magnification will be the magnification in the vertical direction, and the radial 
magnification will be given in the horizontal direction. A reader familiar with lensing models may find this definition of tangential and radial magnifications counterintuitive, 
since we are assigning the radial magnification to a direction that is tangential with respect to the source of the deflection field. This, however, is the case for some tangential 
arcs that stretch in directions perpendicular to the CC, like for instance the arc containing the Icarus event discussed by K17.

If we place a microlens (point source) with Einstein radius $r_e$ at a position ($x_o$, $y_o$), the deflection field around the point source 
is $\alpha_x=xr_e^2/r^2$ and $\alpha_y=yr_e^2/r^2$, where $r^2 = x^2 + y^2$ and locations $(x,y)$ have their origin at $(x_o,y_o)$. The mapping between the image 
plane and the source plane is no longer given by Eq.~\ref{Eq_beta_theta} (in the main text); instead, it can be described by 
\begin{equation}
\beta = 
\left( \begin{array}{c}
  (x+x_o)/\mu_r\\
  (y+y_o)^2/r_E \\ \end{array} \right)
- r_e^2 \left( \begin{array}{c}
  x/r^2 \\
  y/r^2 \\ \end{array} \right), 
\label{Eq_App1}
\end{equation}
where the first term accounts for the deflection field from the cluster ($r_E=\Theta$ and $y=\theta$ in Eq.~\ref{Eq_beta_theta}) and the second term corresponds to the deflection field from the point source. 
We have assumed that the deflection field of the cluster is such that it stretches the images in the $y$ direction by some large factor $\mu_t=1/a_1$ and in the $x$ direction by a smaller factor $\mu_r=1/a_2$ (i.e., $\mu = \mu_t\mu_r$). The constant $\mu_r$ is the small eigenvalue (for a tangential CC) of the magnification (i.e., $a_2=\mu_r^{-1}=1-\kappa+\gamma$). Similarly, $a_1=\mu_t^{-1}=1-\kappa-\gamma$. 
The inverse of the magnification, $\mu^{-1}$, can be computed as the determinant of the Jacobian between the source-plane positions and the image-plane positions:
\begin{equation}
\frac{d\beta}{d\theta} = 
\left( \begin{array}{cc}
a_2 + (x^2-y^2)\frac{r_e^2}{r^4}     &                   2xy\frac{r_e^2}{r^4}  \\
      2xy\frac{r_e^2}{r^4}           &  2\frac{(y+y_o)}{r_E}-(x^2-y^2)\frac{r_e^2}{r^4}  \\ \end{array} \right).
\end{equation}
Computing the determinant of the expression above results in
\begin{equation}
\mu^{-1}=a_1a_2 - (a_2-a_1)\frac{x^2-y^2}{x^2+y^2}\frac{r_e^2}{r^2} - \frac{r_e^4}{r^4},
\label{eq_Nick1}
\end{equation}
where $a_1=2(y+y_o)/r_E$. When there is no microlens ($r_e=0$), $\mu^{-1}$ is equal to the original $\mu^{-1}=\mu_t^{-1}\mu_r^{-1}=a_1a_2$. 
The term $(x^2-y^2)/(x^2+y^2)$ is equal to $\cos(2\delta)=Q$, where $\tan(\delta)=y/x$. $Q$ changes sign in different quadrants as shown in Fig.~\ref{Fig_NickEq}. 

\begin{figure}  
 \centerline{ \includegraphics[width=8cm]{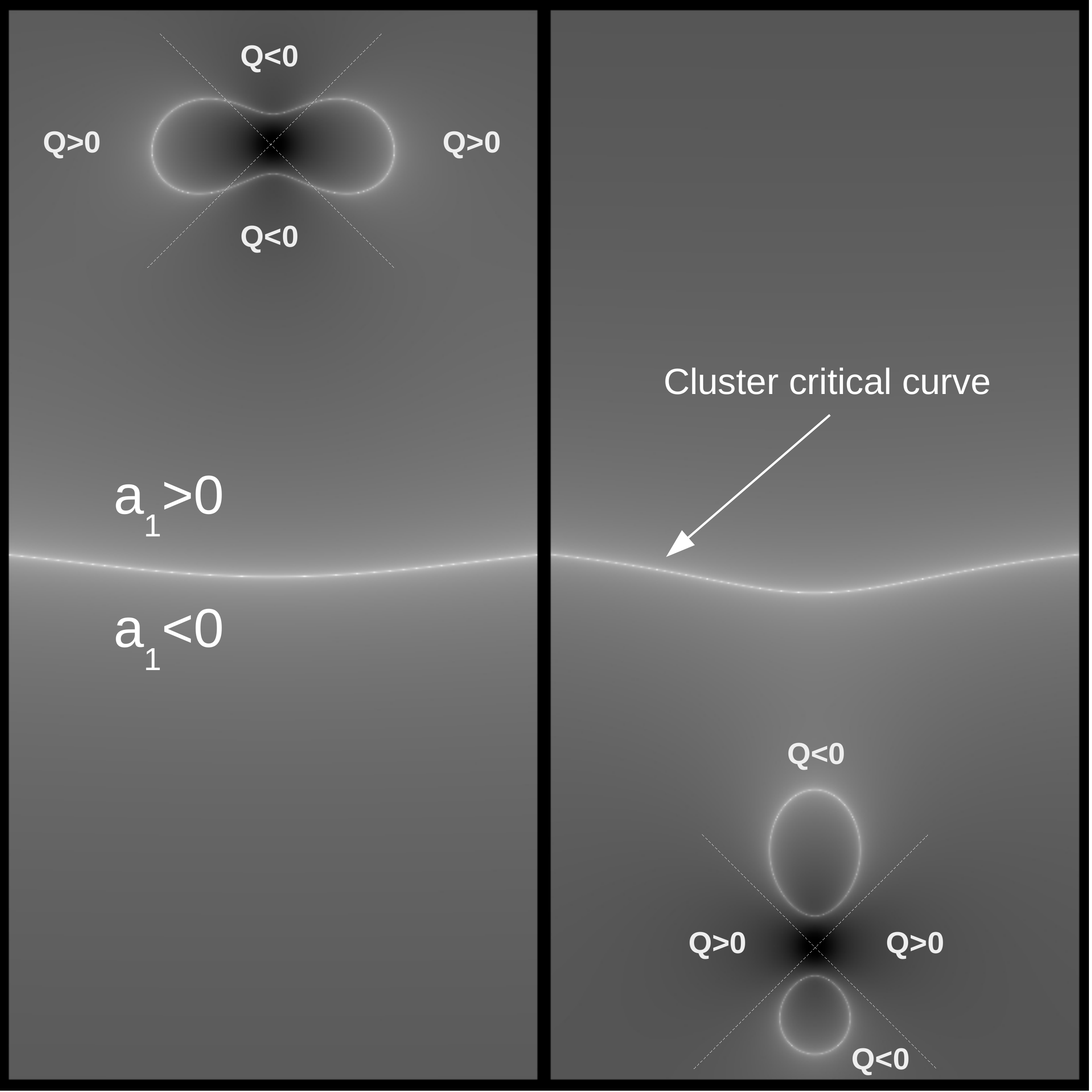}}
  \caption{Magnification expected for a microlens near a CC. The inverse of the magnification is given by Eq.~\ref{eq_Nick1}. 
           The left panel corresponds to the case  where the microlens is on the $a_1>0$ side (Icarus) discussed in the text, while the right panel is for the case where the 
           microlens is on the side with $a_1<0$ (Iapyx). In both panels, the CC from the cluster is shown as the (nearly) horizontal line. 
   } 
   \label{Fig_NickEq}  
\end{figure}  

Approximate solutions to Eq.~\ref{eq_Nick1} can be found after a few simple assumptions are made. Analytical solutions, even though approximate, are very useful 
for unveiling scalings with the microlens mass or distance to the CC of the cluster that can be exploited later in statistical analyses. 
To find an analytical solution of Eq.~\ref{eq_Nick1}, we assume that $a_1=2(y+y_o)/r_E \approx 2y_o/r_E={\rm constant}$. 
This approximation is valid when $y_o \gg y$, which corresponds to a distance from the cluster CC much larger than the Einstein radius of the microlens. 
A simple solution (CCs) can be found after setting this equation to zero, 
multiplying by $r^4$, and doing the variable change $x=r^2/r_e^2$.  
This results in the following general solution,  
\begin{equation}
r^2=r_e^2\frac{(a_2-a_1)Q}{2a_1a_2}\left[1\pm\sqrt{1+\frac{4a_1a_2}{(a_2-a_1)^2Q^2}}  \ \right].
\label{eq_Nick2}
\end{equation}
Depending on the values of $a_1$, $a_2$, and $Q$, the quantity $r^2$ can be positive or negative. If $r^2$ is negative there is no real solution. 
On each side of the CC, $a_1=1-\kappa-\gamma$ changes sign (and equals zero on the CC). The solutions of Eq.~\ref{eq_Nick1} can be divided 
into two sets, one having $a_1>0$ and the second with $a_1<0$. Below we discuss these solutions based on simple approximations.

For simplicity we assume that $\mu_r$ is constant and positive. Near a tangential CC, $\mu_r$ (or similarly $a_2$) changes very slowly around 
a given point, so this is a good approximation for our purposes.
Also, in our case (microlenses near a tangential CC), $\mu_t \gg \mu_r$ (or $a_2 \gg a_1$)  
and the term $(a_2-a_1)$ is always positive. For convenience we define 
\begin{equation} 
 F=1\pm\sqrt{1+4a_1a_2/((a_2-a_1)Q)^2}, 
\end{equation}
and in particular,
\begin{equation} 
 F_{+}=1+\sqrt{1+\frac{4a_1a_2}{(a_2-a_1)^2Q^2}} 
\end{equation}
and 
\begin{equation} 
 F_{-}=1-\sqrt{1+\frac{4a_1a_2}{(a_2-a_1)^2Q^2}}. 
\end{equation}
Since  $a_2 \gg a_1$, we can approximate $F_{+}\approx 2$ and  $F_{-} \approx -2a_1/(a_2Q^2)$. 
The term $Q=\cos(2\delta)$ can be either positive or negative depending on the quadrant: 
$Q<0$ in $\pi /4 < \delta < 3 \pi /4$ (hereafter quadrant 1) and $5\pi /4 < \delta  < 7\pi /4$ (hereafter quadrant 3), and  $Q>0$ in  $-\pi /4 < \delta < \pi /4$ (hereafter quadrant 4) 
and  $3\pi /4 < \delta <  5\pi /4$ (hereafter quadrant 2). For clarity we mark these quadrants in Fig.~\ref{Fig_NickEq}.  

\subsection{Solutions for $a_1>0$}
If $a_1>0$, then $Q$ and $F$ must be the same sign in order to have $r^2>0$. If $a_1>0$, then $F_{+}>0$ and  $F_{-}<0$. 
In quadrants 1 and 3, $Q<0$ and only the solution with $F_{-}$ is real. The corresponding solution is $r \approx r_e\sqrt{\mu_r/|Q|} \approx r_e$ for $Q \approx -1$. These solutions can be used to estimate the size of the critical curves, and the caustics, associated with the  microlenses and by extension to estimate the cross-section for microlensing \citep[see][]{Oguri2017}. 
In quadrants 2 and 4, $Q>0$ and only the solution with $F_{+}$ is real. The corresponding solution is $r \approx r_e\sqrt{\mu_t|Q|} \gg r_e$ for $Q \approx 1$. 

When $a_1>0$, the micro-CC reaches a maximum distance much larger than $r_e$ when $Q=1$ (horizontal direction or $\delta \approx 0, \delta \approx \pi$) and decreases toward a distance comparable to $r_e$ at $Q=-1$ (vertical direction or $\delta \approx \pm \pi/$2). For smaller values of $|Q|$, $r$ moves between these two extreme values. 
The CC in this case resembles an hourglass on its side.  

\subsection{Solutions for $a_1<0$}
If $a_1<0$, both  $F_{+}$ and  $F_{-}$ are positive, so if $Q>0$ there is no real solution since $r^2<0$. 
A solution exists only in quadrants 1 and 3 where $Q<0$. In this case ($Q<0$) there are two solutions, one for $F_{+}$ and one for $F_{-}$. 
The solutions have magnitudes similar to the case  $a_1>0$ discussed above --- that is, $r_{+} \approx r_e\sqrt{\mu_r/|Q|} \approx r_e$ and $r_{-} \approx r_e\sqrt{\mu_t|Q|} \gg r_e$ (for  $Q \approx -1$). 

When $a_1<0$, there are no CCs in quadrants 2 and 4, and the CC is oriented in the direction of the deflection field (i.e., in the vertical direction in our configuration). 
In quadrants 3 and 4 there are two solutions, and for $Q\approx-1$ the smallest solution is close to the position of the microlens ($r \approx r_e$) while the larger solution extends much farther 
($r \gg r_e$). For smaller values of $|Q|$ there are still two solutions, but the separation between them is smaller.
The CC in this case resembles an hourglass standing up. 
The hourglass configuration is also found in earlier work; see \citet{Chang1984}, \citet{Mao1998}, and also Fig. 8.8 of \cite{SchneiderBook}. The same early work shows how the caustics are different for each parity. Caustics for a microlens on the side with positive party ($a_1>0$) resemble the usual diamond shape but significantly stretched in the direction of the main caustic. On the side with negative parity, the caustics develop a gap in the central region of the diamond shape. This gap is responsible for the low magnification periods observed in the counterimage with negative parity (Iapyx). 

The approximate solutions shown above can be tested by solving Eq.~\ref{eq_Nick1} numerically, and without any approximations. 
The magnification $\mu$ (Eq.~\ref{eq_Nick1}) is visualized graphically in Fig.~\ref{Fig_NickEq} for a configuration where $\mu_t=5$ and $r_E/r_e=5000$. 
The two hourglass shapes are evident, as well as the lack of solution in quadrants 1 and 3 when $a_1<0$. 

It is interesting to highlight the scaling of the solutions above with the magnification. The maximum dimension of the CC is  $r \approx r_e\sqrt{\mu_t}$. 
This is the same result we found in Section \ref{sect_S2} (see Eq.~\ref{Eq_thetaE}), where we show how a microlens with mass $M$ behaves like an microlens with mass $M\mu_t$. 

As noted by \cite{Saha2011}, the minima in the arrival-time surface cannot be demagnified, while this is possible in a saddle point (i.e., on the side where $a_1<0$). 
Even though the magnification pattern can be very complex when the critical curves, or caustics, of microlenses start to overlap, the properties and scalings presented above for a single microlens are still accurate provided the optical depth for lensing is not very high. We find that at optical depths similar to the ones found in the Icarus and Iapyx events, the single-microlens formalism presented above is still accurate for describing individual events.

\newpage


\begin{figure}
\centerline{ \includegraphics[width=18cm]{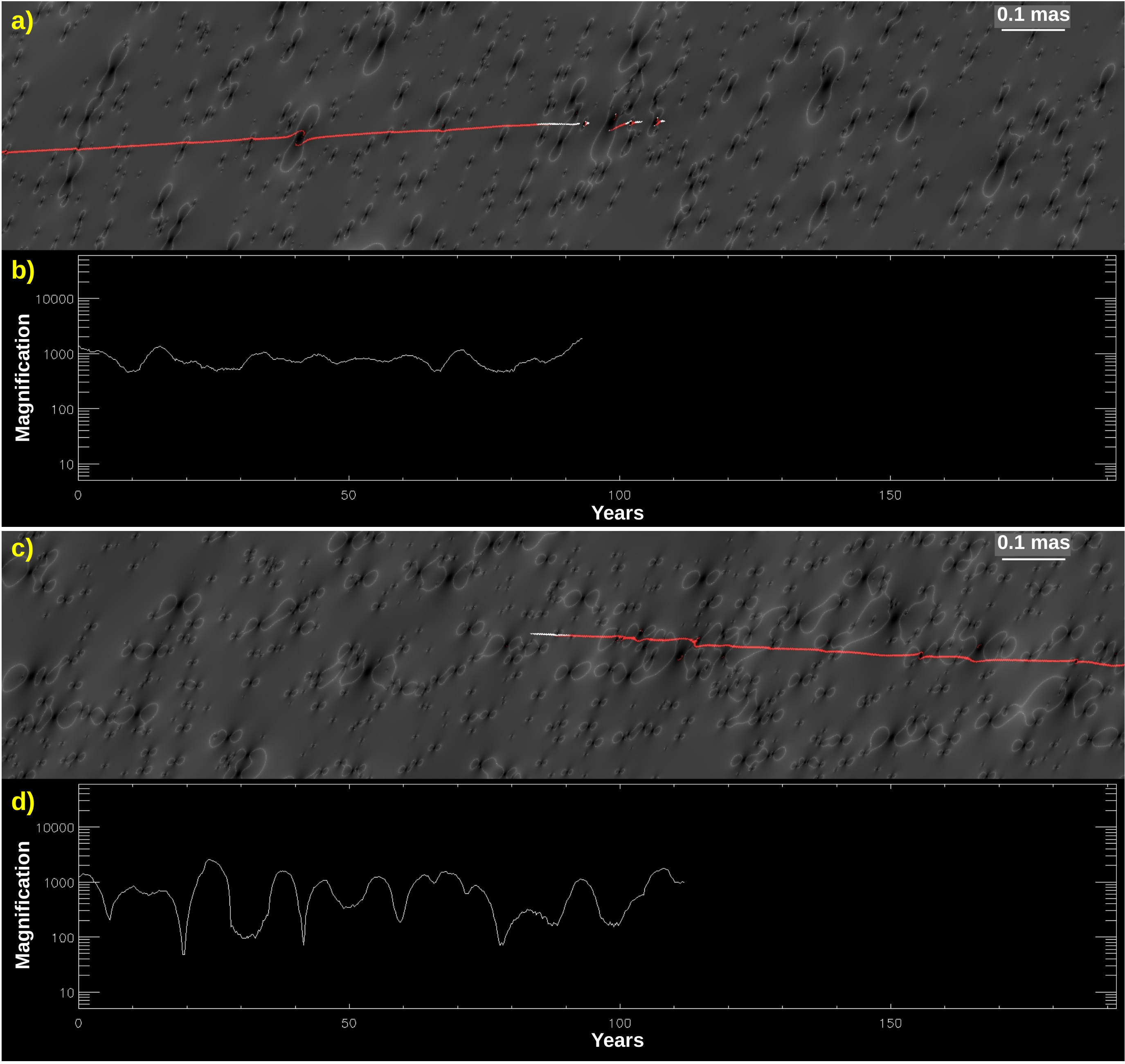}}
\caption{Formation and destruction of counterimages (and corresponding magnification) produced by a moving star behind a field of micolenses near the critical curve of the cluster. The top two panels (a and b) correspond to counterimages on the side with positive parity. The bottom two panels (c and d) are for counterimages in the side with negative parity. The microlenses are stars and remnants from the intracluster medium. In order to better visualize the counterimages, the size of the background star has a very large radius of $7\times10^4 {\rm R}_{\odot}$. The star is moving towards the cluster main caustic at a relative velocity of 1000 km s$^{-1}$. The animation runs from 0 to 190 years and has a duration of 1 minute and 7 seconds.}
\label{fig_movie1}
\end{figure}

\section{Animations}\label{sect_animations}
This section includes various animations of a background star moving between the network of caustics. Figure~\ref{fig_movie2} and Fig.~\ref{fig_movie1} show how counterimages are being formed and destroyed as a very large background star moves with 1000 km s$^{-1}$ relative velocity towards the main caustic. 
The radius of the star is very large ($70000 {\rm R}_{\odot}$) in order to better visualize the counterimages that otherwise would be unresolved at this resolution. This large radius is comparable to the X-ray emitting region of a QSO so the movies are useful also to see how these QSO regions would form and disappear if they could be resolved. Figure~\ref{fig_movie1} is for a case where the microlenses are the stars and remnants from the intracluster medium.  Figure~\ref{fig_movie2} adds to these microlenses PBH with 30 ${\rm M}_{\odot}$. The surface mass density of the PBH makes up to 1\% of the total surface mass density of the cluster. In both cases, the top panel shows counterimages in the side with positive parity and the bottom panel is for counterimages in the side with negative parity. Note how in the case where PBH are present counterimages can form relatively far away from each other. Future observations with milliarcsec or sub-milliarcsec resolution can potentially resolve the different locations and set
strong limits on the fraction of primordial black holes. 

\begin{figure}
\centerline{ \includegraphics[width=18cm]{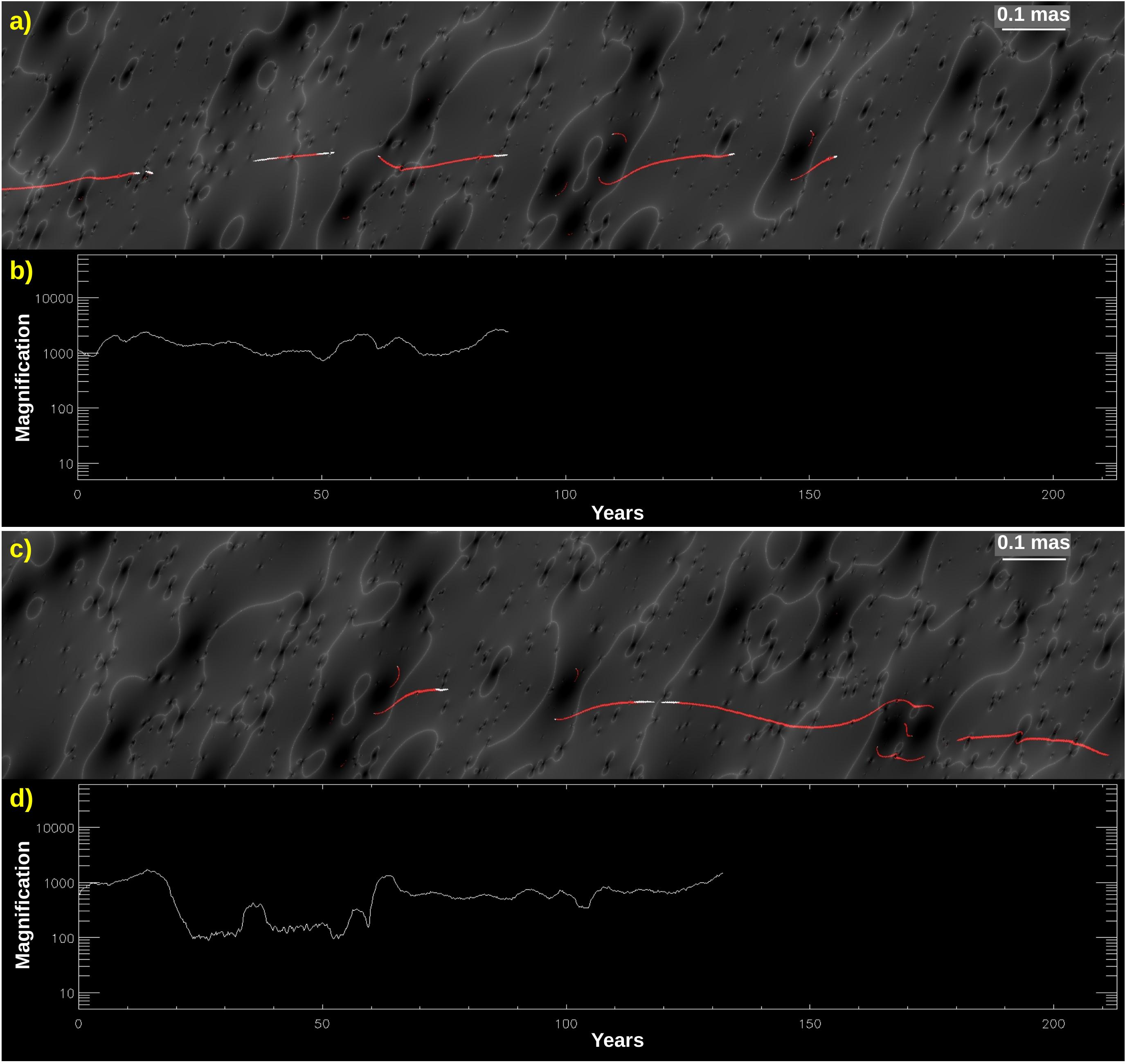}}
\caption{Like in the previous figure but, in addition to the microlenses from the intracluster medium, the simulation contains also a number of PBH equivalent to 1\% of the total mass with a mass per PBH of 30 ${\rm M}_{\odot}$.  The animation runs from 0 to 210 years and has a duration of 1 minute and 14 seconds.}
\label{fig_movie2}
\end{figure}

\newpage

\bibliographystyle{aasjournal} 
\bibliography{MyBiblio} 

\end{document}